\documentclass[12pt]{article} 
\usepackage{amsmath,amssymb,mathrsfs,bbm,bm,hyperref}
\usepackage{comment}
\usepackage[sort]{natbib}
\usepackage[usenames, dvipsnames]{color}
\usepackage{float}
\hypersetup{pdfborder = {0 0 0},colorlinks=true,linkcolor=blue,citecolor=blue,urlcolor=blue}
\usepackage{multirow}
\usepackage{setspace}
\usepackage{subcaption}
\usepackage{booktabs} 
\usepackage{fancyvrb}
\usepackage[framemethod=TikZ]{mdframed}
\usepackage{fancyhdr}
\usepackage[margin = 1in]{geometry}
\usepackage[space]{grffile}
\usepackage[nottoc]{tocbibind}

\usepackage[fit]{truncate} 
\numberwithin{equation}{section}

\newcommand{\E}{\mathbb{E}}
\renewcommand{\P}{\mathbb{P}}
\newcommand{\V}{\mathbb{V}}
\newcommand{\I}{\mathbbm{1}}

\newcommand{\bgamma}{\boldsymbol{\gamma}}

\newcommand{\markedsection}[2]{\section[#2]{#2%
\sectionmark{#1}}
\sectionmark{#1}}

\newcommand{\C}{c}

\newcommand{\T}{T}

\newcounter{cajita}[section]
\renewcommand{\thecajita}{\arabic{section}.\arabic{cajita}}

\newcommand{\pathInputs}{./inputs}

\newcounter{snippet}
\newcounter{figuras}
\newcounter{tablas}[section]

\newcommand{\labelsnippet}[1]{%
  \newcounter{#1}%
  \setcounter{#1}{\value{snippet}}%
  \stepcounter{snippet}%
}%

\newcommand{\labelfiguras}[1]{%
  \newcounter{#1}%
  \setcounter{#1}{\value{figuras}}%
  \stepcounter{figuras}%
}%

\newcommand{\labeltablas}[1]{%
  \newcounter{#1}%
  \setcounter{#1}{\value{tablas}}%
  \stepcounter{tablas}%
}%

\usepackage{listings}
\definecolor{codegreen}{rgb}{0,0.6,0}
\definecolor{codegray}{rgb}{0.5,0.5,0.5}
\definecolor{codepurple}{rgb}{0.58,0,0.82}
\definecolor{backcolour}{rgb}{0.95,0.95,0.95}
\definecolor{maincolour}{rgb}{0,0.1,0.1}

\lstdefinestyle{mystyle}{
	language=R,
    backgroundcolor=\color{backcolour},   
    commentstyle=\color{codegreen},
    keywordstyle=\color{magenta},
    keywords={},
    numberstyle=\tiny\color{codegray},
    stringstyle=\color{codepurple},
    basicstyle=\footnotesize\ttfamily\color{maincolour},
    breakatwhitespace=false,         
    breaklines=true,                 
    captionpos=b,                    
    keepspaces=true,                 
    numbers=none,                    
    numbersep=5pt,                  
    showspaces=false,                
    showstringspaces=false,
    showtabs=false,
    tabsize=2,
		xleftmargin=-15pt
}
\newcommand{\repliWeb}{https://sites.google.com/site/rdpackages/replication/cit-2018-cup/snippets}

\newcommand{\rsnip}[2] {
\VerbatimInput[frame=single,rulecolor=\color{JungleGreen},framesep=1.5mm, framerule=0.5mm,label=\fbox{#2}, fontsize=\small, xleftmargin=-0.6cm,xrightmargin=-0.6cm,baselinestretch=1.1,samepage=TRUE]{\pathInputs/#1}
}

\newcommand{\statasnip}[2] {
\VerbatimInput[frame=single,rulecolor=\color{Rhodamine},framesep=1.5mm, framerule=0.5mm, label=\fbox{#2}, fontsize=\small, xleftmargin=-0.6cm,xrightmargin=-0.6cm,baselinestretch=1.1,samepage=TRUE]{\pathInputs/#1-COMMAND-ONLY.txt}
}	

\newcommand{\Rlink}[1]{
R Snippet #1
}

\newcommand{\Slink}[1]{
Stata Snippet #1
}

\begin{document}
\pagestyle{fancy}
\frenchspacing
\onehalfspacing

\begin{titlepage}
  \title{\bf A Practical Introduction to Regression Discontinuity Designs: Foundations}
  \author{
    Matias D. Cattaneo\thanks{Department of Operations Research and Financial Engineering, Princeton University.}
    \and Nicol\'as Idrobo\thanks{Department of Political Science, University of Pennsylvania.}
    \and Roc\'{i}o Titiunik\thanks{Department of Politics, Princeton University.}
}

\date{\large{\vspace{0.5in} \today}}
\end{titlepage}
\maketitle

\thispagestyle{empty}

\begin{center}
Element prepared for\medskip\\
\textit{Cambridge Elements: Quantitative and Computational Methods for Social Science}\medskip\\
Cambridge University Press\bigskip\\
Published version:\medskip\\
\url{https://doi.org/10.1017/9781108684606}

\end{center}
\thispagestyle{empty}

\newpage
\begin{abstract}
	In this Element and its accompanying Element, Matias D. 
	Cattaneo, Nicol\'{a}s Idrobo, and Roc\'{i}o Titiunik provide an accessible and practical guide for the analysis and interpretation of Regression Discontinuity (RD) designs that encourages the use of a common set of practices and facilitates the accumulation of RD-based empirical evidence. In this Element, the authors discuss the foundations of the canonical Sharp RD design, which has the following features: (i) the score is continuously distributed and has only one dimension, (ii) there is only one cutoff, and (iii) compliance with the treatment assignment is perfect. In the accompanying Element, the authors discuss practical and conceptual extensions to the basic RD setup.
	
\end{abstract}

\thispagestyle{empty}

\newpage\pagenumbering{roman}\setcounter{page}{1}
{\singlespacing\tableofcontents}

\newpage\pagenumbering{arabic}\setcounter{page}{1}
\setcounter{secnumdepth}{0}\section*{Acknowledgments}\setcounter{secnumdepth}{3}

This Element, together with its accompanying Element (\textit{A Practical Introduction to Regression Discontinuity Designs: Extensions}, \citeauthor*{Cattaneo-Idrobo-Titiunik_2019_Vol2}), collects and expands the instructional materials we prepared for more than $30$ short courses and workshops on Regression Discontinuity (RD) methodology taught over the years 2014--2018. These teaching materials were used at various institutions and programs, including the Asian Development Bank, the Philippine Institute for Development Studies, the International Food Policy Research Institute, the ICPSR's Summer Program in Quantitative Methods of Social Research, the Abdul Latif Jameel Poverty Action Lab, the Inter-American Development Bank, the Georgetown Center for Econometric Practice, and the Universidad Cat\'{o}lica del Uruguay's Winter School in Methodology and Data Analysis. The materials were also employed for teaching at the undergraduate and graduate level at Brigham Young University, Cornell University, Instituto Tecnol\'ogico Aut\'onomo de M\'exico, Pennsylvania State University, Pontificia Universidad Cat\'{o}lica de Chile, University of Michigan, and  Universidad Torcuato Di Tella. We thank all these institutions and programs, as well as their many audiences, for the support, feedback, and encouragement we received over the years.

The work collected in our two Elements evolved and benefited from many insightful discussions with our present and former collaborators: Sebasti\'an Calonico, Robert Erikson, Juan Carlos Escanciano, Max Farrell, Yingjie Feng, Brigham Frandsen, Sebasti\'an Galiani, Michael Jansson, Luke Keele, Marko Kla\v{s}nja, Xinwei Ma, Kenichi Nagasawa, Brendan Nyhan, Jasjeet Sekhon, Gonzalo Vazquez-Bare, and Jos\'e Zubizarreta. Their intellectual contribution to our research program on RD designs has been invaluable, and certainly made our Elements much better than they would have been otherwise. We also thank Alberto Abadie, Joshua Angrist, Ivan Canay, Richard Crump, David Drukker, Sebastian Galiani, Guido Imbens, Patrick Kline, Justin McCrary, David McKenzie, Douglas Miller, Aniceto Orbeta, Zhuan Pei, and Andres Santos for the many stimulating discussions and criticisms we received from them over the years, which also shaped the work presented here in important ways. The co-editors Michael Alvarez and Nathaniel Beck offered useful and constructive comments on a preliminary draft of our manuscript, including the suggestion of splitting the content into two stand-alone Elements. We also thank the anonymous reviewers, who provided very valuable feedback. Last but not least, we gratefully acknowledge the support of the National Science Foundation through grant \href{http://www.nsf.gov/awardsearch/showAward?AWD_ID=1357561}{SES-1357561}.

The goal of our two Elements is purposely practical and hence we focus on the empirical analysis of RD designs. We do not seek to provide a comprehensive literature review on RD designs nor discuss theoretical aspects in detail. In this first Element, we employ the data of \citet*{Meyersson_2014_ECMA} as the main running example for empirical illustration. We thank this author for making his data and codes publicly available. We provide complete replication codes in both \texttt{R} and \texttt{Stata} for the entire empirical analysis discussed throughout the Element and, in addition, we provide replication codes for a second empirical illustration using the data of \citet*{Cattaneo-Frandsen-Titiunik_2015_JCI}. This second empirical example is not discussed in the text to conserve space, and because it is already analyzed in our companion software articles. 

\texttt{R} and \texttt{Stata} scripts replicating all the numerical results are available at \url{http://www.cambridge.org/introRDD}, and can be run interactively on-line via \textsf{CODE-OCEAN} (hyperlinks for each chapter are given below). Finally, the latest version of the general-purpose, open-source software we use, as well as other related materials, can be found at:
\begin{center}\url{https://sites.google.com/site/rdpackages/}\end{center}

\newpage
\section{Introduction}

An important goal in the social sciences is to understand the causal effect of a treatment on outcomes of interest. As social scientists, we are interested in questions as varied as the effect of minimum wage increases on unemployment, the role of information dissemination on political participation, the impact of educational reforms on student achievement, and the effects of conditional cash transfers on children's health. The analysis of such effects is relatively straightforward when the treatment of interest is randomly assigned, as this ensures the comparability of units assigned to the treatment and control conditions. However, by its very nature, many interventions of interest to social scientists cannot be randomly assigned for either ethical or practical reasons---often both.

In the absence of randomized treatment assignment, research designs that allow for the rigorous study of non-experimental interventions are particularly promising. One of these is the Regression Discontinuity (RD) design, which has emerged as one of the most credible non-experimental strategies for the analysis of causal effects. In the RD design, all units have a score, and a treatment is assigned to those units whose value of the score exceeds a known cutoff or threshold, and not assigned to those units whose value of the score is below the cutoff. The key feature of the design is that the probability of receiving the treatment changes abruptly at the known threshold. If units are unable to perfectly ``sort'' around this threshold, the discontinuous change in this probability can be used to learn about the local causal effect of the treatment on an outcome of interest, because units with scores barely below the cutoff can be used as a comparison group for units with scores barely above it.

The first step to employ the RD design in practice is to learn how to recognize it. There are three fundamental components in the RD design---a score, a cutoff, and a treatment. Without these three basic defining features, RD methodology cannot be employed. Therefore, an RD analysis is not always applicable to data, unlike other non-experimental methods such as those based on regression adjustments or more sophisticated selection-on-observables approaches, which can always be used to describe the conditional relationship between outcomes and treatments. The difference arises because RD is a research design, not an estimation strategy. In order to study causal effects with an RD design, the score, treatment, and cutoff must exist and be well defined, and the relationship between them must satisfy particular conditions that are objective and verifiable. The key defining feature of any canonical RD design is that the probability of treatment assignment as a function of the score changes discontinuously at the cutoff---a condition that is directly testable. In addition, the RD design comes with an extensive array of falsification tests and related empirical approaches that can be used to offer empirical support for its validity, enhancing the credibility of particular applications. These features give the RD design an objective basis for implementation and testing that is usually lacking in other non-experimental empirical strategies, and endow it with superior credibility among non-experimental methods.

The popularity of the RD design has grown markedly over recent decades, and it is now used frequently in Economics, Political Science, Education, Epidemiology, Criminology, and many other disciplines. The RD design is also commonly used for impact and policy evaluation outside academia. This recent proliferation of RD applications has been accompanied by great disparities in how RD analysis is implemented, interpreted, and evaluated. RD applications often differ significantly in how authors estimate the effects of interest, make statistical inferences, present their results, evaluate the plausibility of the underlying assumptions, and interpret the estimated effects. The lack of consensus about best practices for validation, estimation, inference, and interpretation of RD results makes it hard for scholars and policy-makers to judge the plausibility of the evidence and to compare results from different RD studies. 

In both this Element and the accompanying Element, \textit{A Practical Introduction to Regression Discontinuity Designs: Extensions} (\citeauthor*{Cattaneo-Idrobo-Titiunik_2019_Vol2}, forthcoming), our goal is to provide an accessible and practical guide for the analysis and interpretation of RD designs that encourages the use of a common set of practices and facilitates the accumulation of RD-based empirical evidence. In this Element, our focus is on the canonical RD setup that has the following features: (i) the score is continuously distributed and has only one dimension, (ii) there is only one cutoff, and (iii) compliance with treatment assignment is perfect, i.e., all units with score equal to or greater than the cutoff actually receive the treatment, and all units with score below the cutoff fail to receive the treatment and instead receive the control condition. We call this setup the Sharp RD design, and assume it throughout this Element. In the accompanying Element, we discuss extensions and departures from the basic Sharp RD design, including Fuzzy RD designs where compliance is imperfect, RD designs with multiple cutoffs, RD designs with multiple scores, geographic RD designs, and RD designs with discrete running variables.

In addition to the existence of a treatment assignment rule based on a score and a cutoff, the formal interpretation, estimation, and inference of RD treatment effects requires several other assumptions. First, we need to define the parameter of interest and provide assumptions under which this parameter is identifiable, i.e., conditions under which it is uniquely estimable in some objective sense (finite sample or super-population). Second, we must impose additional assumptions to ensure that the parameter can be estimated; these assumptions will naturally vary according to the estimation/inference method employed and the parameter under consideration. There are two main frameworks for RD analysis, one based on continuity assumptions and another based on local randomization assumptions. Each of these defines different parameters of interest, relies on different identification assumptions, and employs different estimation and inference methods. These two alternative frameworks also generate different testable implications, which can be used to assess their validity in specific applications; see \citet*{Cattaneo-Titiunik-VazquezBare_2017_JPAM} for more discussion.

In this Element, we discuss the standard or \textit{continuity-based} framework for RD analysis. This approach is based on conditions that ensure the smoothness of the regression functions, and is the framework most commonly employed in practice. We discuss the alternative \textit{local randomization} framework for RD analysis in the second Element. This latter approach is based on conditions that ensure that the treatment can be interpreted as being randomly assigned for units near the cutoff. Both the continuity-based approach and the local randomization approach rely on the assumption that units that receive very similar score values on opposite sides of the cutoff are comparable to each other in all relevant aspects, except for their treatment status. The main distinction between these frameworks is how the idea of comparability is formalized: in the continuity-based framework, comparability is conceptualized as continuity of average (or some other feature of) potential outcomes near the cutoff, while in the local randomization framework, comparability is conceptualized as conditions that mimic a randomized experiment in a neighborhood around the cutoff.

Our upcoming discussion of the continuity-based approach focuses on the required assumptions, the adequate interpretation of the target parameters, the graphical illustration of the design, the appropriate methods to estimate treatment effects and conduct statistical inference, and the available strategies to evaluate the plausibility of the design. Our presentation of the topics is intentionally geared towards practitioners: our main goal is to clarify conceptual issues in the analysis of RD designs, and offer an accessible guide for applied researchers and policy-makers who wish to implement RD analysis. For this reason, we omit most technical discussions, but provide references for the technically inclined reader at the end of each section.

To ensure that our discussion is most useful to practitioners, we illustrate all methods by revisiting a study conducted by \citet*{Meyersson_2014_ECMA}, who analyzed the effect of Islamic political representation in Turkey's municipal elections on the educational attainment of women. The score in this RD design is the margin of victory of the largest Islamic party in the municipality, a (nearly) continuous random variable, which makes the example suitable to illustrate both the continuity-based methods in this Element, and also the local randomization methods in our second Element.

All the RD methods we discuss and illustrate are implemented using various general-purpose software packages, which are free and available for both \texttt{R} and \texttt{Stata}, two leading statistical software environments widely used in the social sciences. Each numerical illustration we present includes an \texttt{R} command with its output, and the analogous \texttt{Stata} command that reproduces the same analysis, though we omit the \texttt{Stata} output to avoid repetition. The local polynomial methods for continuity-based RD analysis are implemented in the package \texttt{rdrobust}, which is presented and illustrated in three companion software articles: \citet*{Calonico-Cattaneo-Titiunik_2014_Stata}, \citet*{Calonico-Cattaneo-Titiunik_2015_R} and \citet*{Calonico-Cattaneo-Farrell-Titiunik_2017_Stata}. This package has three functions specifically designed for continuity-based RD analysis: \texttt{rdbwselect} for data-driven bandwidth selection methods, \texttt{rdrobust} for local polynomial point estimation and inference, and \texttt{rdplot} for graphical RD analysis. In addition, the package \texttt{rddensity}, discussed by \citet*{Cattaneo-Jansson-Ma_2018_Stata}, provides manipulation tests of density discontinuity based on local polynomial density estimation methods. The accompanying package \texttt{rdlocrand}, which is presented and illustrated by \citet*{Cattaneo-Titiunik-VazquezBare_2016_Stata}, implements the local randomization methods discussed in the second Element.

\texttt{R} and \texttt{Stata} software, replication codes, and other supplementary materials, are available at \url{https://sites.google.com/site/rdpackages/}. In that website, we also provide replication codes for two other empirical applications, both following closely our discussion. One employs the data on US Senate incumbency advantage originally analyzed by \citet*{Cattaneo-Frandsen-Titiunik_2015_JCI}, while the other uses the Head Start data originally analyzed by \citet*{Ludwig-Miller_2007_QJE} and employed in \citet*{Cattaneo-Titiunik-VazquezBare_2017_JPAM}. Furthermore, a third distinct empirical illustration of the methods discussed in this Element, using the data of \citet*{Klasnja-Titiunik_2017_APSR}, is also available, and further discussed in \citet*{Cattaneo-Titiunik-VazquezBare_2019_Sage}.

To conclude, we emphasize that our main goal is to provide a succinct practical guide for empirical RD analysis, not to offer a comprehensive review of the literature on RD methodology---though we do offer references after each topic is presented for those interested in further reading. For early review articles see \citet*{Imbens-Lemieux_2008_JoE} and \citet*{Lee-Lemieux_2010_JEL}, and for an edited volume with a contemporaneous overview of the RD literature see \citet*{Cattaneo-Escanciano_2017_AIE}. We are currently working on a literature review  \citep*{Cattaneo-Titiunik_2019_RDreview} that complements these two practical Elements. See also \citet{Abadie-Cattaneo_2018_ARE} for an overview of program evaluation methods, and further references on RD designs.

\newpage
\section{The Sharp RD Design}
\label{sec:recognize}
\setcounter{figuras}{1}
\setcounter{snippet}{1}
\setcounter{tablas}{1}

In the RD design, all units in the study receive a \textit{score} (also known as \textit{running variable}, \textit{forcing variable}, or \textit{index}), and a treatment is assigned to those units whose score is above a known cutoff and not assigned to those units whose score is below the cutoff. Our running example is based on the study by \citet*{Meyersson_2014_ECMA}, who explored the effect of Islamic political representation in Turkey's municipal elections on the educational attainment of women. In this study, the units are municipalities and the score is the margin of victory of the (largest) Islamic party in the $1994$ Turkish mayoral elections. The treatment is the Islamic party's electoral victory, and the cutoff is zero: municipalities elect an Islamic mayor when the Islamic vote margin is above zero, and elect a secular mayor otherwise. 

These three components---score, cutoff, and treatment---define the RD design in general, and characterize its most important feature: in the RD design, unlike in other non-experimental studies, the assignment of the treatment follows a rule that is known (at least to the researcher) and hence empirically verifiable. To formalize, we assume that there are $n$ units, indexed by $i=1,2,\ldots,n$, each unit has a score or running variable $X_i$, and $\C$ is a known cutoff. Units with $X_i \geq \C$ are assigned to the treatment condition, and units with $X_i < \C$ are assigned to the control condition. This treatment assignment, denoted $\T_i$, is defined as $\T_i=\I(X_i \geq \C)$, where $\I(\cdot)$ is the indicator function, and it implies that the probability of treatment assignment as a function of the score changes discontinuously at the cutoff. 

Being \textit{assigned} to the treatment condition, however, is not the same as \textit{receiving} or \textit{complying with} the treatment. As in experimental and other non-experimental settings, this distinction is important in RD designs because non-compliance introduces complications and typically requires stronger assumptions to learn about treatment effects of interest. Following prior literature, we call Sharp RD design any RD design where the treatment condition assigned is identical to the treatment condition actually received for all units. Any RD design where compliance with treatment assignment is imperfect is referred to as Fuzzy RD design. In this Element, we focus exclusively on the Sharp RD design with a single score and a single cutoff. In the second Element (\textit{A Practical Introduction to Regression Discontinuity Designs: Extensions}, \citeauthor*{Cattaneo-Idrobo-Titiunik_2019_Vol2}, forthcoming), we discuss and illustrate the Fuzzy RD design, extending the basic Sharp RD setup to settings where compliance with treatment is imperfect. (The second Element also discusses other extensions, including settings with multiple scores and multiple cutoffs.)

\begin{figure}[ht]
\vspace{-0.2in}
\begin{center}
		\includegraphics[scale=0.70]{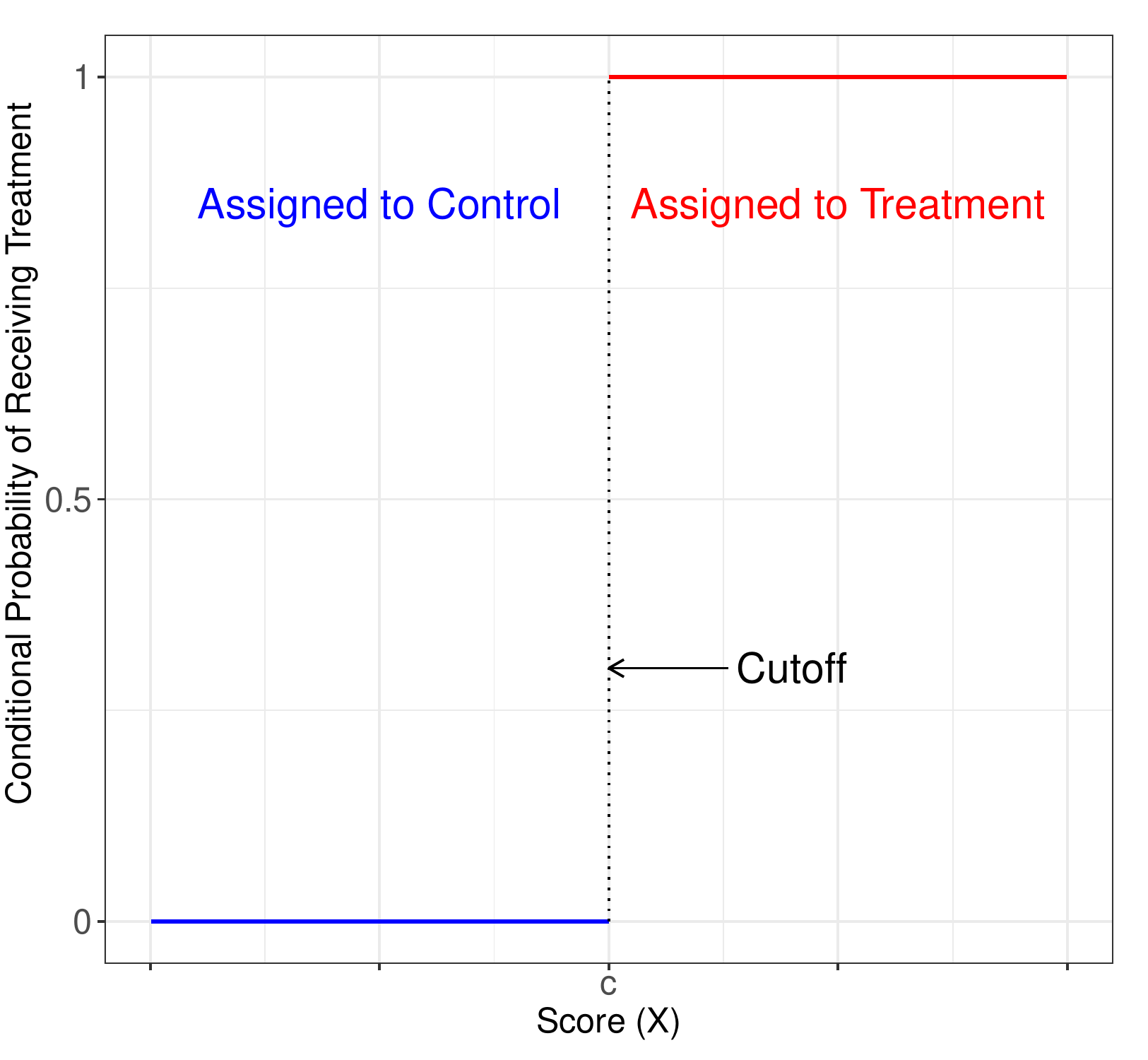}
\end{center}
\caption{Conditional Probability of Receiving Treatment in the Sharp RD Design}
\label{fig:sharpRD}
\end{figure}

Regardless of whether we have perfect or imperfect compliance, a defining feature of all RD designs is that the conditional probability of actually receiving treatment given the score changes discontinuously at the cutoff. We illustrate this for the Sharp RD design in Figure \ref{fig:sharpRD}, where we plot the conditional probability of receiving treatment given the score, $\P(\T_i=1 | X_i = x)$, for different values of the running variable $X_i$. As shown in the figure, in a Sharp RD design, this probability changes exactly from zero to one at the cutoff. Since in the Sharp RD design treatment assigned and treatment received are identical, this figure reflects both treatment assignment and treatment take-up.\\

\labelfiguras{figA}

Although it is common to use the language of experimental methods and talk about the RD treatment ``assignment,'' in some RD applications units find themselves in different circumstances depending on their score value, and it is only ex post that the researcher interprets one of those circumstances as a treatment and the other as a control condition. This is different from an experiment, where the treatment and control conditions are always defined ex ante by the researcher, and units are explicitly assigned to one of these conditions. For example, in the RD design studied by \cite{Meyersson_2014_ECMA}, a municipality is treated when it elects a mayor from an Islamic party, and control when it elects a mayor from a secular party. In this case, there is no explicit assignment of municipalities to different ex ante experimental conditions; rather, depending on the outcome of the election, municipalities find themselves in different situations (with or without an Islamic mayor), which can be understood as treatment versus control for some analytic purposes. These conceptual distinctions between experimental and RD treatment assignments do not affect the validity of the RD mathematical expressions. But the reader should keep in mind these caveats when we employ the term ``treatment assignment'' in the RD context. 

Following the causal inference literature, we adopt the potential outcomes framework and assume that each unit has two potential outcomes, $Y_i(1)$ and $Y_i(0)$, corresponding, respectively, to the outcomes that would be observed under the treatment or control conditions. In this framework, treatment effects are defined in terms of comparisons between features of (the distribution of) both potential outcomes, such as their means, variances or quantiles. Although every unit is assumed to have both $Y_i(1)$ and $Y_i(0)$, these outcomes are called potential because only one of them is observed. If unit $i$ receives the treatment, we will observe $Y_i(1)$, the unit's outcome under treatment, and $Y_i(0)$ will remain latent or unobserved. Similarly, if $i$ receives the control condition, we will observe $Y_i(0)$ but not $Y_i(1)$. This results in the so-called fundamental problem of causal inference, and implies that the treatment effect at the individual level is fundamentally unknowable.

The observed outcome is
\begin{equation*}
Y_i = (1-\T_i) \cdot Y_i(0) + \T_i \cdot Y_i(1) =
    \left\{ 
    \begin{array}{ccl}
    Y_i(0) && \text{if } X_i<\C \\ 
    Y_i(1) && \text{if } X_i\geq \C. \\ 
    \end{array}
\right.
\end{equation*}
Throughout this Element, we adopt the usual econometric perspective that sees the data $(Y_i,X_i)_{i=1}^n$ as a random sample from a larger population, taking the potential outcomes $(Y_i(1),Y_i(0))_{i=1}^n$ as random variables. We consider an alternative perspective in the second Element when we discuss inference in the context of the local randomization RD framework.

\begin{figure}[ht]
\centering
\vspace{-0.15in}
\includegraphics[scale=0.70]{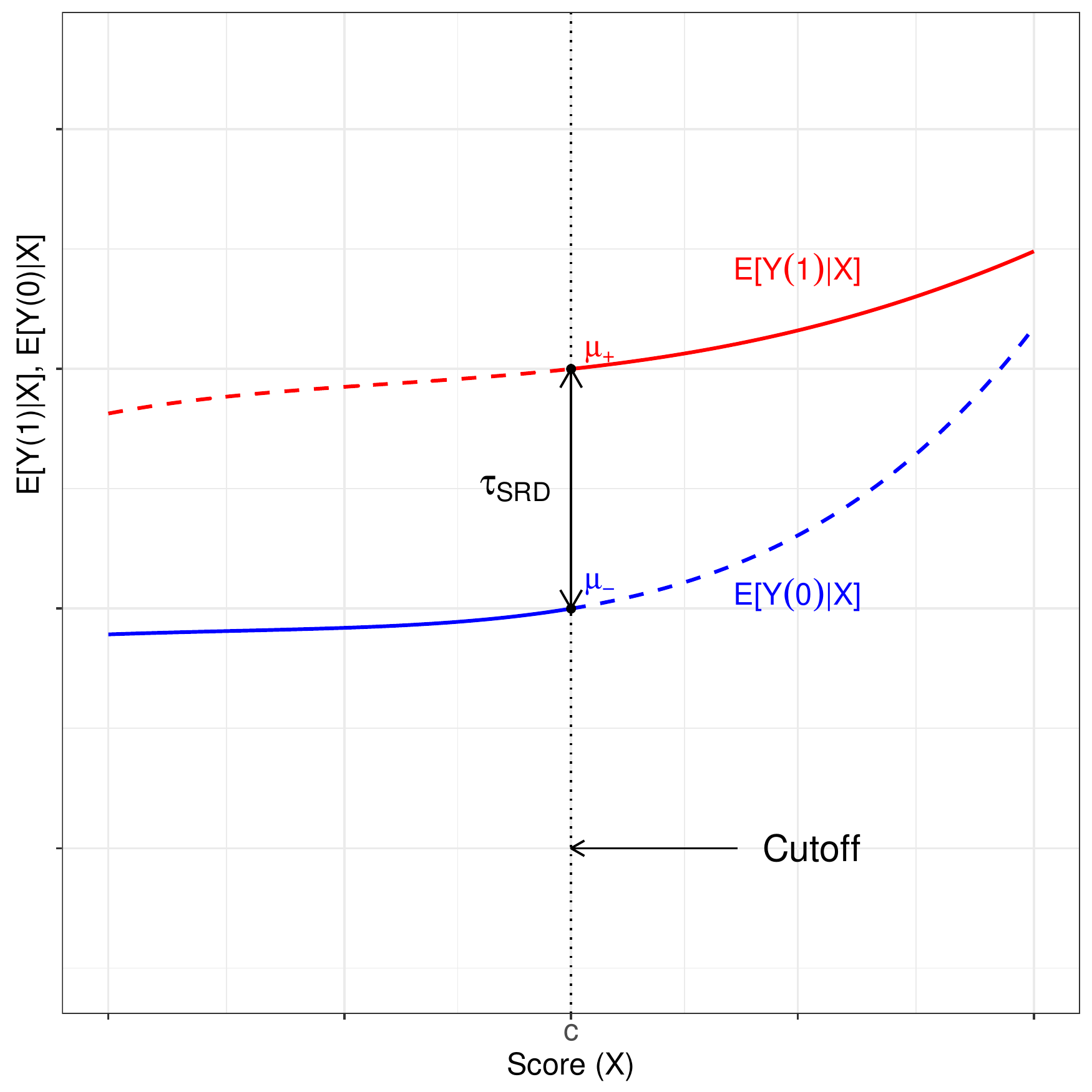}
\caption{RD Treatment Effect in Sharp RD Design}
\label{fig:RDeffect-Model1}
\end{figure}

In the specific context of the Sharp RD design, the fundamental problem of causal inference occurs because we only observe the outcome under control, $Y_i(0)$, for those units whose score is below the cutoff, and we only observe the outcome under treatment, $Y_i(1)$, for those units whose score is above the cutoff. We illustrate this problem in Figure \ref{fig:RDeffect-Model1}, which plots the average potential outcomes given the score, $\E[Y_i(1) | X_i=x]$ and $\E[Y_i(0)|X_i=x]$, against the score. In statistics, conditional expectation functions such as these are usually called \textit{regression functions}. As shown in Figure \ref{fig:RDeffect-Model1}, the regression function $\E[Y_i(1) | X_i]$ is observed for values of the score to the right of the cutoff---because when $X_i\geq \C$, the observed outcome $Y_i$ is equal to the potential outcome under treatment, $Y_i(1)$, for every $i$. This is represented with the solid red line. However, to the left of the cutoff, all units are untreated, and therefore $\E[Y_i(1) | X_i]$ is not observed (represented by a dashed red line). A similar phenomenon occurs for  $\E[Y_i(0) | X_i]$, which is observed for values of the score to the left of the cutoff (solid blue line), $X_i<\C$, but unobserved for $X_i\geq \C$ (dashed blue line). Thus, the observed average outcome given the score is
\begin{equation*}
\E[Y_i | X_i] =\left\{ 
    \begin{array}{ccl}
    \E[Y_i(0) | X_i] && \text{if } X_i<\C, \\ 
    \E[Y_i(1) | X_i] && \text{if } X_i\geq \C. \\ 
    \end{array}
\right.
\end{equation*}

The Sharp RD design exhibits an extreme case of lack of common support, as units in the control ($\T_i=\I(X_i\geq\C)=0$) and treatment ($\T_i=\I(X_i\geq\C)=1$) groups cannot have the same value of the running variable ($X_i$). This feature sets aside RD designs from other non-experimental settings, and highlights that RD analysis fundamentally relies on extrapolation towards the cutoff point. As we discuss throughout this Element, a central goal of empirical RD analysis is to adequately perform (local) extrapolation in order to compare control and treatment units. This unique feature of the RD design also makes causal interpretation of some parameters potentially more difficult; see \citet*{Cattaneo-Titiunik-VazquezBare_2017_JPAM} for further discussion on this point.

\labelfiguras{figB}
 
As shown in Figure \ref{fig:RDeffect-Model1}, the average treatment effect at a given value of the score, $\E[Y_i(1)|X_i=x] - \E[Y_i(0)|X_i=x]$, is the vertical distance between the two regression curves at that value. This distance cannot be directly estimated because we never observe both curves for the same value of $x$. However, a special situation occurs at the cutoff $\C$: this is the only point at which we ``almost'' observe both curves. To see this, we imagine having units with score exactly equal to $\C$, and units with score barely below $\C$ (that is, with score $\C-\varepsilon$ for a small and positive $\varepsilon$). The former units would receive treatment, and the latter would receive control. Yet if the values of the average potential outcomes at $\C$ are not abruptly different from their values at points near $\C$, the units with $X_i=\C$ and $X_i=\C-\varepsilon$ would be very similar except for their treatment status, and we could approximately calculate the vertical distance at $\C$ using observed outcomes. In the figure, the vertical distance at $\C$ is $\E[Y_i(1)|X_i=\C] - \E[Y_i(0)|X_i=\C] \equiv \mu_{+} - \mu_{-}$; this is precisely the treatment effect that can be estimated with a Sharp RD design. The \textit{Sharp RD treatment effect} is thus formally defined as
\[\tau_{\mathtt{SRD}} \equiv \E[Y_i(1) - Y_i(0) | X_i=\C].\]

This parameter captures the (reduced form) treatment effect for units with score values $X_i=\C$. It answers the following question: what would be the average outcome change for units with score level $X_i=\C$ if we switched their status from control to treated? As we discuss below, this treatment effect is, by construction, local in nature and, in the absence of additional assumptions, not informative about treatment effects at other levels of the score. Moreover, since the definition of a Sharp RD design implies that all units with $X_i=\C$ are treated, $\tau_{\mathtt{SRD}}$ can be interpreted as a (local, RD) average treatment effect on the treated.

The assumption of comparability between units with very similar values of the score but on opposite sides of the cutoff is the fundamental concept on which all RD designs are based. This idea was first formalized by \citet*{Hahn-Todd-vanderKlaauw_2001_ECMA} using continuity assumptions. These authors showed that, among other conditions, if the regression functions $\E[Y_i(1) | X_i=x]$ and $\E[Y_i(0) | X_i=x]$, seen as functions of $x$, are continuous at $x=\C$, then in a Sharp RD design we have
\begin{equation}\label{HTV}
\E[Y_i(1)-Y_i(0)|X_i=\C] = \lim_{x\downarrow{\C}} \E[Y_i|X_i=x] - \lim_{x\uparrow{\C}} \E[Y_i|X_i=x]\text{.}
\end{equation}

The result in Equation (\ref{HTV}) says that, if the average potential outcomes are continuous functions of the score at $\C$, the difference between the limits of the treated and control average \textit{observed} outcomes as the score converges to the cutoff is equal to the average treatment effect at the cutoff. Informally, a function $g(x)$ is continuous at the point $x=a$ if the values of $g(x)$ and $g(a)$ get close to each other as $x$ gets close to $a$. In the RD context, continuity means that as the score $x$ gets closer and closer to the cutoff $\C$, the average potential outcome function $\E[Y_i(0) | X_i=x]$ gets closer and closer to its value at the cutoff, $\E[Y_i(0) | X_i=\C]$ (and analogously for $\E[Y_i(1) | X_i=x]$). Thus, continuity gives a formal justification for estimating the Sharp RD effect by focusing on observations above and below the cutoff in a very small neighborhood around it. By virtue of being very close to the cutoff, the observations in this neighborhood will have very similar score values; and by virtue of continuity, their average potential outcomes will also be similar. Therefore, continuity offers one justification for using observations just below the cutoff to approximate the average outcome that units just above the cutoff would have had if they had received the control condition instead of the treatment.

\subsection{The Effect of Islamic Political Representation on Women's Education}
\label{subsec:intro-Meyersson}
We now introduce in more detail the empirical example that we employ throughout this Element, originally analyzed by \citet*{Meyersson_2014_ECMA}, henceforth Meyersson. This example employs a Sharp RD design, based on close elections in Turkey, to study the impact of having a mayor from an Islamic party on various outcomes. The running variable is based on vote shares, as popularized by the work of \citet*{Lee_2008_JoE}.

Meyersson is broadly interested in the effect of Islamic parties' control of local governments on women's rights, in particular on the educational attainment of young women. The methodological challenge is that municipalities where the support for Islamic parties is high enough to result in the election of an Islamic mayor may differ systematically from municipalities where the support for Islamic parties is more tenuous and results in the election of a secular mayor. (For brevity, we refer to a mayor who belongs to one of the Islamic parties as an ``Islamic mayor,'' and to a mayor who belongs to a non-Islamic party as a ``secular mayor.'') If some of the characteristics on which both types of municipalities differ affect (or are correlated with) the educational outcomes of women, a simple comparison of municipalities with an Islamic versus a secular mayor will be misleading. For example, municipalities where an Islamic mayor wins in $1994$ may be on average more religiously conservative than municipalities where a secular mayor is elected. If religious conservatism affects the educational outcomes of women, the na\"ive comparison between municipalities controlled by an Islamic versus a secular mayor will not successfully isolate the effect of the Islamic party's control of the local government. Instead, the effect of interest will be contaminated by differences in the degree of religious conservatism between the two groups.

\begin{figure}[ht]
	\vspace{-0.4in}
	\hspace{-0.25in}%
	\begin{subfigure}[t]{0.43\textwidth}
		\centering
		\includegraphics[scale=0.50]{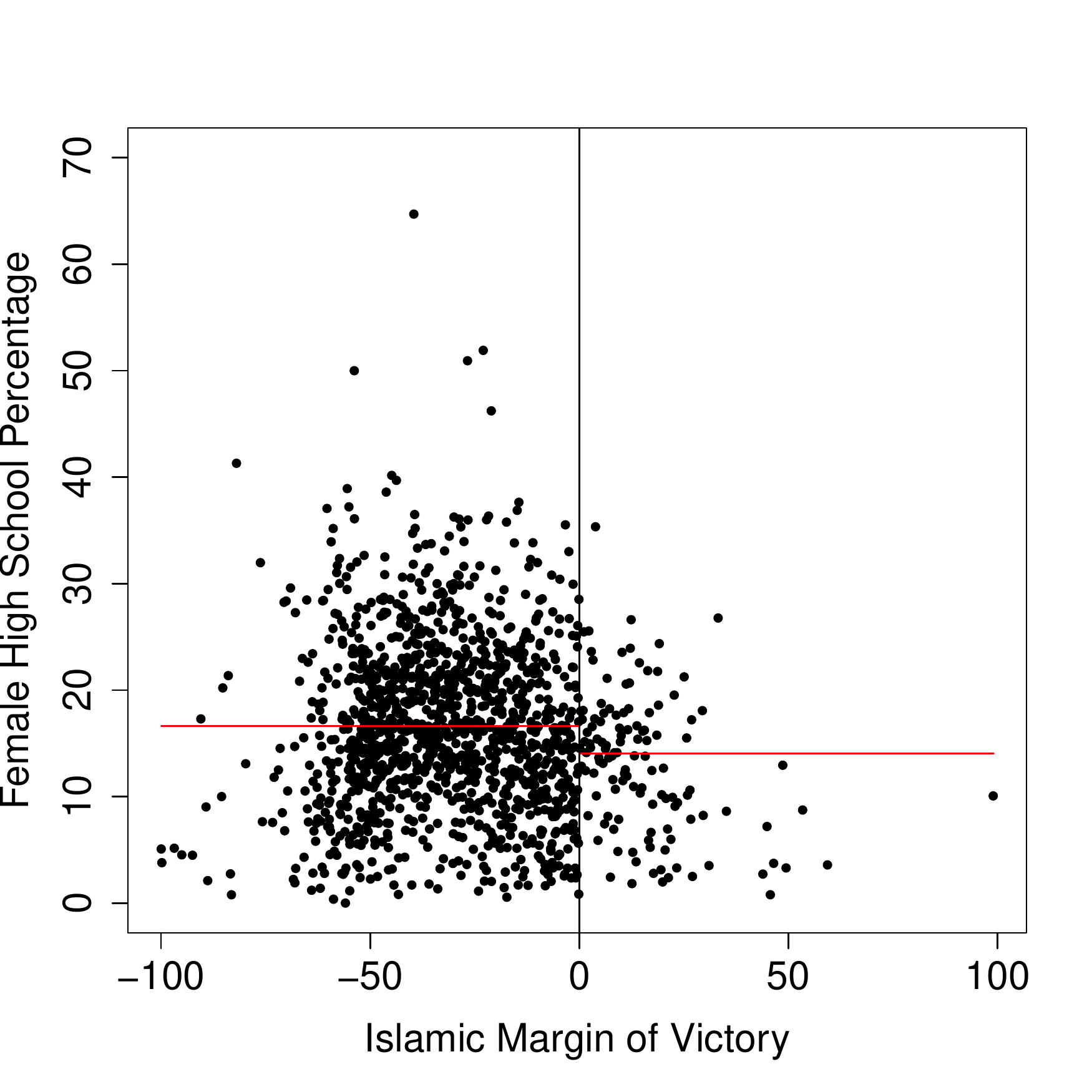}
		\caption{Raw Comparison of Means}\label{p0}		
	\end{subfigure}
	\hspace{0.50in}%
	\begin{subfigure}[t]{0.43\textwidth}
		\centering
		\includegraphics[scale=0.50]{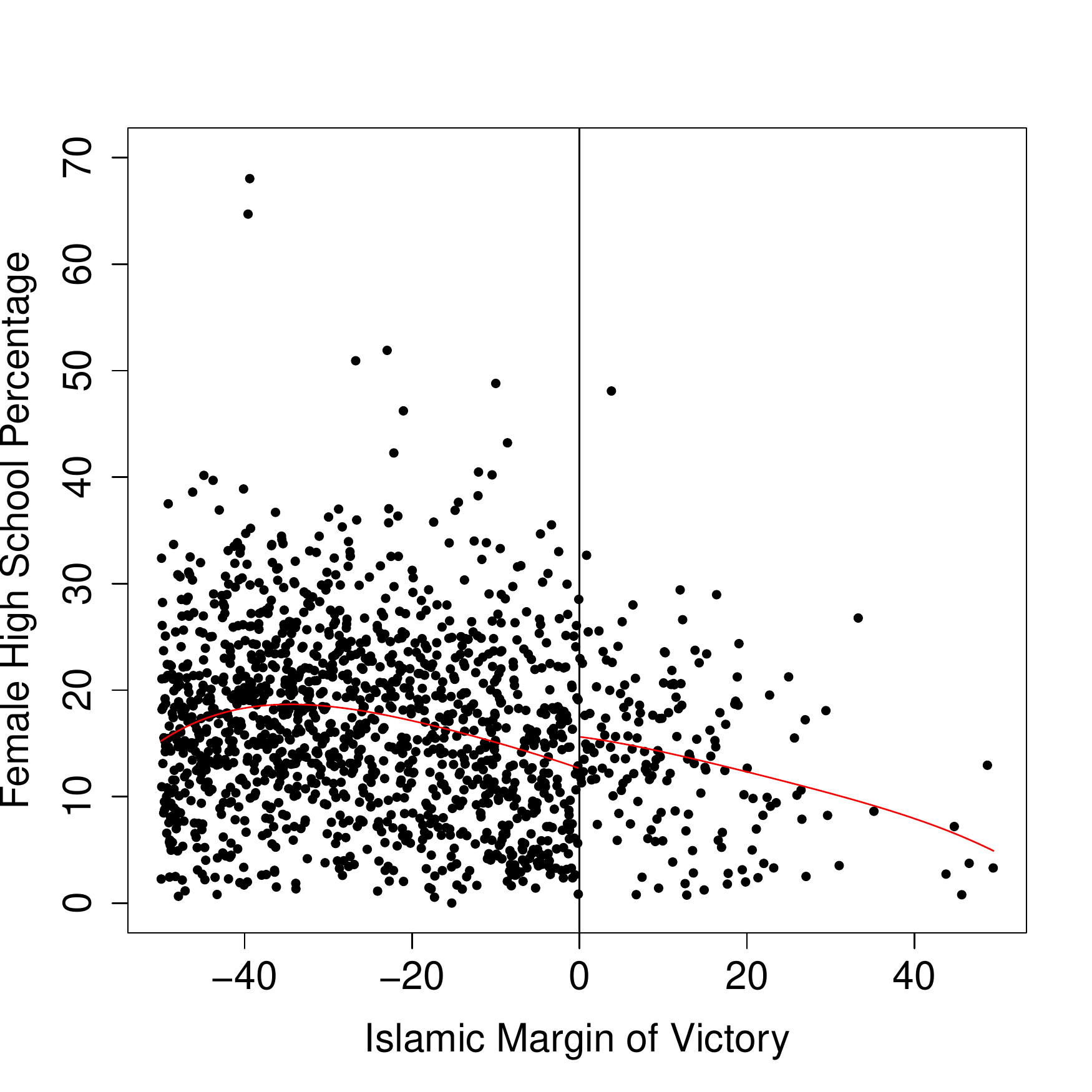}
		\caption{Local Comparison of Means}\label{p4}
	\end{subfigure}
	\caption{Municipalities with Islamic Mayor versus Municipalities with Secular Mayor (Meyersson data)} 
	\label{fig:Meyersson-naive}
\end{figure}

This challenge is illustrated in Figure \ref{fig:Meyersson-naive}, where we plot the percentage of young women who had completed high school by $2000$ against the Islamic margin of victory in the $1994$ mayoral elections (more information on these variables is given below). These figures are examples of so-called RD plots, which we discuss in detail in Section \ref{sec:graph}. In Figure \ref{fig:Meyersson-naive}(\subref{p0}), we show the scatter plot of the raw data (where each point is an observation), superimposing the overall sample mean for each group; treated observations (municipalities where an Islamic mayor is elected) are located to the right of zero, and control observations (municipalities where a secular mayor is elected) are located to the left of zero. The raw comparison reveals a negative average difference: municipalities with an Islamic mayor have, on average, lower educational attainment of women. Figure \ref{fig:Meyersson-naive}(\subref{p4}), shows the scatter plot for the subset of municipalities where the Islamic margin of victory is within $50$ percentage points, a range that includes $83\%$ of the total observations; this second figure superimposes a fourth-order polynomial fit separately on either side of the cutoff. Figure \ref{fig:Meyersson-naive}(\subref{p4}) reveals that the negative average effect in Figure \ref{fig:Meyersson-naive}(\subref{p0}) arises because there is an overall negative relationship or slope between Islamic vote percentage and educational attainment of women for the majority of the observations, so that the higher the Islamic margin of victory, the lower the educational attainment of women. Thus, a na\"ive comparison of treated and control municipalities, which differ systematically in the Islamic vote percentage, will mask systematic differences and may lead to incorrect inferences about the effect of electing an Islamic mayor.

The RD design can be used in cases such as these to isolate a treatment effect of interest from all other systematic differences between treated and control groups. Under appropriate assumptions, a comparison of municipalities where the Islamic party barely wins the election versus municipalities where the Islamic party barely loses will reveal the causal (local) effect of Islamic party control of the local government on the educational attainment of women. If parties cannot systematically manipulate the vote percentage they obtain, observations just above and just below the cutoff will tend to be comparable in terms of all characteristics with the exception of the party that won the 1994 election. Thus, right at the cutoff, the comparison is free of the complications introduced by systematic observed and unobserved differences between the groups. This strategy is illustrated in Figure \ref{fig:Meyersson-naive}(\subref{p4}), where we see that, despite the negative slope on either side, right near the cutoff the effect of an Islamic victory on the educational attainment of women is positive, in stark contrast to the negative difference-in-means in Figure \ref{fig:Meyersson-naive}(\subref{p0}).

\labelfiguras{figC}

Meyersson's original study employs an RD design to circumvent these methodological challenges and to estimate a causal effect of local Islamic rule. The design is focused exclusively on the $1994$ Turkish mayoral elections. The unit of analysis is the municipality, and the score is the Islamic margin of victory, defined as the difference between the vote percentage obtained by the largest Islamic party, and the vote percentage obtained by the largest secular party opponent. Two Islamic parties competed in the $1994$ mayoral elections, \textit{Refah} and \textit{B\"uy\"uk Birlik Partisi} (BBP). However, the results essentially capture the effect of a victory by \textit{Refah}, as the BBP received only 0.94\% of the national vote and won in only 11 of the 329 municipalities where an Islamic mayor was elected. As defined, the Islamic margin of victory can be positive or negative, and the cutoff that determines an Islamic party victory is located at zero. Given this setup, the treatment group consists of municipalities that elected a mayor from an Islamic party in $1994$, and the control group consists of municipalities that elected a mayor from a secular party. The outcome we re-analyze is the educational attainment of women who were (potentially) in high school during the period 1994-2000, calculated as the percentage of the cohort of women aged $15$ to $20$ in $2000$ who had completed high school by $2000$ according to the $2000$ Turkish census. For brevity, we refer to this outcome as the educational attainment of women.

In order to streamline the computer code for our analysis, we rename the variables in the following way.
\begin{itemize}
	\item \texttt{Y}: educational attainment of women, measured as the percentage of women aged $15$ to $20$ in $2000$ who had completed high school by $2000$. 
	\item \texttt{X}: vote margin obtained by the Islamic party in the $1994$ Turkish mayoral elections, measured as the vote percentage obtained by the Islamic party minus the vote percentage obtained by its strongest secular party opponent.
	
	\item \texttt{T}: electoral victory of the Islamic party in $1994$, equal to $1$ if the Islamic party won the mayoral election and $0$ otherwise.
\end{itemize}

The Meyersson dataset also contains several predetermined covariates that we use in subsequent sections to investigate the plausibility of the RD design, and also to illustrate covariate-adjusted estimation methods. The covariates that we include in our analysis are the Islamic vote percentage in $1994$ (\texttt{vshr\_islam1994}), the number of parties receiving votes in $1994$ (\texttt{partycount}), the logarithm of the population in $1994$ (\texttt{lpop1994}), an indicator equal to one if the municipality elected an Islamic party in the previous election in $1989$ (\texttt{i89}), a district center indicator (\texttt{merkezi}), a province center indicator (\texttt{merkezp}), a sub-metro center indicator (\texttt{subbuyuk}), and a metro center indicator (\texttt{buyuk}). 

\labeltablas{tableA}
\begin{table}[h!]
	\centering
	\caption{Descriptive Statistics for Meyersson}
	\resizebox{\textwidth}{!}{\begin{tabular}{lccccc}
			\hline \hline
			Variable & Mean & Median & Std. Dev. & Min. & Max.\\
			\hline
			Y & 16.306 & 15.523 & 9.584 & 0.000 & 68.038\\ 
X & -28.141 & -31.426 & 22.115 & -100.000 & 99.051\\ 
T & 0.120 & 0.000 & 0.325 & 0.000 & 1.000\\ 
Percentage of men aged 15-20 with high school education & 19.238 & 18.724 & 7.737 & 0.000 & 68.307\\ 
Islamic percentage of votes in 1994 & 13.872 & 7.029 & 15.385 & 0.000 & 99.526\\ 
Number of parties receiving votes 1994 & 5.541 & 5.000 & 2.192 & 1.000 & 14.000\\ 
Log population in 1994 & 7.840 & 7.479 & 1.188 & 5.493 & 15.338\\ 
Percentage of population below 19 in 2000 & 40.511 & 39.721 & 8.297 & 6.544 & 68.764\\ 
Percentage of population above 60 in 2000 & 9.222 & 8.461 & 3.960 & 1.665 & 27.225\\ 
Gender ratio in 2000 & 107.325 & 103.209 & 25.293 & 74.987 & 1033.636\\ 
Household size in 2000 & 5.835 & 5.274 & 2.360 & 2.823 & 33.634\\ 
District center & 0.345 & 0.000 & 0.475 & 0.000 & 1.000\\ 
Province center & 0.023 & 0.000 & 0.149 & 0.000 & 1.000\\ 
Sub-metro center & 0.022 & 0.000 & 0.146 & 0.000 & 1.000\\ 

			\hline \hline
			Note: the number of observations for all variables is 2,629
	\end{tabular}}
	\label{tab:meyersson-descriptive}
\end{table}

Table \ref{tab:meyersson-descriptive} presents descriptive statistics for the three RD variables (\texttt{Y}, \texttt{X}, and \texttt{T}), and the municipality-level predetermined covariates. The outcome of interest (\texttt{Y}) has a minimum of $0$ and a maximum of $68.04$, with a mean of $16.31$. This implies that there is at least one municipality in $2000$ where no women in the 15-to-20 age cohort had completed high school, and on average $16.31\%$ of women in this cohort had completed high school by the year $2000$. The Islamic vote margin (\texttt{X}) ranges from $-100$ (party receives zero votes) to $100$ (party receives $100\%$ of the vote), and it has a mean of $-28.14$, implying that on average the Islamic party loses by $28.14$ percentage points. This explains why the mean of the treatment variable (\texttt{T}) is $0.120$, since this indicates that in $1994$ an Islamic mayor was elected in only $12.0\%$ of the municipalities. This small proportion of victories is consistent with the finding that the average margin of victory is negative and thus leads to electoral loss. 

\subsection{The Local Nature of RD Effects}
\label{sec:RDLocal}

The Sharp RD parameter presented above can be interpreted as causal in the sense that it captures the average difference in potential outcomes under treatment versus control. However, in contrast to other causal parameters in the potential outcomes framework, this average difference is calculated at a single point on the support of a continuous random variable (the score $X_i$), and as a result captures a causal effect that is local in nature. According to some perspectives, this parameter cannot even be interpreted as causal because it cannot be reproduced via manipulation or experimentation. 

Regardless of its status as a causal parameter, the RD treatment effect tends to have limited external validity, that is, the RD effect is generally not representative of the treatment effects that would occur for units with scores away from the cutoff. As discussed above, in the canonical Sharp RD design, the RD effect can be interpreted graphically as the vertical difference between $\E[Y_i(1) | X_i=x]$ and $\E[Y_i(0) | X_i=x]$ at the point where the score equals the cutoff, $x=\C$. In the general case where the average treatment effect varies as a function of the score $X_i$, as is very common in applications, this effect may not be informative of the average effect of treatment at values of $x$ different from $\C$. For this reason, in the absence of specific (usually restrictive) assumptions about the global shape of the regression functions, the effect recovered by the RD design is only the average effect of treatment for units \textit{local to the cutoff}, i.e., for units with score values $X_i=\C$.

In the context of the Meyersson application, the lack of external validity is reflected in the focus on close, as opposed to all, elections. As illustrated in Figure \ref{fig:Meyersson-naive}(\subref{p0}), it seems that the educational attainment of women is higher in municipalities where the Islamic party barely wins than in municipalities where the party barely loses the election. By definition, the sample of municipalities near the cutoff comprises constituencies where the Islamic party is very competitive. It is likely that the political preferences and religious affiliation of Turkish citizens in these municipalities differ systematically from those in municipalities where the Islamic party wins or loses by very large margins. This means that, although the RD results indicate that Islamic mayors lead to an increase in the educational attainment of women in competitive municipalities, it is not possible to know whether the same positive effect in female education would be seen if a mayor from the Islamic party governed a municipality with strong preferences for secular political parties. Figure \ref{fig:Meyersson-naive} reveals that the vast majority  of observations in the sample of 1994 Turkish mayoral elections is composed of municipalities where the Islamic party lost by a very large margin; without further assumptions, the RD effect is not informative about the educational effect of an Islamic party victory in these municipalities.

\begin{figure}[H]
\hspace{-0.2in}%
	\begin{subfigure}[t]{0.49\textwidth}
\centering
		\includegraphics[scale=0.65]{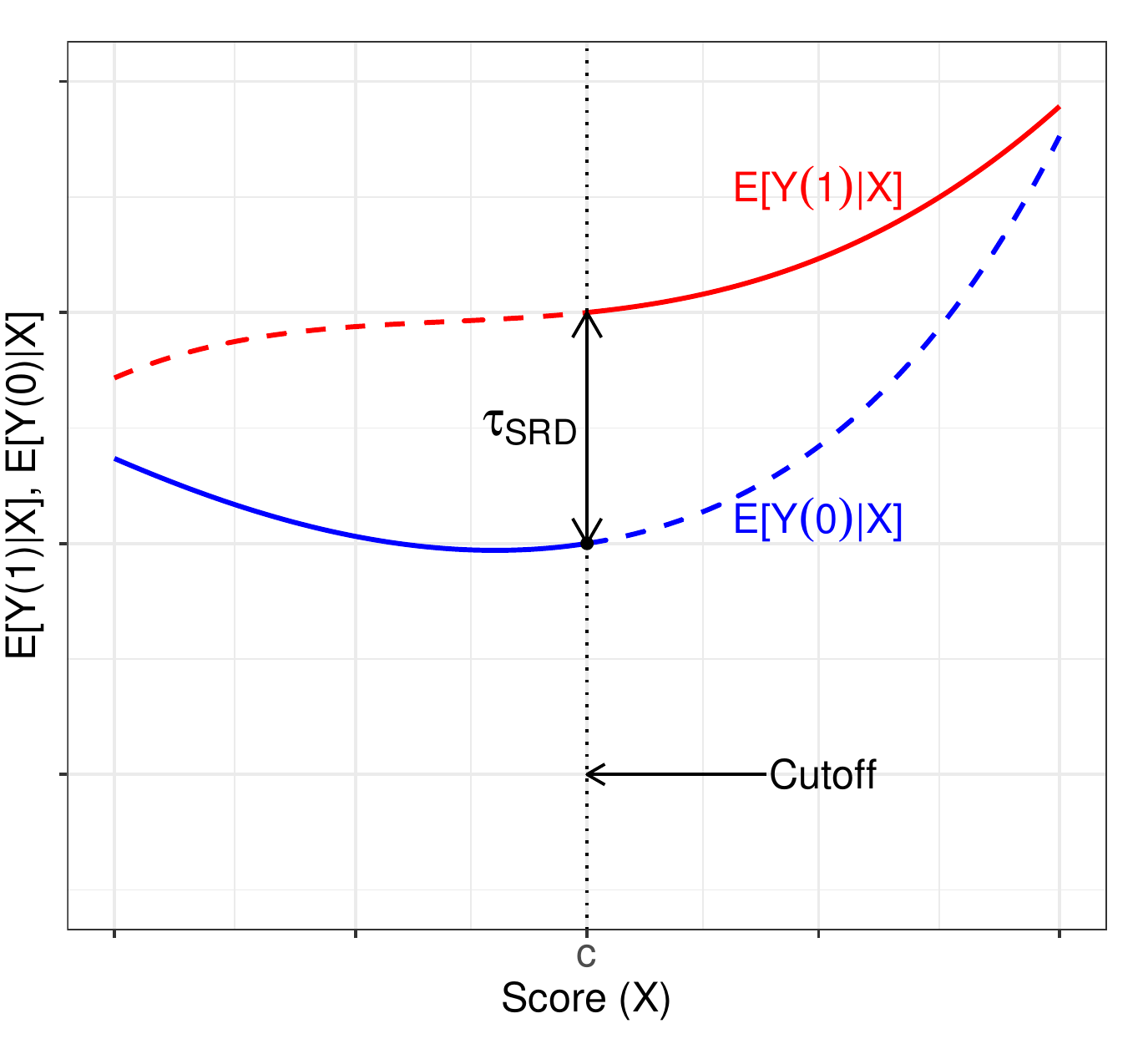}
		\caption{Mild Heterogeneity}
 	    \label{heter1}		
	\end{subfigure}
\hspace{0.25in}%
	\begin{subfigure}[t]{0.49\textwidth}
\centering
		\includegraphics[scale=0.65]{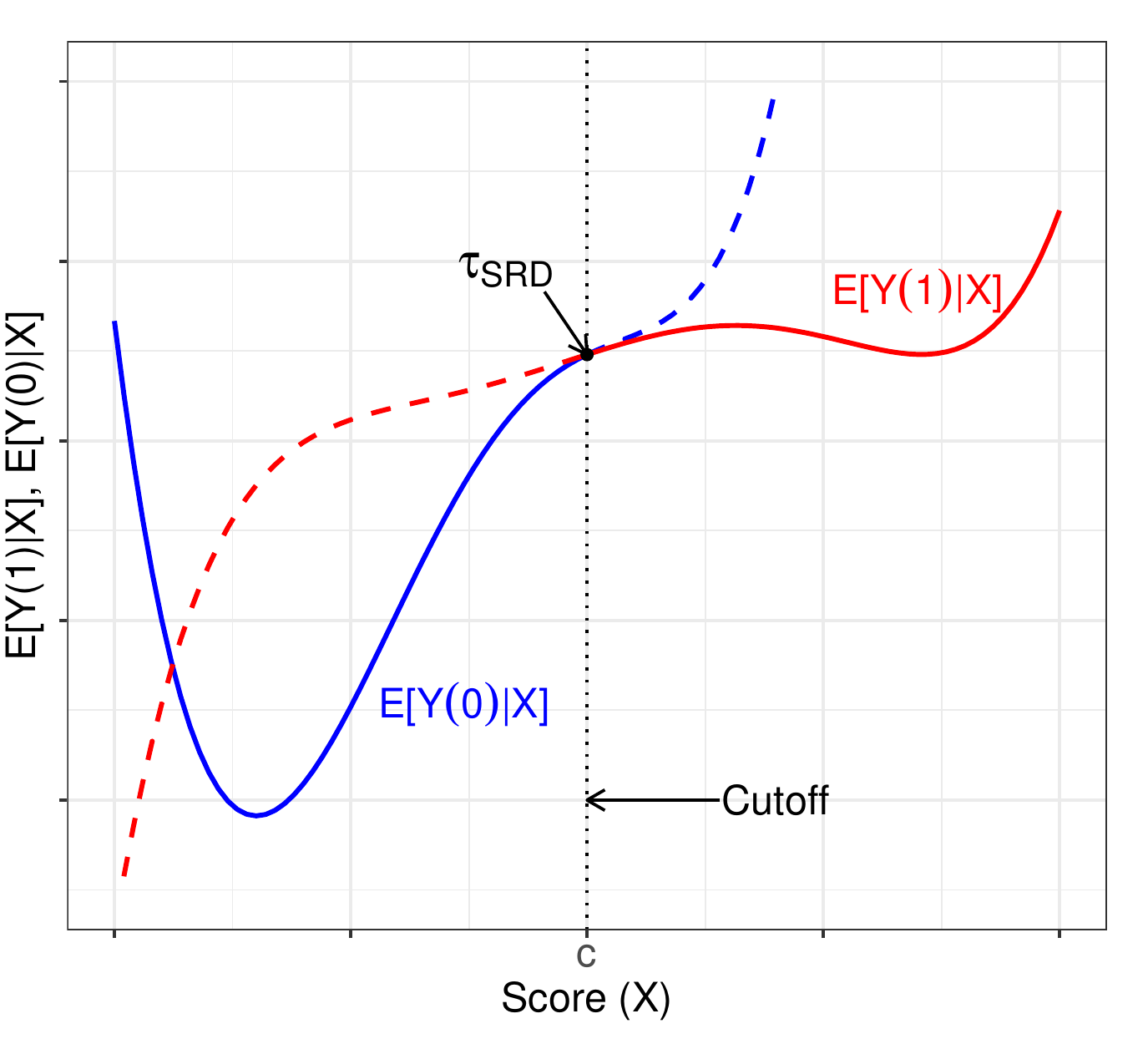}
		\caption{Severe Heterogeneity}
		\label{heter2}
	\end{subfigure}
\caption{Local Nature of the RD Effect}
\label{fig:RDheter}
\end{figure}

In general, the degree of representativeness or external validity of the RD treatment effect will depend on the specific application under consideration. For example, in the hypothetical scenario illustrated in Figure \ref{fig:RDheter}(\subref{heter1}), the vertical distance between $\E[Y_i(1) | X_i=x]$ and $\E[Y_i(0) | X_i=x]$ at $x=\C$ is considerably higher than at other points, but the effect is positive everywhere. A much more heterogeneous hypothetical scenario is shown in Figure \ref{fig:RDheter}(\subref{heter2}), where the effect is zero at the cutoff but ranges from positive to negative at other points. Since the counterfactual (dotted) regression functions are never observed in real examples, it is not possible to know with certainty the degree of external validity of any given RD application.

\labelfiguras{figD}

Increasing the external validity of RD estimates and estimands is a topic of active research and, regardless of the approach taken,  necessarily requires more assumptions. For example, extrapolation of RD treatment effects can be done by imposing additional assumptions about (i) the regression functions near the cutoff \citep*{Dong-Lewbel_2015_Restat,Wing-Cook_2013_JPAM}, (ii) local independence assumptions \citep*{Angrist-Rokkanen_2015_JASA} (iii) exploiting specific features of the design such as imperfect compliance \citep*{Bertanha-Imbens_2019_JBES}, or (iv) the presence of multiple cutoffs \citep*{Cattaneo-Keele-Titiunik-VazquezBare_2016_JOP,Cattaneo-Keele-Titiunik-VazquezBare_2019_wp}. On this regard, RD designs are not different from randomized experiments: they both require additional assumptions to map internally valid estimates into externally valid ones.

\subsection{Further Reading}

For an introduction to causal inference based on potential outcomes see \citet{Imbens-Rubin_2015_Book} and references therein. For a review on causal inference and program evaluation methods see \citet{Abadie-Cattaneo_2018_ARE} and references therein. The RD design was originally proposed by \citet{Thistlethwaite-Campbell_1960_JEP}, and historical as well as early review articles are given by \citet{Cook_2008_JoE}, \citet{Imbens-Lemieux_2008_JoE}, and \citet{Lee-Lemieux_2010_JEL}. \citet{Lee_2008_JoE} provided an influential contribution to the identification of RD effects; \citet{Lee_2008_JoE} and \cite{Pettersson-Lidbom2008-JEEA} were the first to apply the RD design to close elections. The edited volume by \citet{Cattaneo-Escanciano_2017_AIE} provides a recent overview of the RD literature and includes several recent methodological and practical contributions.

\newpage
\section{RD Plots}
\label{sec:graph}

An appealing feature of the RD design is that it can be illustrated graphically. This graphical representation, in combination with the formal approaches to estimation, inference, and falsification discussed below, adds transparency to the analysis by displaying the observations used for estimation and inference. RD plots also allow researchers to readily summarize the main empirical findings as well as other important features of the work conducted. We now discuss the most transparent and effective methods to graphically illustrate the RD design.

At first glance, it seems that one should be able to illustrate the relationship between the outcome and the running variable by simply constructing a scatter plot of the observed outcome against the score, clearly identifying the points above and below the cutoff. However, this strategy is rarely useful, as it is often hard to see ``jumps'' or discontinuities in the outcome-score relationship by simply looking at the raw data. We illustrate this point with the Meyersson application, plotting the educational attainment of women against the Islamic vote margin using the raw observations. We create this scatter plot in \texttt{R} with the \texttt{plot} command.

\labelsnippet{snippetLrawplots}
\rsnip{Vol-1-R_meyersson_rdplot_raw.txt}{\Rlink{\thesnippetLrawplots}}
\statasnip{Vol-1-STATA_meyersson_rdplot_raw}{\Slink{\thesnippetLrawplots}}

\labelfiguras{figE}
\begin{figure}[h]
\centering
\includegraphics[scale=0.45]{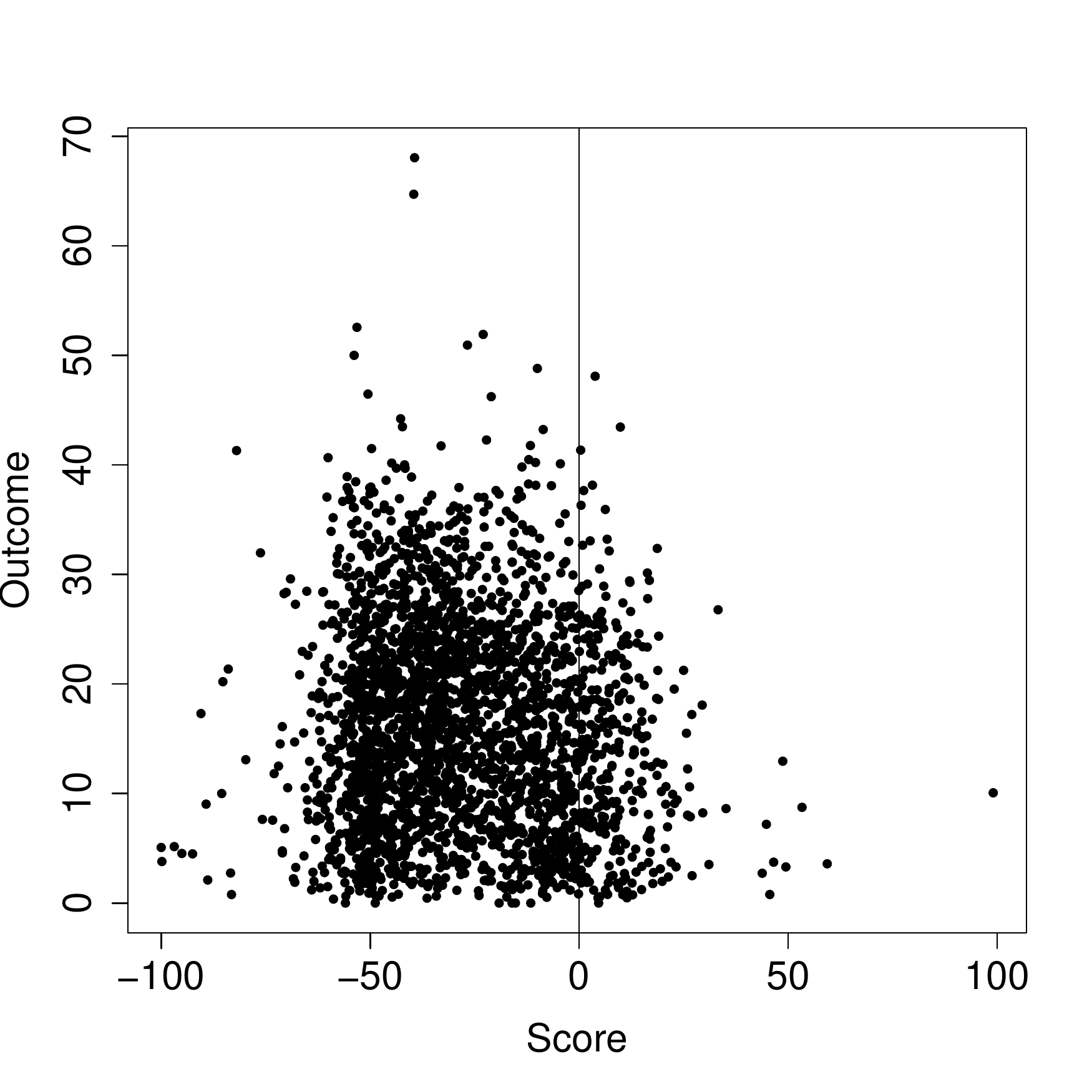}
\caption{Scatter Plot (Meyersson Data)}
\label{fig:ScatterMeyersson}
\end{figure}

Each point in Figure \ref{fig:ScatterMeyersson} corresponds to one raw municipality-level observation in the dataset, so there are 2,629 points in the scatter plot (see Table \ref{tab:meyersson-descriptive}). Although this plot is helpful to visualize the raw observations, detect outliers, etc., its effectiveness for visualizing the RD design is limited. In this application, there is empirical evidence that the Islamic party's victory translates into a small increase in women's educational attainment. Despite this evidence of a positive RD effect, a jump in the values of the outcome at the cutoff cannot be seen by simply looking at the raw cloud of points around the cutoff in Figure \ref{fig:ScatterMeyersson}. In general, raw scatter plots do not allow for easy visualization of the RD effect even when the effect is large.
       
A more useful approach is to aggregate or ``smooth'' the data before plotting. The typical RD plot presents two summaries: (i) a global polynomial fit, represented by a solid line, and (ii) local sample means, represented by dots. The global polynomial fit is simply a smooth approximation to the unknown regression functions based on a fourth- or fifth-order polynomial regression fit of the outcome on the score, fitted separately above and below the cutoff, and using the original raw data. In contrast, the local sample means are created by first choosing disjoint (i.e., non-overlapping) intervals or ``bins'' of the score, calculating the mean of the outcome for the observations falling within each bin, and then plotting the average outcome in each bin against the mid point of the bin. Local sample means can be interpreted as a non-smooth approximation to the unknown regression functions. The combination of these two ingredients in the same plot allows researchers to visualize the global or overall shape of the regression functions for treated and control observations, while at the same time retaining enough information about the local behavior of the data to observe the RD treatment effect and the variability of the data around the global fit. Note that, in the standard RD plot, the global polynomial is calculated using the original observations, not the binned observations.

For example, in the Meyersson application, if we use $20$ bins of equal length on each side of the cutoff, we partition the support of the Islamic margin of victory into $40$ disjoint intervals of length $5$. Recall that a party's margin of victory ranges from $-100$ to $100$, and that the Islamic margin of victory in the Meyersson data ranges from $-100$ to $99.051$. Table \ref{tab:MeyerssonBinsAdHoc} shows the bins and the corresponding average outcomes in this case, where we denote the bins by $\mathcal{B}_{-,1}, \mathcal{B}_{-,2}, \ldots, \mathcal{B}_{-,20}$ (control group) and $\mathcal{B}_{+,1}, \mathcal{B}_{+,2}, \ldots, \mathcal{B}_{+,20}$ (treatment group), using the subscripts $-$ and $+$ to indicate, respectively, bins located to the left and right of the cutoff. In this table, each local sample average is computed as
\[\bar{Y}_{-,j} = \frac{1}{\text{\#}\{X_i \in \mathcal{B}_{-,j}\}}\sum_{i:X_i \in \mathcal{B}_{-,j} } Y_i \qquad\text{and}\qquad
  \bar{Y}_{+,j} = \frac{1}{\text{\#}\{X_i \in \mathcal{B}_{+,j}\}}\sum_{i:X_i \in \mathcal{B}_{+,j} } Y_i,\]
where $j=1,2,\cdots,20$.

\labeltablas{tableB}
\begin{table}[H]
  \caption{Partition of Islamic Margin of Victory into 40 Bins of Equal Length (Meyersson Data)}
  \label{tab:MeyerssonBinsAdHoc}
  \centering
  \resizebox{.9\textwidth}{!}{\begin{tabular}{lccc}
  \toprule
  \multicolumn{1}{c}{\multirow{2}{*}{Bin}} & \multirow{2}{*}{Average Outcome in Bin} & Number of    & Group \\
	                                         &                                         & Observations & Assignment \\
  \midrule
  $\mathcal{B}_{-,1} = [-100, -95)$ & $\bar{Y}_{-,1} = 4.6366 $ & 4 & Control \\
  $\mathcal{B}_{-,2} = [-95, -90)$  & $\bar{Y}_{-,2} = 10.8942 $ & 2 & Control \\
  $\vdots$                           & $\vdots$ & $\vdots$ & $\vdots$ \\
  $\mathcal{B}_{-,19} = [-10, -5)$    & $\bar{Y}_{-,19} = 12.9518$ & 149 & Control \\
  $\mathcal{B}_{-,20} = [-5, 0)$      & $\bar{Y}_{-,20} = 13.8267$ & 148 & Control \\
  \midrule
  $\mathcal{B}_{+,1} = [0, 5)$       & $\bar{Y}_{+,1} = 15.3678$ & 109 &Treatment \\
   $\mathcal{B}_{+,2} = [5, 10)$     & $\bar{Y}_{+,2} = 13.9640$ & 83 & Treatment \\
  $\vdots$                           & $\vdots$ & $\vdots$ & $\vdots$ \\
  $\mathcal{B}_{+,19} = [90, 95)$    & $\bar{Y}_{+,19} = \text{NA} $ & 0 & Treatment \\
  $\mathcal{B}_{+,20} = [95, 100]$   & $\bar{Y}_{+,20} = 10.0629$ & 1 & Treatment \\
\bottomrule        
\end{tabular}}
\end{table}

\begin{figure}[H]
\centering
\includegraphics[scale=0.5]{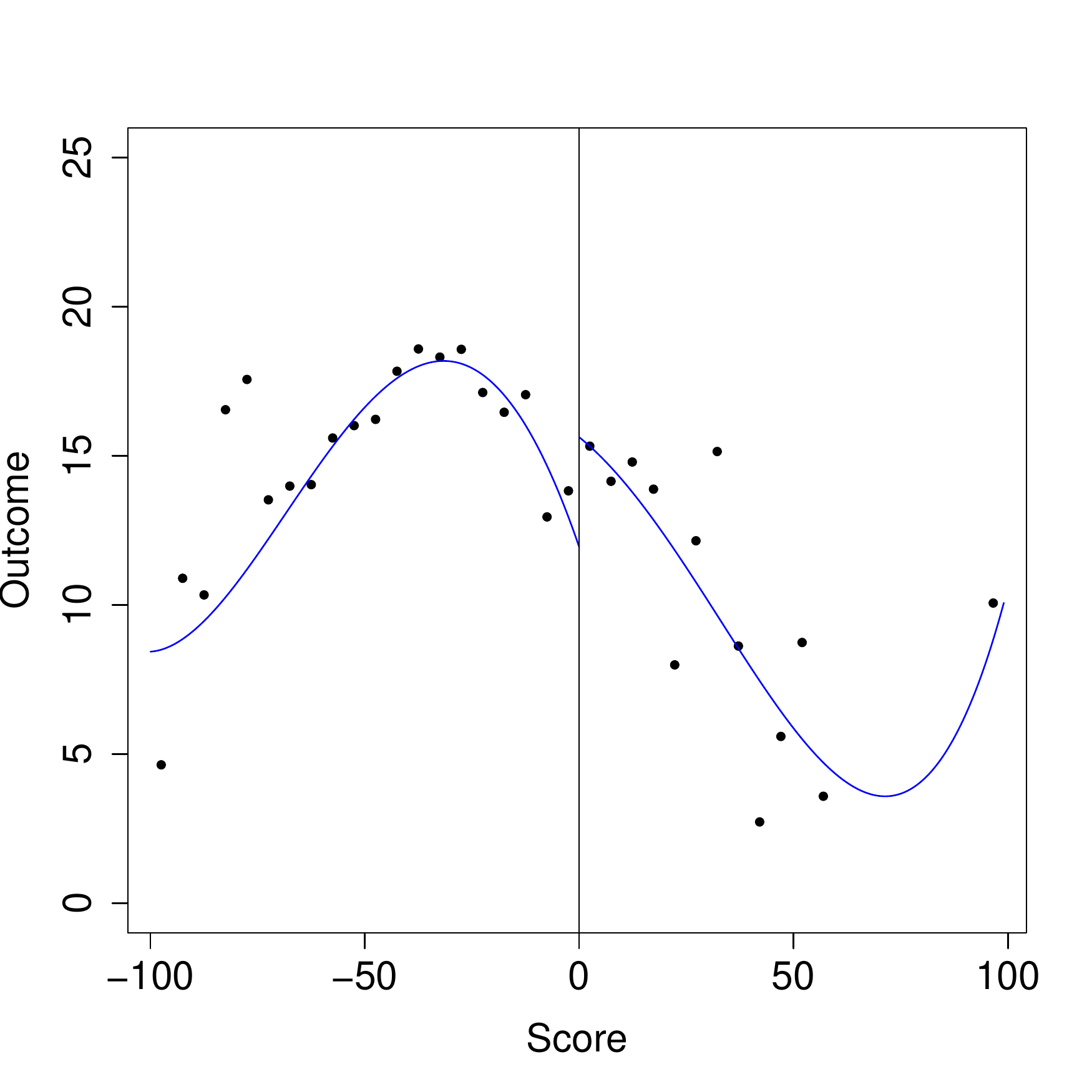}
\caption{RD Plot for Meyersson Data Using 40 Bins of Equal Length}
\label{fig:MeyerssonBinsAdHoc}
\end{figure}

In Figure  \ref{fig:MeyerssonBinsAdHoc}, we plot the binned outcome means shown in Table \ref{tab:MeyerssonBinsAdHoc} against the score, adding a fourth-order global polynomial fit estimated separately for treated and control observations. (Below we show how to create this plot using the \texttt{rdplot} command.) The global fit reveals that the observed regression function seems to be non-linear, particularly on the control (left) side. At the same time, the binned means let us see the local behavior of the average response variable around the global fit. The plot also reveals a positive jump at the cutoff: the average educational attainment of women seems to be higher in those municipalities where the Islamic party obtained a barely positive margin of victory than in those municipalities where the Islamic party barely lost. 

\labelfiguras{figF}

The types of information conveyed by Figures \ref{fig:ScatterMeyersson} and \ref{fig:MeyerssonBinsAdHoc} are very different. In the raw scatter plot in Figure \ref{fig:ScatterMeyersson}, it is difficult to see any systematic pattern, and there is no visible discontinuity in the average outcome at the cutoff. In contrast, when we bin the data and include a global polynomial fit in Figure \ref{fig:MeyerssonBinsAdHoc}, the plot now allows us to see a discontinuity at the cutoff and to better understand the shape of the underlying regression function over the whole support of the running variable.  Binning the data may reveal striking patterns that can remain hidden in a simple scatter plot. Since binning often leads to drastically different patterns from those seen in the raw data, we now discuss how to choose the type and number of bins in a data-driven, transparent, and (sometimes) optimal way.

\subsection{Choosing the Location of Bins}
There are two different types of bins that can be used in the construction of RD plots: bins that have equal length, as in Table \ref{tab:MeyerssonBinsAdHoc}, or bins that contain (roughly) the same number of observations but whose length may differ. We refer to these two types as evenly-spaced and quantile-spaced bins, respectively.

In order to define the bins more precisely, we assume that the running variable takes values inside the interval $[x_l,x_u]$. In the Meyersson application, $x_l=-100$ and $x_u=100$. We continue to use the subscripts $+$ and $-$ to denote treated and control observations, respectively. The bins are constructed separately for treated and control observations, partitioning the support in non-overlapping intervals. We use $J_{-}$ and $J_{+}$ to denote the total number of bins chosen to the left and right of the cutoff, respectively. 

We define the bins generally as follows:
  \begin{align*}
    \mathtt{Control}\; \mathtt{Bins}\text{:} \;\;\;\; \mathcal{B}_{-,j}=\left\{\begin{tabular}{lll}
                             $[x_l~,~b_{-,1})$ & & $j=1$\\ 
                             $[b_{-,j-1}~,~b_{-,j})$ & & $j=2,\cdots ,J_{-}-1$\\ 
                             $[b_{-,J_{-}-1}~,~\C)$ & & $j=J_{-}$
                      \end{tabular}\right. 
  \end{align*}
  \begin{align*}
    \mathtt{Treated}\; \mathtt{Bins}\text{:} \; \;\;\; \mathcal{B}_{+,j}=\left\{\begin{tabular}{lll}
                             $[\C~,~b_{+,1})$ & & $j=1$\\ 
                             $[b_{+,j-1}~,~b_{+,j})$ & & $j=2,\cdots ,J_{+}-1$\\ 
                             $[b_{+,J_{+}-1}~,~x_u]$ & & $j=J_{+}$,
                                    \end{tabular}\right. 
\end{align*}    
with $b_{-,0}<b_{-,1}<\cdots<b_{-,J_{-}}$ and $b_{+,0}<b_{+,1}<\cdots<b_{+,J_{+}}$. In other words, the union of the control and treated bins, $\mathcal{B}_{-,1} \cup \mathcal{B}_{-,2} \cup \ldots \cup  \mathcal{B}_{J_{-}} \cup \mathcal{B}_{+,1} \cup \mathcal{B}_{+,2} \cup \ldots \cup \mathcal{B}_{J_{+}}$, forms a disjoint partition of the support of the running variable,  $[x_l,x_u]$, centered at the cutoff $\C$.  

Letting $X_{-,(i)}$ and $X_{+,(i)}$ denote the $i$th quantiles of the control and treatment subsamples, respectively, and $\lfloor\cdot\rfloor$ denote the floor function, we can now formally define evenly-spaced (ES) and quantile-spaced (QS) bins.
\begin{itemize}
\item \textbf{Evenly-spaced (ES) bins}: non-overlapping intervals that partition the entire support of the running variable, all of the same length within each treatment assignment status:
\[b_{-,j}=x_l+\frac{j\;(\C-x_l)}{J_{-}} \qquad\text{and}\qquad b_{+,j}=\C+\frac{j\;(x_u-\C)}{J_{+}}.\]
Note that all ES bins in the control side have length $\frac{\C-x_l}{J_{-}}$ and all bins in the treated side have length $\frac{x_u-\C}{J_{+}}$. 

\item \textbf{Quantile-spaced (QS) bins}: non-overlapping intervals that partition the entire support of the running variable, all containing (roughly) the same number of observations within each treatment assignment status:
\[b_{-,j}=X_{-,(\lfloor j/J_{-}\rfloor)} \qquad\text{and}\qquad b_{+,j}=X_{+,(\lfloor j/J_{+}\rfloor)}. \]
Note that the length of QS bins may differ even within treatment assignment status; the bins will be larger in regions of the support where there are fewer observations.
\end{itemize}

In practical terms, the most important difference between ES and QS bins is the underlying variability of the local mean estimate in every bin. Although ES bins have equal length, if the observations are not uniformly distributed on $[x_l, x_u]$, each bin may contain a different number of observations. In an RD plot with ES bins, each of the local means represented by a dot may be computed using a different number of observations and thus may be more or less precisely calculated than the other local means in the plot, affecting comparability. For example, Table \ref{tab:MeyerssonBinsAdHoc} shows that there are only 4 observations in $[-100,-95]$, and only 2 observations in $[-95,-90]$; thus, the variance of these local mean estimates is very high because they are constructed with very few observations.

In contrast, QS bins contain approximately the same number of observations by construction. Moreover, a quantile-spaced RD plot has the advantage of providing a quick visual representation of the density of observations over the support of the running variable. For example, if there are very few observations far from the cutoff, an RD plot with quantile-spaced bins will tend to be ``empty'' near the extremes of  $[x_l,x_u]$, and will quickly convey the message that there are no observations with values of the score near $x_l$ or $x_u$.

We now use the \texttt{rdplot} command to produce different RD plots and illustrate the differences between binning strategies, using the \texttt{binselect} option to choose between ES and QS methods. First, we reproduce the RD plot above, using 20 evenly-spaced bins on each side via the option \texttt{nbins}. 

\labelsnippet{snippetLrdplotB}
\rsnip{Vol-1-R_meyersson_rdplot_es_20bins.txt}{\Rlink{\thesnippetLrdplotB}}
\statasnip{Vol-1-STATA_meyersson_rdplot_es_20bins}{\Slink{\thesnippetLrdplotB}}

\labelfiguras{figG}
\begin{figure}[h]
	\begin{subfigure}[t]{0.485\textwidth}
		\centering
		\includegraphics[scale=0.45]{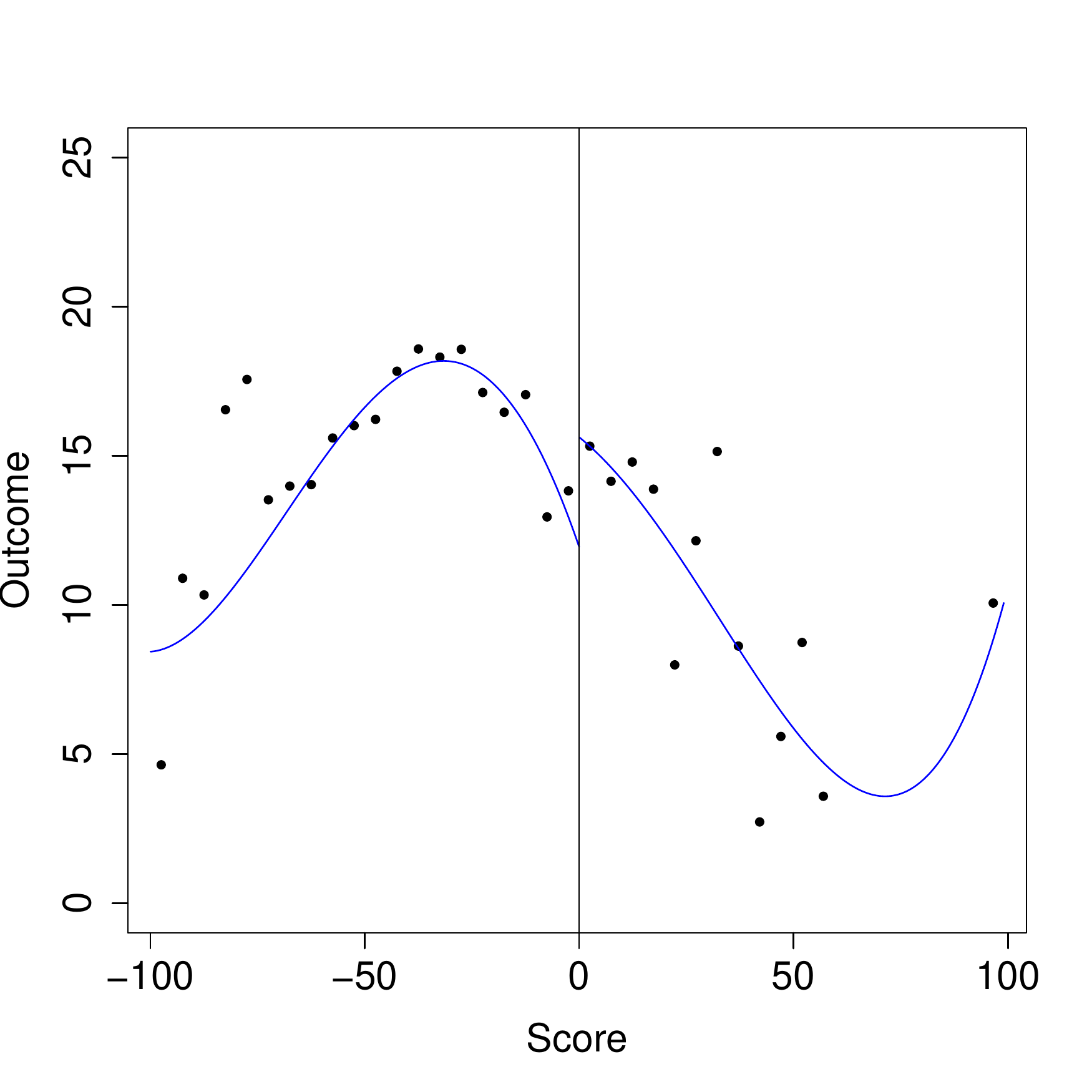}
		\caption{40 Evenly-Spaced Bins}\label{fig:MeyerssonAdHocES}
	\end{subfigure}
	\hspace{0.40in}%
	\begin{subfigure}[t]{0.485\textwidth}
		\centering
		\includegraphics[scale=0.45]{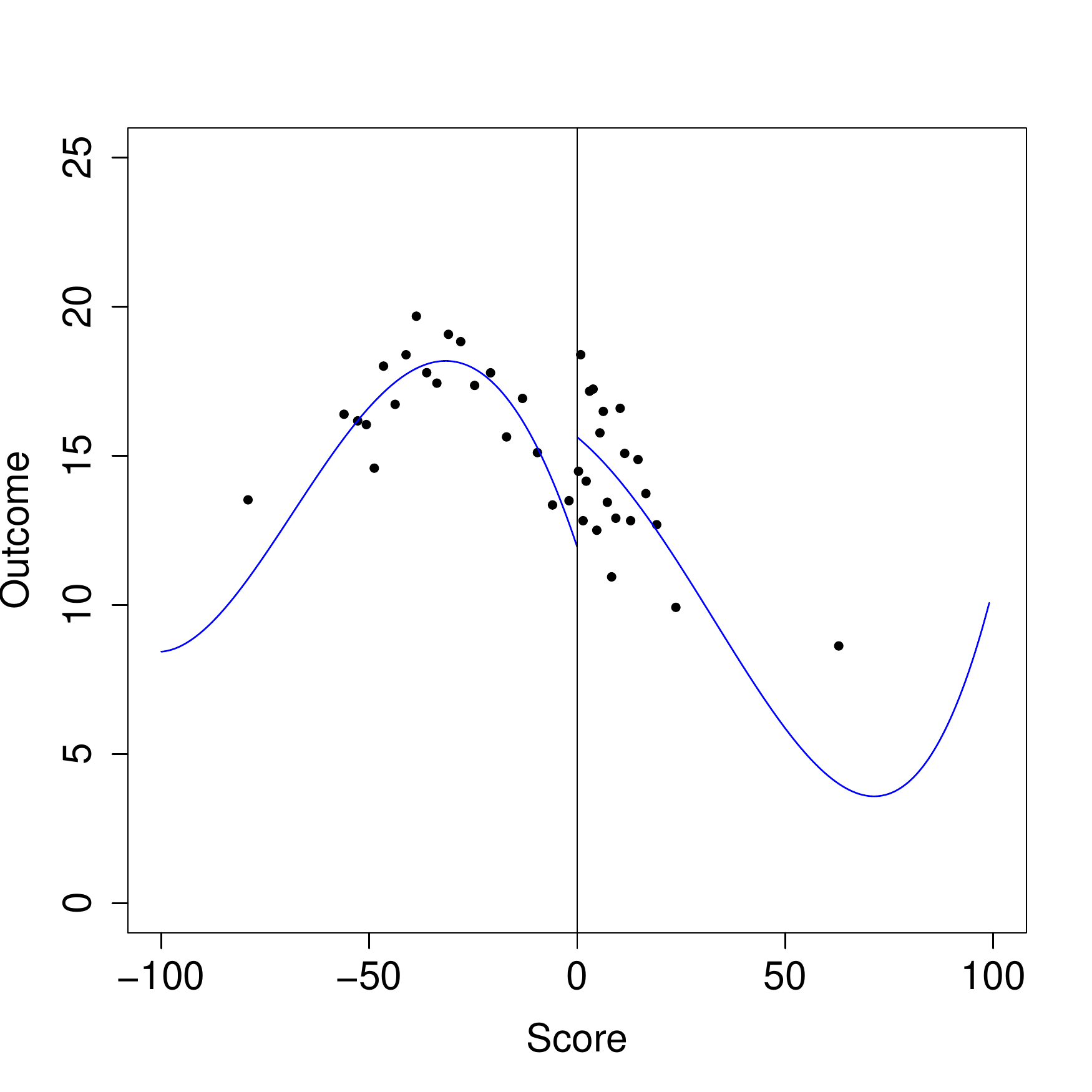}
		\caption{40 Quantile-Spaced Bins}\label{fig:MeyerssonAdHocQS}
	\end{subfigure}
	\caption{RD Plots---Meyersson Data} 
	\label{fig:RDplots2}
\end{figure}

The full output of \texttt{rdplot} includes several descriptive statistics in addition to the actual plot, which is shown in Figure \ref{fig:RDplots2}(\subref{fig:MeyerssonAdHocES}). The total number of observations is shown in the very top row, where we can also see the type of weights used to plot the observations. We have $2,629$ observations in total, which by default are all given equal or uniform weight, as is indicated by the output \texttt{Kernel = Uniform}. The rest of the output is divided in two columns, corresponding to observations located to the left or right of the cutoff, respectively. 

The output shows that there are $2,314$ observations to the left of the cutoff, and $315$ to the right, consistent with our descriptive analysis indicating that the Islamic party lost the majority of these electoral races. The third row in the top panel indicates that the global polynomial fit used in the RD plot is of order 4 on both sides of the cutoff. The fourth row indicates the window or bandwidth $h$ where the global polynomial fit was conducted; the global fit uses all observations in $[\C-h,\C)$ on the control side, and all observations in $[\C,\C+h]$ on the treated side. By default, all control and treated observations are included in the control and treated fit, respectively. Since the range of the Islamic margin of victory is $[-100,99.051]$, the bandwidth on the right is slightly smaller than $100$. Finally, the last row in the top panel shows the scale selected, which is an optional factor by which the chosen number of bins can be multiplied to either increase or decrease the original choice; by default, this factor is one and no scaling is performed.

The lower part of the output shows results on the number and type of bins selected. The top two rows show that we have selected 20 bins to the left of the cutoff, and 20 bins to the right of the cutoff. On the control side, the length of each bin is exactly $5 = \frac{\C-x_l}{J_{-}}=\frac{0-(-100)}{20}=100/20$. However, the actual length of the ES bins to the right of the cutoff is slightly smaller than $5$, as the edge of the support on the treated side is $99.051$ instead of $100$. The actual length of the bins to the right of the cutoff is $\frac{x_u-\C}{J_{+}}=\frac{99.051-0}{20}=99.051/20=4.9526$. We postpone discussion of the five bottom rows until the next subsection where we discuss optimal bin number selection. 

We now compare this plot to an RD plot that also uses 20 bins on each side, but with quantile-spaced bins instead of evenly-spaced bins selected by setting the option \texttt{binselect = "qs"}. The resulting plot is shown in Figure \ref{fig:RDplots2}(\subref{fig:MeyerssonAdHocQS}).\\

\labelsnippet{snippetLrdplotC}
\rsnip{Vol-1-R_meyersson_rdplot_qs_20bins.txt}{\Rlink{\thesnippetLrdplotC}}
\statasnip{Vol-1-STATA_meyersson_rdplot_qs_20bins}{\Slink{\thesnippetLrdplotC}}

A comparison of the two RD plots in Figure \ref{fig:RDplots2} reveals where the observations are located. In the evenly-spaced RD plot in Figure \ref{fig:RDplots2}(\subref{fig:MeyerssonAdHocES}), there are five bins in the interval [-100,-75] of the running variable. In contrast, in the quantile-spaced RD plot in Figure \ref{fig:RDplots2}(\subref{fig:MeyerssonAdHocQS}), this interval is entirely contained in the first bin. The vast difference in the length of QS and ES bins occurs because, as shown in Table \ref{tab:MeyerssonBinsAdHoc}, there are very few observations near $-100$, which leads to local mean estimates with high variance. This problem is avoided when we choose QS bins, which ensures that each bin has the same number of observations. 

\subsection{Choosing the Number of Bins}
Once the positioning of the bins has been decided by choosing either QS or ES bins, the only remaining choice is the total number of bins on either side of the cutoff---the quantities $J_{-}$ and $J_{+}$. Below we discuss two methods to produce data-driven, automatic RD plots by selecting $J_{-}$ and $J_{+}$, given a choice of QS or ES bins.

\subsubsection{Integrated Mean Squared Error (IMSE) Method}
The first method we discuss selects the values of $J_{-}$ and $J_{+}$ that minimize an asymptotic approximation to the integrated mean-squared error (IMSE) of the local means estimator, that is, the sum of the expansions of the (integrated) variance and squared bias. If we choose a large number of bins, we have a small bias because the bins are smaller and the local constant fit is better; but this reduction in bias comes at a cost, as increasing the number of bins leads to fewer observations per bin and thus more variability within bin. The IMSE-optimal $J_{-}$ and $J_{+}$ are the numbers of bins that balance squared-bias and variance so that the IMSE is (approximately) minimized.

By construction, choosing an IMSE-optimal number of bins will result in binned sample means that ``trace out'' the underlying regression function; this is useful to assess the overall shape of the regression function, perhaps to identify potential discontinuities in these functions that occur far from the cutoff. However, the IMSE-optimal method often results in a very smooth plot where the local means nearly overlap with the global polynomial fit, and may not be appropriate to capture the local variability of the data near the cutoff.

The IMSE-optimal values of $J_{-}$ and $J_{+}$ are, respectively, 
\[J_{-}^{\mathtt{IMSE}} = \left\lceil \mathscr{C}_-^{\mathtt{IMSE}}\;n^{1/3}\right\rceil \qquad \text{and} \qquad
J_{+}^{\mathtt{IMSE}} = \left\lceil \mathscr{C}_+^{\mathtt{IMSE}}\;n^{1/3}\right\rceil,\]
where $n$ is the total number of observations, $\lceil\cdot\rceil$ denotes the ceiling operator, and the exact form of the constants $\mathscr{C}_-^{\mathtt{IMSE}}$ and $\mathscr{C}_+^{\mathtt{IMSE}}$ depends on whether ES or QS bins are used (and some features of the underlying data generating process). In practice, the unknown constants $\mathscr{C}_-^{\mathtt{IMSE}}$ and $\mathscr{C}_+^{\mathtt{IMSE}}$ are estimated using preliminary, objective data-driven procedures.

In order to produce an RD plot that uses an IMSE-optimal number of evenly-spaced bins, we use the command \texttt{rdplot} with the option \texttt{binselect = "es"}, but this time omitting the \texttt{nbins = c(20 20)} option. When the number of bins is omitted, \texttt{rdplot} automatically chooses the number of bins according to the criterion specified with  \texttt{binselect}. We now produce an RD plot that uses ES bins and chooses the total number of bins on either side of the cutoff to be IMSE-optimal.

\labelsnippet{snippetLrdplotD}
\rsnip{Vol-1-R_meyersson_rdplot_es.txt}{\Rlink{\thesnippetLrdplotD}}
\statasnip{Vol-1-STATA_meyersson_rdplot_es}{\Slink{\thesnippetLrdplotD}}

\labelfiguras{figH}
\begin{figure}[H]
    \centering
    \includegraphics[scale=0.45]{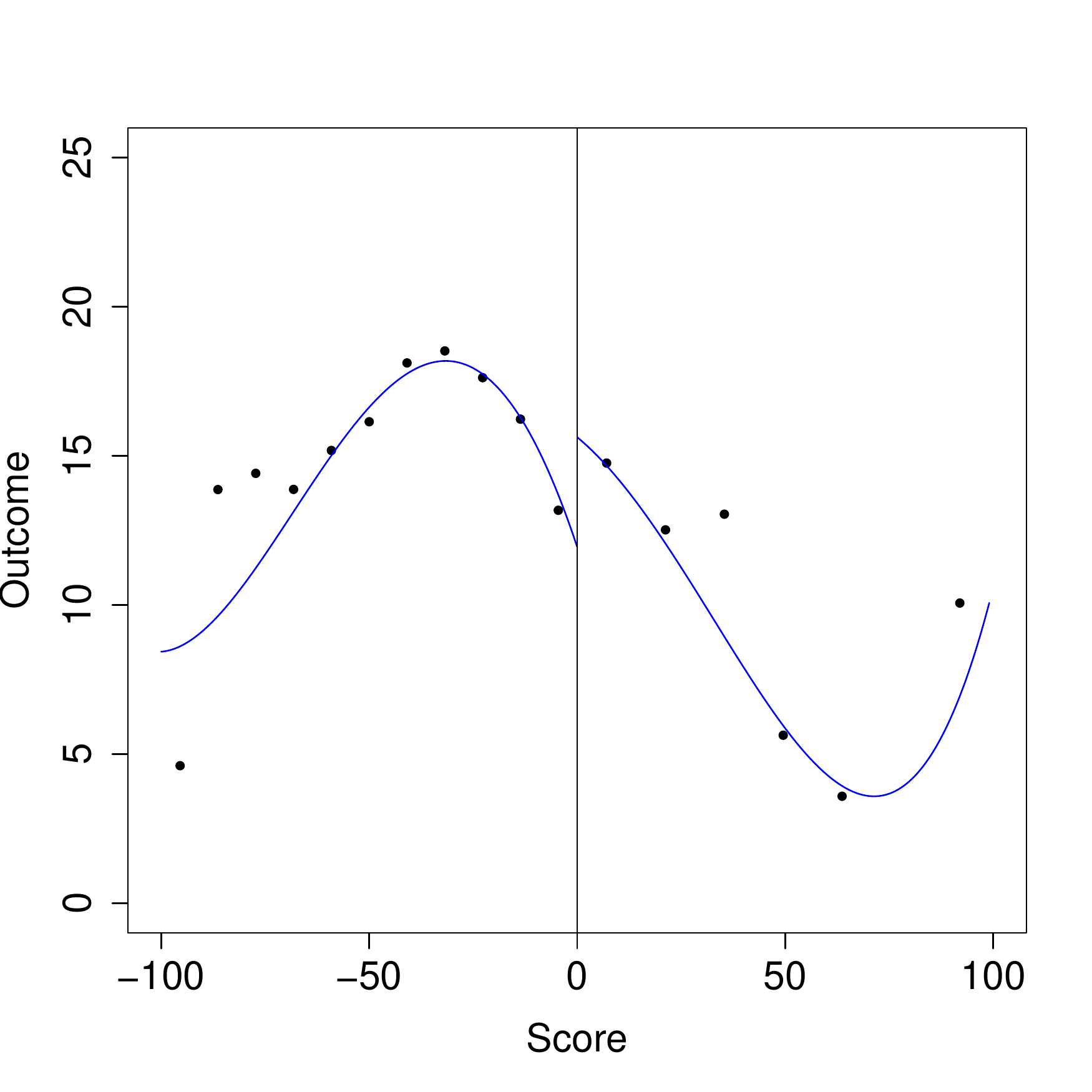}
    \caption{IMSE RD Plot with Evenly-Spaced Bins (Meyersson Data)}\label{fig:MeyerssonIMSE-ES}
\end{figure}

The plot is shown in Figure \ref{fig:MeyerssonIMSE-ES}. The output reports both the average and the median length of the bins. In the ES case, since each bin has the same length, each bin has length equal to both the average and the median length on each side. The IMSE criterion leads to different numbers of ES bins above and below the cutoff. As shown in the \texttt{Bins Selected} row, the IMSE-optimal number of bins is 11 below the cutoff and 7 above it. As a result, the lengths of the bins above and below the cutoff are different: above the cutoff, each bin has a length of 14.150 percentage points, while below the cutoff the bins are smaller, with a length of 9.091. The middle rows show the optimal number of bins according to both the IMSE criterion and the mimicking variance criterion (we discuss the latter in the next subsection). The bottom three rows show the bias and variance weights implied by the chosen number of bins in the IMSE objective function. When the IMSE criterion is used, these weights are always equal to 1/2.

To produce an RD plot that uses an IMSE-optimal number of quantile-spaced bins, we use the option \texttt{binselect = "qs"} instead of \texttt{binselect = "es"}.

\labelsnippet{snippetLrdplotE}
\rsnip{Vol-1-R_meyersson_rdplot_qs.txt}{\Rlink{\thesnippetLrdplotE}}
\statasnip{Vol-1-STATA_meyersson_rdplot_qs}{\Slink{\thesnippetLrdplotE}}

\labelfiguras{figI}
\begin{figure}[H]
    \centering
    \includegraphics[scale=0.45]{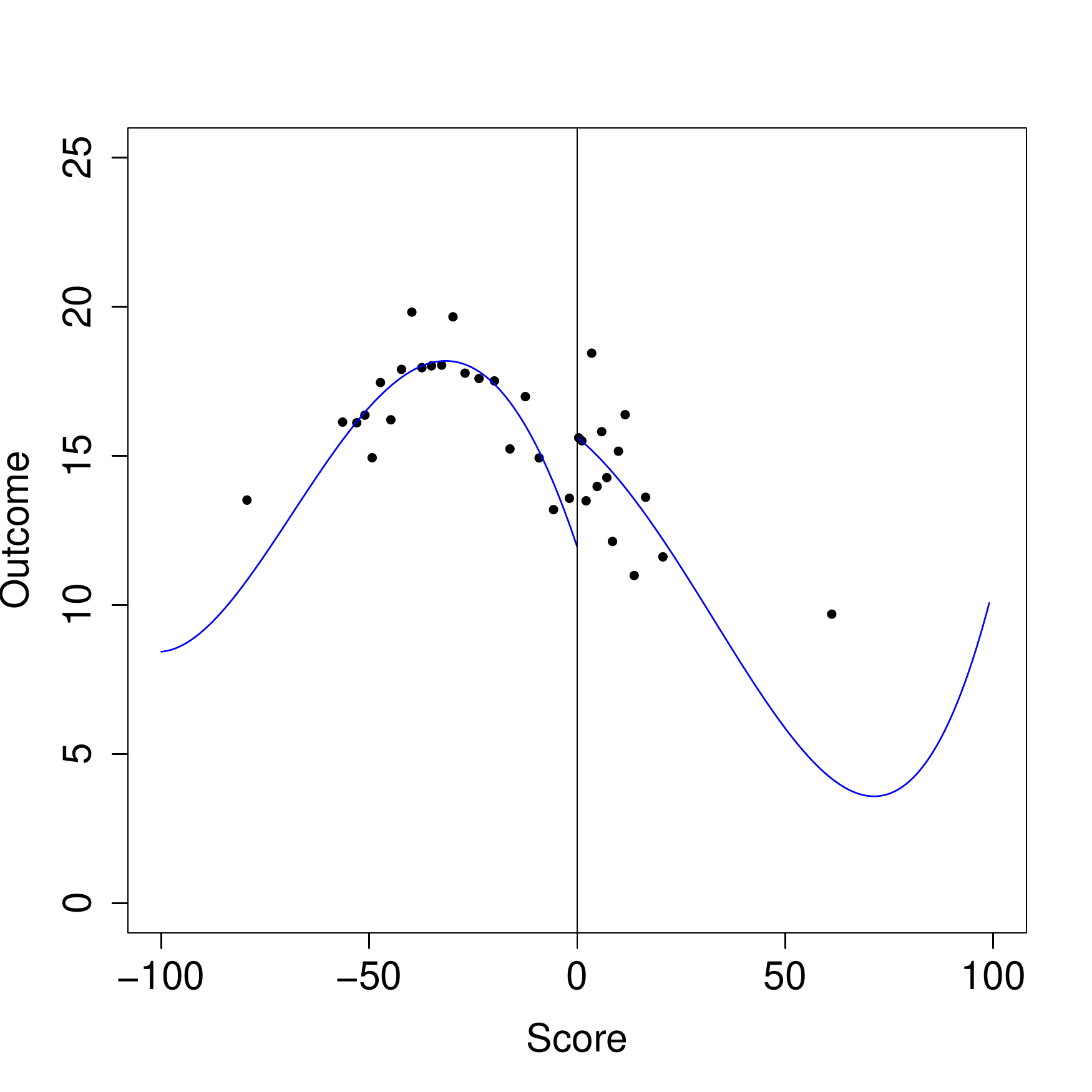}
    \caption{IMSE RD Plot with Quantile-Spaced Bins (Meyersson Data)}\label{fig:MeyerssonIMSE-QS}
\end{figure}

The resulting plot is shown in Figure \ref{fig:MeyerssonIMSE-QS}. Note that the IMSE-optimal number of QS bins is much larger on both sides, with 21 bins below the cutoff and 14 above it, versus 11 and 7 in the analogous ES plot in Figure \ref{fig:MeyerssonIMSE-ES}. The average bin length is 4.7572 below the cutoff, and 7.0821 above it. As expected, the median length of the bins is much smaller than the average length on both sides of the cutoff, particularly above. Since there are very few observations where the Islamic vote margin is above 50\%, the length of the last bin above the cutoff must be very large in order to ensure that it contains $315/14 \approx 22$ observations.

\subsubsection{Mimicking Variance Method}
The second method to select the number of bins chooses the vales of $J_{-}$ and $J_{+}$ so that the binned means have an asymptotic (integrated) variability that is approximately equal to the variability of the raw data. In other words, the number of bins is chosen so that the overall variability of the binned means ``mimics'' the overall variability in the raw scatter plot of the data. In the Meyersson application, this method involves choosing $J_{-}$ and $J_{+}$ so that the binned means have a total variability approximately equal to the variability illustrated in Figure \ref{fig:ScatterMeyersson}. We refer to this choice of total number of bins as a mimicking variance (MV) choice.

The mimicking-variance values of $J_{-}$ and $J_{+}$ are
\[J_{-}^\mathtt{MV} = \left\lceil\mathscr{C}_-^{\mathtt{MV}}\;\frac{n}{\log(n)^2}\right\rceil, \qquad \text{and} \qquad
J_{+}^{\mathtt{MV}} = \left\lceil\mathscr{C}_+^{\mathtt{MV}}\;\frac{n}{\log(n)^2}\right\rceil,\]
where again $n$ is the sample size and the exact form of the constants $\mathscr{C}_-^{\mathtt{MV}}$ and $\mathscr{C}_+^{\mathtt{MV}}$ depends on whether ES or QS bins are used (and some features of the underlying data generating process). These constants are different from those appearing in the IMSE-optimal choices and, in practice, are also estimated using preliminary, objective data-driven procedures.

In general, $J_{-}^\mathtt{MV} > J_{-}^\mathtt{ES}$ and $J_{+}^\mathtt{MV} > J_{+}^\mathtt{ES}$. That is, the MV method leads to a larger number of bins than the IMSE method, resulting in an RD plot with more dots representing local means and thus giving a better sense of the variability of the data. In order to produce an RD plot with ES bins and an MV total number of bins on either side, we use the option \texttt{binselect ="esmv"}.

\labelsnippet{snippetLrdplotF}
\rsnip{Vol-1-R_meyersson_rdplot_esmv.txt}{\Rlink{\thesnippetLrdplotF}}
\statasnip{Vol-1-STATA_meyersson_rdplot_esmv}{\Slink{\thesnippetLrdplotF}}

\labelfiguras{figJ}
\begin{figure}[H]
    \centering
    \includegraphics[scale=0.45]{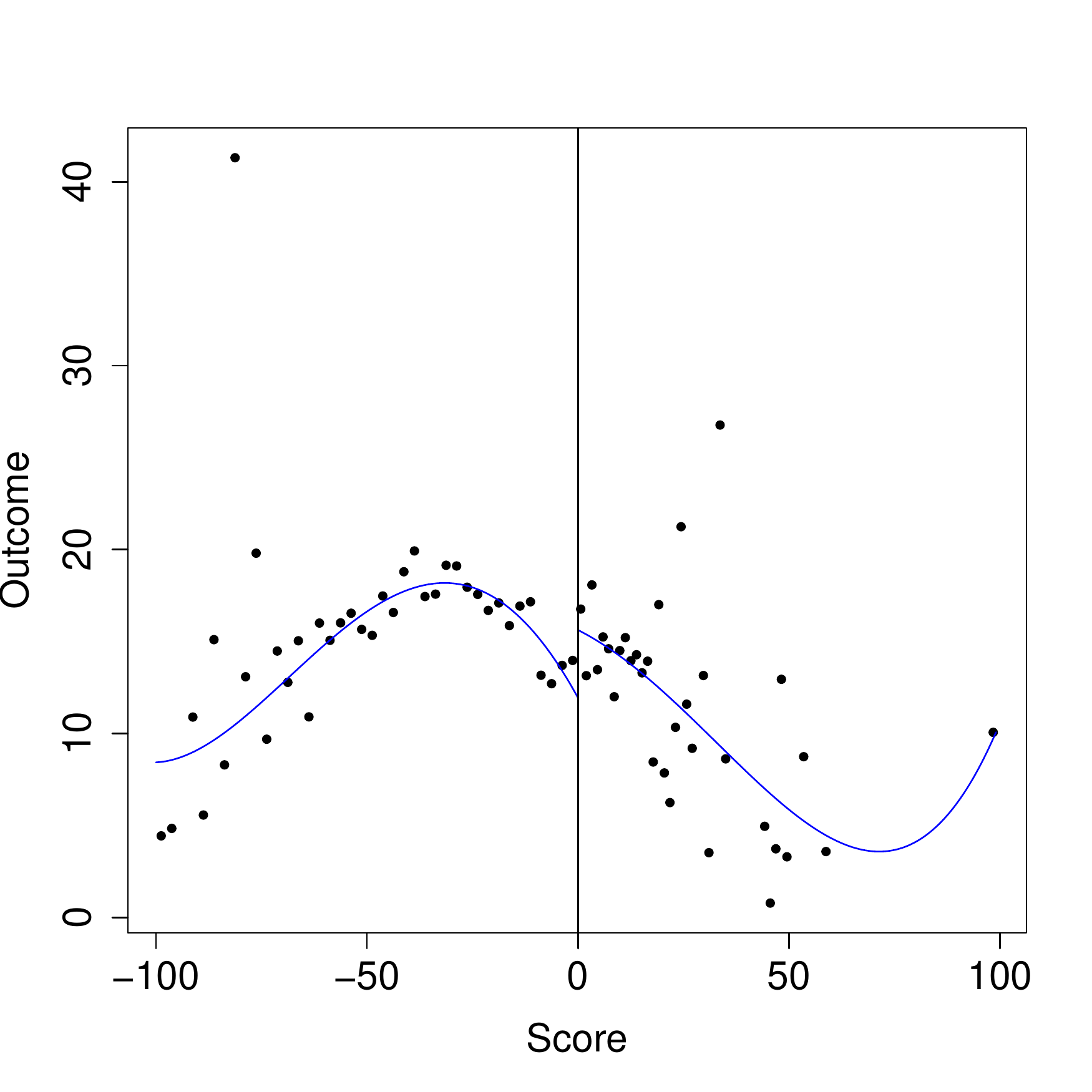}
    \caption{Mimicking Variance RD Plot with Evenly-Spaced Bins (Meyersson Data)}
		\label{fig:MeyerssonMV-ES}
\end{figure}

As shown in the output and illustrated in Figure \ref{fig:MeyerssonMV-ES}, this produces a much higher number of bins than we obtained with the IMSE criterion for both ES and QS bins. The MV total number of bins is 40 below the cutoff and 75 above the cutoff, with length 2.5 and 1.321, respectively. The difference in the chosen number of bins between the IMSE and the MV criteria is dramatic. The middle rows show the number of bins that would have been produced according to the IMSE criterion (11 and 7) and the number of bins that would have been produced according to the MV criterion (40 and 75). This allows for a quick comparison between both methods. Finally, the bottom rows indicate that the chosen number of MV bins on both sides of the cutoff is equivalent to the number of bins that would have been chosen according to an IMSE criterion where, instead of giving the bias and the variance each a weight of 1/2, these weights had been, respectively, 0.020 and 0.980 below the cutoff, and 0.001 and 0.999 above the cutoff. Thus, we see that if we want to justify the MV choice in terms of the IMSE criterion, we must weigh the bias much more than the variance.

Finally, to create an RD plot that chooses the total number of bins according to the MV criterion but uses QS bins, we use the option \texttt{binselect = "qsmv"}.

\labelsnippet{snippetLrdplotG}
\rsnip{Vol-1-R_meyersson_rdplot_qsmv.txt}{\Rlink{\thesnippetLrdplotG}}
\statasnip{Vol-1-STATA_meyersson_rdplot_qsmv}{\Slink{\thesnippetLrdplotG}}

\labelfiguras{figK}
\begin{figure}[H]
    \centering
    \includegraphics[scale=0.45]{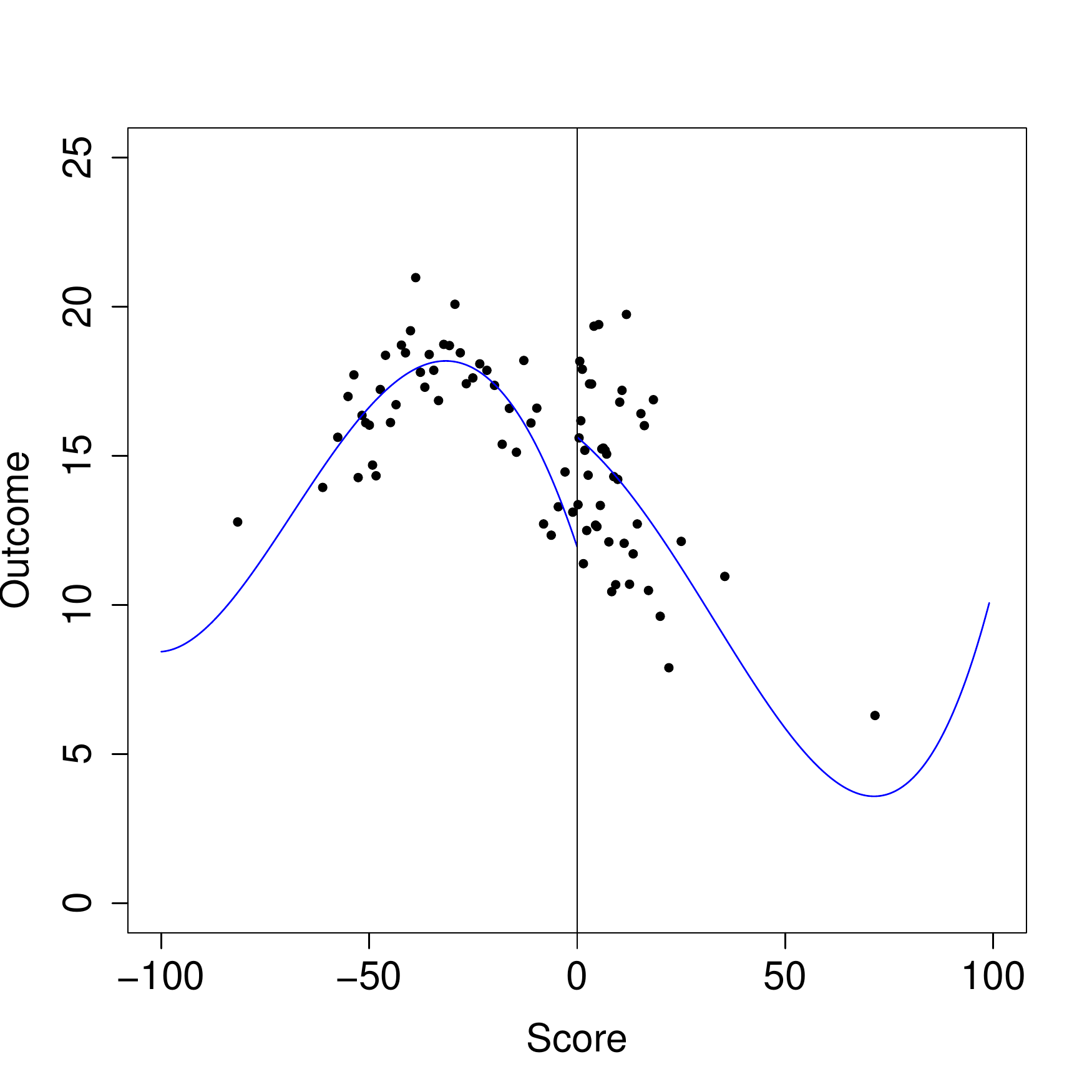}
    \caption{Mimicking Variance RD Plot with Quantile-Spaced Bins---Meyersson Data}
		\label{fig:MeyerssonMV-QS}
\end{figure}

The resulting plot is shown in Figure \ref{fig:MeyerssonMV-QS}. Below the cutoff, the MV number of QS bins is very similar to the MV choice for ES bins (44 versus 40). However, above the cutoff, the MV number of QS bins is much lower than for ES bins (41 versus 75). This occurs because, although the range of the running variable is $[-100,99.051]$, there are very few observations in the intervals  $[-100,-50]$ and $[50,100]$ far from the cutoff. Since ES bins force the length of the bins to be the same everywhere in the support, the number of ES bins has to be large in order to produce small enough bins to adequately mimic the overall variability of the scatter plot in regions with few observations. In contrast, QS bins can be short near the cutoff and long away from the cutoff, so they can mimic the overall variability by adapting their length to the density of the data. 

In sum, bins can be chosen in many different ways. Which method of implementation is most appropriate depends on the researcher's particular goal, for example, illustrating/testing for the overall functional form versus showing the variability of the data. We recommend to start with MV bins to better illustrate the variability of the outcome as a function of the score, ideally comparing ES to QS bins to highlight the distributional features of the score. Then, if needed, the researcher can select the number of bins to be IMSE-optimal in order to explore the global features of the regression function.

\subsection{Further Reading}

A detailed discussion of RD plots and formal methods for automatic data-driven bin selection are given by \citet*{Calonico-Cattaneo-Titiunik_2015_JASA}. This paper formalized the commonly used RD plots with evenly-spaced binning, introduced RD plots with quantile-spaced binning, and developed optimal choices for the number of bins in terms of both integrated mean squared error and mimicking variance targets. See also \citet*{Calonico-Cattaneo-Farrell-Titiunik_2017_Stata} for other features of RD plots, including confidence intervals for the local means in each bin. RD plots are special cases of nonparametric partitioning estimators---see \cite{Cattaneo-Farrell_2013_JoE}, \citet*{Cattaneo-Farrell-Feng_2019_AoS}, and references therein. Finally, see \citet*{Cattaneo-Crump-Farrell-Feng_2019_wp} for closely related binscatter methods.

\newpage
\markedsection{Continuity-based RD Approach}{The Continuity-Based Approach to RD Analysis}
\label{sec:localpoly}

We now discuss empirical methods for estimation and inference in RD designs based on continuity assumptions and extrapolation towards the cutoff point, which rely on large-sample approximations with random potential outcomes under repeated sampling. These methods offer tools useful not only for the analysis of main treatment effects, but also for falsification and validation of the design, which we discuss in Section \ref{sec:falsification}. The approach discussed here is based on formal statistical methods and hence leads to disciplined and objective empirical analysis, which typically has two related but distinct goals: point estimation of RD treatment effect (i.e., give a scalar estimate of the vertical distance between the regression functions at the cutoff), and statistical inference about the RD treatment effect (i.e., construct valid statistical hypothesis tests and confidence intervals).

The methods discussed in this section are based on the continuity conditions underlying Equation (\ref{HTV}), and generalizations thereof. This framework for RD analysis, which we call the \textit{continuity-based} RD framework, uses methodological tools that directly rely on continuity (and differentiability) assumptions and define $\tau_{\mathtt{SRD}}$ as the parameter of interest. In this framework, estimation typically proceeds by using (local to the cutoff) polynomial methods to approximate the regression function $\E[Y_i|X_i=x]$ on each side of the cutoff separately. In practical terms, this involves using least-squares methods to fit a polynomial of the observed outcome on the score. When all the observations are used for estimation, these polynomial fits are global or parametric in nature, like those used in the default RD plots discussed in the previous section. In contrast, when estimation employs only observations with scores near the cutoff, the polynomial fits are local, ``flexible,'' or ``non-parametric.'' Our upcoming discussion focuses exclusively on local polynomial methods, which are by now the standard framework for RD empirical analysis because they offer a good compromise between flexibility and simplicity.

In the second Element (\textit{A Practical Introduction to Regression Discontinuity Designs: Extensions}; \citeauthor*{Cattaneo-Idrobo-Titiunik_2019_Vol2}, forthcoming), we discuss an alternative framework for RD analysis that relies on assumptions of local random assignment of the treatment near the cutoff, and employs tools and ideas from the literature on the analysis of experiments. This alternative approach offers a complement to, and a robustness check for, the local polynomial methods based on continuity assumptions that we discuss in the remainder of this Element. Furthermore, the local randomization RD approach can be used in cases where local polynomial methods are invalid or difficult to justify. 

\subsection{Local Polynomial Approach: Overview}

A fundamental feature of the RD design is that, in general, there are no observations for which the score $X_i$ is exactly equal to the cutoff value $\C$: because the running variable is assumed continuous, there are no (or sometimes in practice very few) observations whose score is $\C$ or very nearly so. Thus, local extrapolation in RD designs is unavoidable in general. In other words, in order to form estimates of the average control response at the cutoff, $\E[Y_i(0) | X_i = \C]$, and of the average treatment response at the cutoff, $\E[Y_i(1) | X_i = \C]$, we must rely on observations further away from the cutoff. In the Sharp RD design, for example, the treatment effect $\tau_{\mathtt{SRD}}$ is the vertical distance between the $\E[Y_i(1) | X_i = x]$ and $\E[Y_i(0) | X_i = x]$ at $x=\C$, as shown in Figure \ref{fig:RDeffect-Model1}, and thus estimation and inference proceed by first \textit{approximating} these unknown regression functions, and then computing the estimated treatment effect and/or the statistical inference procedure of interest. In this context, the key practical issue in RD analysis is how the approximation of the unknown regression functions is done, as this will directly affect the robustness and credibility of the empirical findings.

The problem of approximating an unknown function is well understood: any sufficiently smooth function can be well approximated by a polynomial function, locally or globally, up to an error term. A large literature in statistics has used this principle to develop non-parametric methods based on polynomials or other bases of approximation, relaxing strong parametric assumptions and relying instead on more flexible approximations of the unknown regression function. Applied to the RD point estimation problem, this principle suggests that the unknown regression functions $\E[Y_i(0) | X_i = x]$ and $\E[Y_i(1) | X_i = x]$ can be approximated by a polynomial function of the score. The available statistical results have to be adapted to the RD case, considering the complications that arise because the approximation must occur at the cutoff, which is a boundary point.

Early empirical work employed the idea of polynomial approximation globally, that is, tried to approximate these functions using flexible higher-order polynomials, usually of fourth or fifth order, over the entire support of the data. This global approach is still used in RD plots, as illustrated in the previous section, because the goal there is to illustrate the \textit{entire} unknown regression functions. However, it is now widely recognized that a global polynomial approach does not deliver point estimators and inference procedures with good properties for the RD treatment effect, the main object of interest. The reason is that global polynomial approximations tend to deliver a good approximation overall, but a poor approximation at boundary points---a problem known as Runge's phenomenon in approximation theory. Moreover, global approximations can induce counter-intuitive weighting schemes, for example, when the point estimator is heavily influenced by observations far from the boundary. Since the RD point estimator is defined at a boundary point, global polynomial methods can lead to unreliable RD point estimators, and thus the conclusions from a global parametric RD analysis can be highly misleading. For these reasons, we recommend against using global polynomial methods for formal RD analysis.

Modern RD empirical work employs local polynomial methods, which focus on approximating the regression functions only near the cutoff. Because this approach localizes the polynomial fit to the cutoff (discarding observations sufficiently far away) and employs a low-order polynomial approximation (usually linear or quadratic), it is substantially more robust and less sensitive to boundary and overfitting problems. Furthermore, this approach can be viewed formally as a non-parametric local polynomial approximation, which has also aided the development of a comprehensive toolkit of statistical and econometric results for estimation and inference. In contrast to global higher-order polynomials, local lower-order polynomial approximations can be viewed as intuitive approximations with a potential misspecification of the functional form of the regression function near the cutoff, which can be modeled and understood formally, with the advantage that they are less sensitive to outliers or other extreme features of the data generating process far from the cutoff. Local polynomial methods employ only observations close to the cutoff, and interpret the polynomial used as a local approximation, not necessarily as a correctly specified model. 

The statistical properties of local polynomial estimation and inference depend crucially on the accuracy of the approximation near the cutoff, which is controlled by the size of the neighborhood around the cutoff where the local polynomial is fit. In the upcoming sections, we discuss the modern local polynomial methods for RD analysis, and explain all the steps involved in their implementation for both estimation and inference. We also discuss several extensions and modifications, including the inclusion of predetermined covariates and the use of cluster-robust standard errors.

\subsection{Local Polynomial Point Estimation}

Local polynomial methods implement linear regression fits using only observations near the cutoff point, separately for control and treatment units. Specifically, this approach uses only observations that are between $\C-h$ and $\C+h$, where $h>0$ is a so-called bandwidth that determines the size of the neighborhood around the cutoff where the empirical RD analysis is conducted. Within this bandwidth, it is common to adopt a weighting scheme to ensure that the observations closer to $\C$ receive more weight than those further away; the weights are determined by a kernel function $K(\cdot)$. The local polynomial approach can be understood and analyzed formally as non-parametric, in which case the fit is taken as an approximation to the unknown underlying regression functions within the region determined by the bandwidth.

Local polynomial estimation consists of the following basic steps.
\begin{enumerate}
\item Choose a polynomial order $p$  and a kernel function $K(\cdot)$.
\item Choose a bandwidth $h$. 
\item For observations above the cutoff (i.e., observations with $X_i \geq \C$), fit a weighted least squares regression of the outcome $Y_i$ on a constant and $(X_i-\C), (X_i-\C)^2,\ldots, (X_i-\C)^p$, where $p$ is the chosen polynomial order, with weight $K(\frac{X_i-\C}{h})$ for each observation. The estimated intercept from this local weighted regression, $\hat{\mu}_+$, is an estimate of the point $\mu_+=\E[Y_i(1)|X_i=\C]$: 
\[\hat{\mu}_+  \;\;:\;\;\hat Y_i = \hat{\mu}_+ + \hat\mu_{+,1} (X_i-\C) + \hat\mu_{+,2} (X_i-\C)^2 + \cdots + \hat \mu_{+,p} (X_i-\C)^p\text{.} \]

\item For observations below the cutoff (i.e., observations with $X_i < \C$), fit a weighted least-squares regression of the outcome $Y_i$ on a constant and $(X_i-\C), (X_i-\C)^2,\ldots, (X_i-\C)^p$, where $p$ is the chosen polynomial order, with weight $K(\frac{X_i-\C}{h})$ for each observation. The estimated intercept from this local weighted regression, $\hat{\mu}_-$, is an estimate of the point $\mu_-=\E[Y_i(0)|X_i=\C]$: 
\[\hat{\mu}_-  \;\;:\;\;\hat Y_i = \hat{\mu}_- + \hat\mu_{-,1} (X_i-\C) + \hat\mu_{-,2} (X_i-\C)^2 + \cdots + \hat \mu_{-,p} (X_i-\C)^p \text{.}\]

\item Calculate the Sharp RD point estimate: $\hat{\tau}_\mathtt{SRD} = \hat{\mu}_+ - \hat{\mu}_-$.
\end{enumerate}

\begin{figure}[H]
\centering
\includegraphics[scale=0.75]{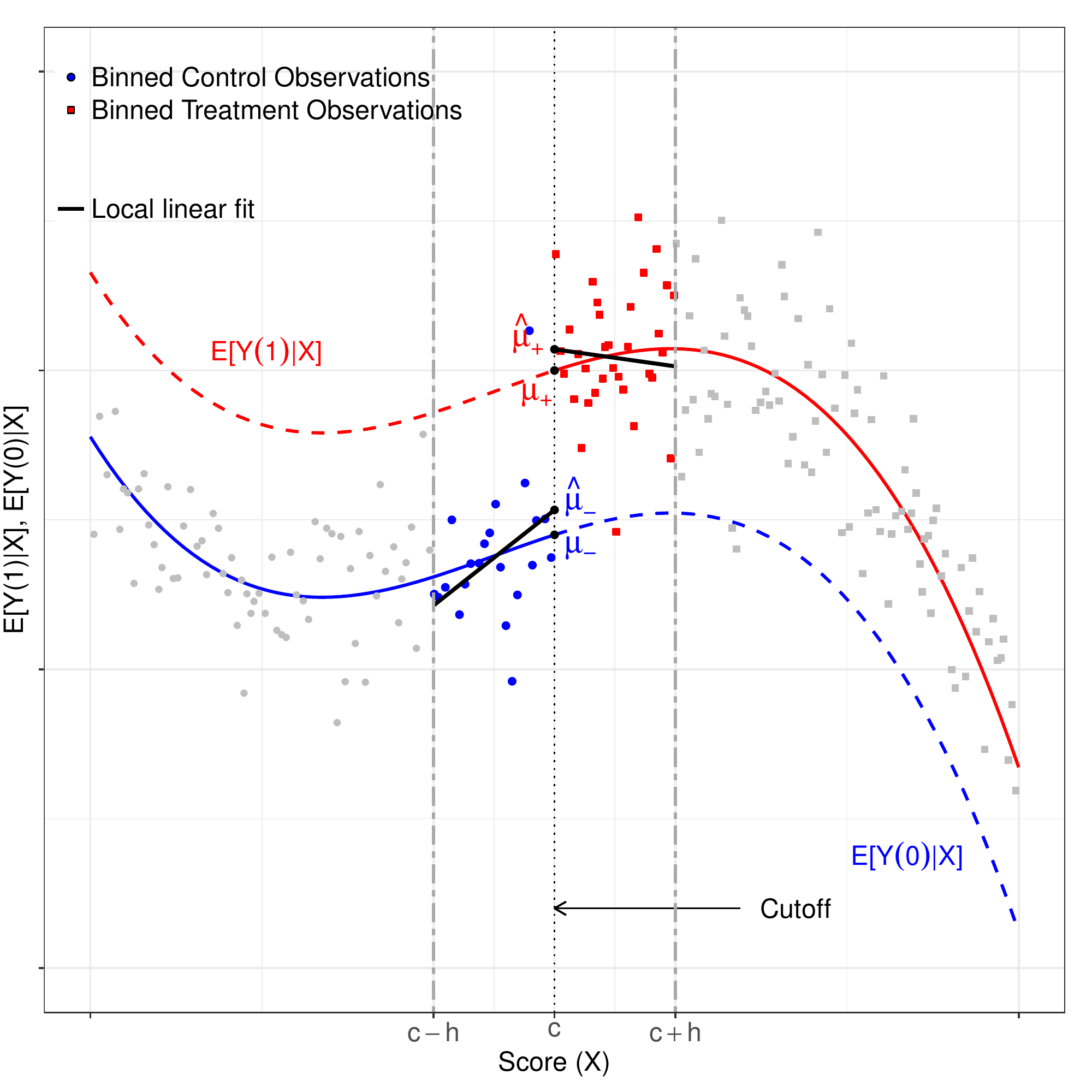}
\caption{RD Estimation with Local Polynomial}
\label{fig:RDest-localpoly}
\end{figure}

A graphical representation of local polynomial RD point estimation is given in Figure \ref{fig:RDest-localpoly}, where a polynomial of order one ($p=1$) is fit within bandwidth $h_1$; observations outside this bandwidth are not used in the estimation. The RD effect is $\tau_{\mathtt{SRD}}=\mu_+ - \mu_-$ and the local polynomial estimator of this effect is $\hat{\mu}_+ - \hat{\mu}_-$. Local polynomial methods produce the fit employing the raw data, not the binned data typically reported in the RD plots.

\labelfiguras{figL}

The implementation of the local polynomial approach thus requires the choice of three main ingredients: the kernel function $K(\cdot)$, the order of the polynomial $p$, and the bandwidth $h$. We now turn to a discussion of each of these choices. \smallskip

\subsubsection{Choice of Kernel Function and Polynomial Order}

The kernel function $K(\cdot)$ assigns non-negative weights to each transformed observation $\frac{X_i-\C}{h}$, based on the distance between the observation's score $X_i$ and the cutoff $\C$. The recommended choice is the triangular kernel function, $K(u) = (1-|u|) \I(|u|\leq 1)$, because when used in conjunction with a bandwidth that optimizes the mean squared error (MSE), it leads to a point estimator with optimal properties (more details about MSE-optimal bandwidths are given below). As illustrated in Figure \ref{fig:kernels}, the triangular kernel function assigns zero weight to all observations with score outside the interval $[\C-h, \C+h]$, and positive weights to all observations within this interval. The weight is maximized at $X_i=\C$, and declines symmetrically and linearly as the value of the score gets farther from the cutoff. 

Despite the desirable asymptotic optimality properties of the triangular kernel, researchers sometimes prefer to use the more simple uniform kernel $K(u) = \I(|u|\leq 1)$, which also gives zero weight to observations with score outside $[\C-h, \C+h]$, but equal weight to all observations whose scores are within this interval, see Figure \ref{fig:kernels}. Employing a local linear estimation with bandwidth $h$ and uniform kernel is therefore equivalent to estimating a simple linear regression without weights using only observations whose distance from the cutoff is at most $h$. A uniform kernel minimizes the asymptotic variance of the local polynomial estimator under some technical conditions. A third weighting scheme sometimes encountered in practice is the Epanechnikov kernel, $K(u) = (1-u^2)\I(|u|\leq 1)$, also depicted in Figure \ref{fig:kernels}, which gives a quadratic decaying weight to observations with $X_i \in [\C-h, \C+h]$ and zero weight to the rest. In practice, estimation and inference results are typically not very sensitive to the particular choice of kernel used.

\labelfiguras{figM}
\begin{figure}[h]
\centering
\includegraphics[scale=0.65]{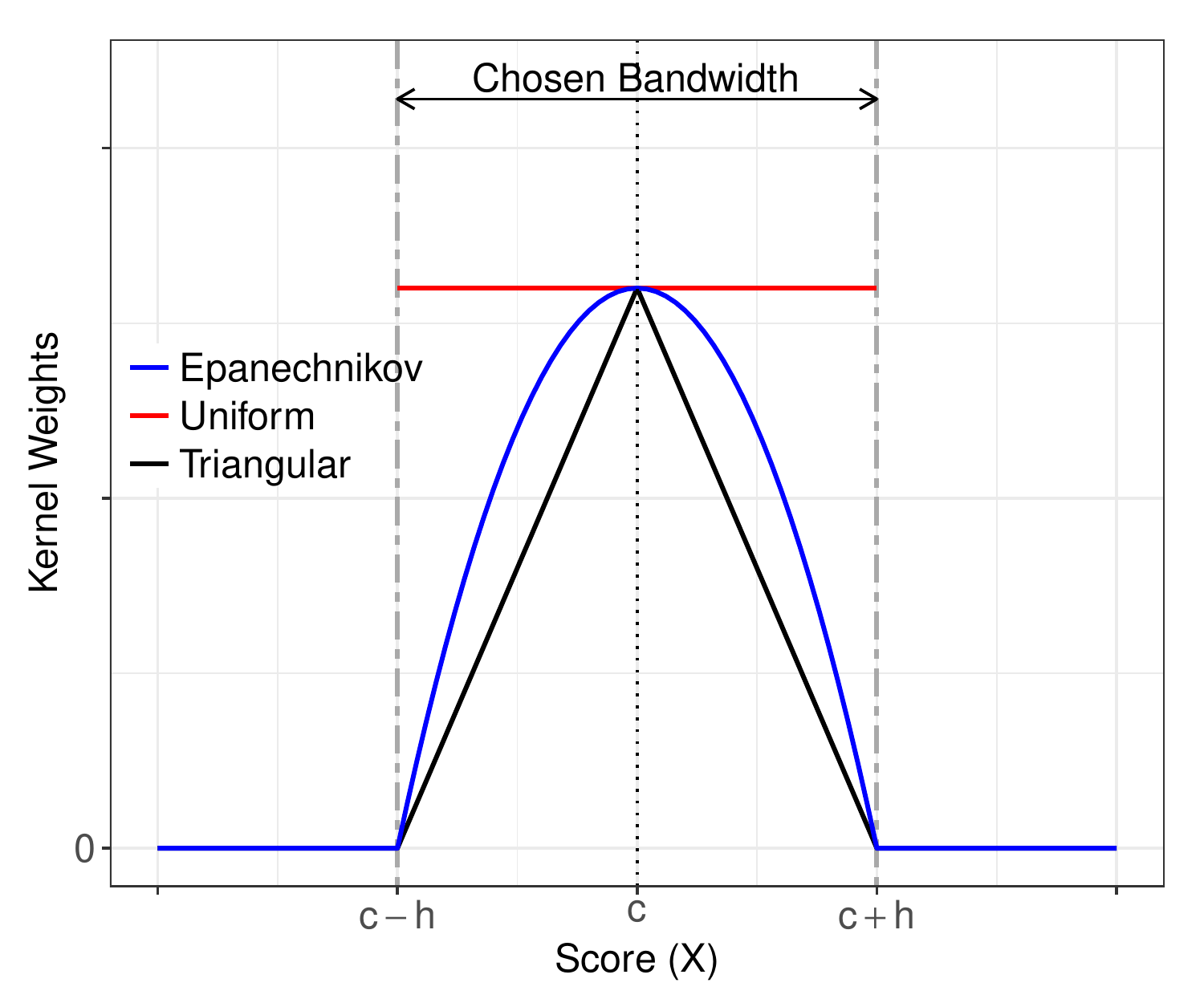}
\caption{Different Kernel Weights for RD Estimation}
\label{fig:kernels}
\end{figure}

A more consequential decision is the choice of the local polynomial order, which must consider various factors. First, a polynomial of order zero---a constant fit---has undesirable theoretical properties at boundary points, which is precisely where RD estimation must occur. Second, for a given bandwidth, increasing the order of the polynomial generally improves the accuracy of the approximation but also increases the variability of the treatment effect estimator. Third, as mentioned above, higher-order polynomials tend to produce overfitting of the data and lead to unreliable results near boundary points. Combined, these factors have led researchers to prefer the local linear RD estimator, which by now is the default point estimator in most applications. In finite samples, of course, the ranking between different local polynomial estimators may be different, but in general the local linear estimator seems to deliver a good trade-off between simplicity, precision, and stability in RD settings.

Although it may seem at first that a linear polynomial is not flexible enough, an appropriately chosen bandwidth  will adjust to the chosen polynomial order so that the linear approximation to the unknown regression functions is reliable. We turn to this issue below.

\subsubsection{Bandwidth Selection and Implementation}
\label{subsec:bw}

The bandwidth $h$ controls the width of the neighborhood around the cutoff that is used to fit the local polynomial that approximates the unknown regression functions. The choice of $h$ is fundamental for the analysis and interpretation of RD designs, as $h$ directly affects the properties of local polynomial  estimation and inference procedures, and empirical findings are often sensitive to its particular value.

\begin{figure}[H]	
\centering
\includegraphics[scale=0.80]{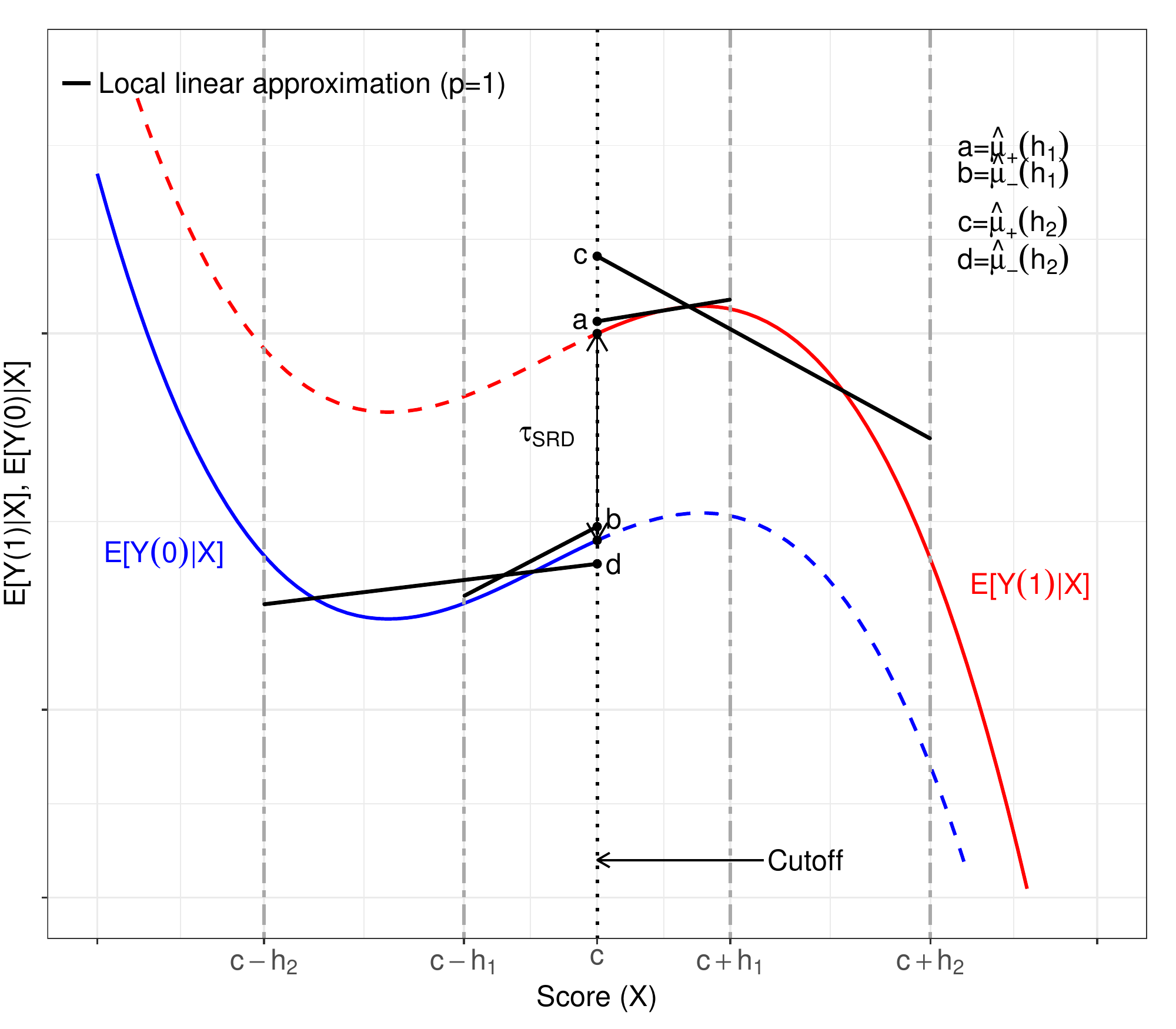}
\caption{Bias in Local Approximations}
\label{fig:linearapprox}
\end{figure}

Figure \ref{fig:linearapprox} illustrates how the error in the approximation is directly related to the bandwidth choice. The unknown regression functions in the figure, $\E[Y_i(1) | X_i = x]$ and $\E[Y_i(0) | X_i = x]$, have considerable curvature. At first, it would seem inappropriate to approximate these functions with a linear polynomial. Indeed, inside the interval $[\C-h_2,\C+h_2]$, a linear approximation yields an estimated RD effect equal to $\hat{\mu}_+(h_2)-\hat{\mu}_-(h_2)$ (distance between points \texttt{c} and \texttt{d}), which is considerably different from the true effect $\tau_{\mathtt{SRD}}$. Thus, a linear fit within bandwidth $h_2$ results in a poor approximation because of misspecification error. However, reducing the bandwidth from $h_2$ to $h_1$ improves the linear approximation considerably, as now the estimated RD effect $\hat{\mu}_+(h_1)-\hat{\mu}_-(h_1)$ (distance between points \texttt{a} and \texttt{b}) is much closer to the population treatment effect $\tau_{\mathtt{SRD}}$. The reason is that the regression functions are nearly linear in the interval $[\C-h_1,\C+h_1]$, and therefore the linear approximation results in a smaller misspecification error when the bandwidth shrinks from $h_2$ to $h_1$. This illustrates the general principle that, given a polynomial order, the accuracy of the approximation can always be improved by reducing the bandwidth. 

\labelfiguras{figN}

Choosing a smaller $h$ will reduce the misspecification error (also known as ``smoothing bias'') of the local polynomial approximation, but will simultaneously tend to increase the variance of the estimated coefficients because fewer observations will be available for estimation. On the other hand, a larger $h$ will result in more smoothing bias if the unknown function differs considerably from the polynomial model used for approximation, but will reduce the variance because the number of observations in the interval $[\C-h,\C+h]$ will be larger. For this reason, the choice of bandwidth is said to involve a ``bias-variance trade-off.''

Since RD empirical results are often sensitive to the choice of bandwidth, it is important to select $h$ in a data-driven, automatic way to avoid specification searching and ad hoc decisions. Most bandwidth selection methods try to balance some form of bias-variance trade-off (sometimes involving other features of the estimator, inference procedure, and data generating process). The most popular approach in practice seeks to minimize the MSE of the local polynomial RD point estimator, $\hat{\tau}_\mathtt{SRD}$, given a choice of polynomial order and kernel function. Since the MSE of an estimator is the sum of its squared bias and its variance, this approach effectively chooses $h$ to optimize a bias-variance trade-off. The precise procedure involves deriving an asymptotic approximation to the MSE of  $\hat{\tau}_\mathtt{SRD}$, optimizing it with respect to $h$, and using data-driven methods to estimate the unknown quantities in the resulting formula of the optimal $h$.

The general form of the approximate (conditional) MSE for the RD treatment effect is
\[\mathsf{MSE}(\hat{\tau}_\mathtt{SRD}) =\mathsf{Bias}^2(\hat{\tau}_\mathtt{SRD}) + \mathsf{Variance}(\hat{\tau}_\mathtt{SRD})
  = \mathscr{B}^2 + \mathscr{V},\]
where the approximate (conditional) bias and variance of the estimator are
\[\mathscr{B} = h^{2(p+1)}\;\mathcal{B} \qquad\text{and}\qquad
  \mathscr{V} = \frac{1}{nh}\;\mathcal{V},
\]
respectively. The quantities $\mathcal{B}$ and $\mathcal{V}$ represent, respectively, the (leading) bias and variance of the RD point estimator $\hat{\tau}_\mathtt{SRD}$, not including the rates controlled by the sample size and bandwidth choice. Although we omit the technical details, we present the general form of $\mathscr{B}$ and $\mathscr{V}$ to clarify the most important trade-offs involved in the choice of an MSE-optimal bandwidth for the local polynomial RD estimate, and because these quantities will be used for inference below.

The general form of the bias $\mathscr{B}$ is determined by the bandwidth $h^{2(p+1)}$ and the quantities
\[\mathcal{B}   = \mathcal{B}_+ - \mathcal{B}_-, \qquad
  \mathcal{B}_- \approx \mu _{-}^{(p+1)}B_{-}, \qquad
  \mathcal{B}_+ \approx \mu _{+}^{(p+1)}B_{+}
\]
where the derivatives
\[\mu_{+}^{(p+1)} = \lim_{x \downarrow \C}\frac{d^{p+1}\E[Y_i(1)|X=x]}{dx^{p+1}}
  \qquad\text{and}\qquad
  \mu_{-}^{(p+1)} = \lim_{x \uparrow \C}\frac{d^{p+1}\E[Y_i(0)|X=x]}{dx^{p+1}}
\]
are related to the ``curvature'' of the unknown regression functions for treatment and control units, respectively, and the known constants $B_{+}$ and $B_{-}$ are related to the kernel function and the order of the polynomial used. These calculations assume a common bandwidth $h$, but the expressions can be extended to allow for different bandwidths on the left and right of the cutoff. 

The bias term $\mathscr{B}$ associated with the local polynomial RD point estimator of order $p$, $\hat{\tau}_{\mathtt{SRD}}$, depends on the $(p+1)$th derivatives of the regression functions $\E[Y_i(1)|X=x]$ and $\E[Y_i(0)|X=x]$ with respect to the running variable. This is a more formal characterization of the phenomenon we illustrated in Figure \ref{fig:linearapprox}. When we approximate $\E[Y_i(1)|X=x]$ and $\E[Y_i(0)|X=x]$ with a local polynomial of order $p$, that approximation has an error (unless $\E[Y_i(1)|X=x]$ and $\E[Y_i(0)|X=x]$ happen to be polynomials of at most order $p$). The leading term of the approximation error is related to the derivative of order $p+1$, that is, the order following the polynomial order used to estimate $\tau_{\mathtt{SRD}}$. For example, as illustrated in Figure \ref{fig:linearapprox}, if we use a local linear polynomial to estimate $\tau_{\mathtt{SRD}}$, our approximation by construction ignores the second-order term (which depends on the second derivative of the function), and all higher-order terms (which depend on the higher-order derivatives). Thus, the leading bias associated with a local linear estimator depends on the second derivatives of the regression functions, which are the leading terms in the error of approximation incurred when we set $p=1$. Similarly, if we use a local quadratic polynomial to estimate $\hat{\tau}_{\mathtt{SRD}}$, the leading bias will depend on the third derivatives of the regression function. 

The variance term $\mathscr{V}$ depends on the sample size and bandwidth through the expression $\frac{1}{nh}$ and also involves the quantities
\[\mathcal{V} = \mathcal{V}_- + \mathcal{V}_+,\qquad
  \mathcal{V}_- \approx \frac{\sigma^2_-}{f} V_-, \qquad
  \mathcal{V}_+ \approx \frac{\sigma^2_+}{f} V_+ 
\]
where
\[\sigma^2_+ = \lim_{x \downarrow \C}\V[Y_i(1)|X_i=x]
  \qquad\text{and}\qquad
  \sigma^2_- = \lim_{x \uparrow \C}\V[Y_i(0)|X_i=x]
\]
capture the conditional variability of the outcome given the score at the cutoff for treatment and control units, respectively, $f$ denotes the density of the score variable at the cutoff, and the known constants $V_-$ and $V_+$ are related to the kernel function and the order of the polynomial used.

As the number of observations near the cutoff decreases (e.g., as the density $f$ decreases), the contribution of the variance term to the MSE increases, and vice versa as the number of observations near the cutoff increases. This captures the intuition that the variability of the RD point estimator will partly depend on the density of observations near the cutoff. Similarly, an increase (decrease) in the conditional variability of the outcome given the score will increase (decrease) the MSE of the RD point estimators.

In order to obtain an MSE-optimal point estimator $\hat{\tau}_\mathtt{SRD}$, we choose the bandwidth that minimizes the MSE approximation:
\[\min_{h>0}  \left(h^{2(p+1)}\;\mathcal{B}^2 + \frac{1}{nh}\;\mathcal{V} \right),\]
which leads to the MSE-optimal bandwidth choice
\[h_{\mathtt{MSE}} = \left(\frac{\mathcal{V}}{2(p+1)\mathcal{B}^2}\right)^{1/(2p+3)} \; n^{-1/(2p+3)}.\]
This formula formally incorporates the bias-variance trade-off mentioned above. It follows that $h_{\mathtt{MSE}}$ is proportional to $n^{-1/(2p+3)}$, and that this MSE-optimal bandwidth increases with $\mathcal{V}$ and decreases with $\mathcal{B}$. In other words, a larger asymptotic variance will lead to a larger MSE-optimal bandwidth; this is intuitive, as a larger bandwidth will include more observations in the estimation and thus reduce the variance of the resulting point estimator. In contrast, a larger asymptotic bias will lead to a smaller bandwidth, as a smaller bandwidth will reduce the approximation error and reduce the bias of the resulting point estimator.

Another way to see this trade-off is to note that if we chose a bandwidth $h>h_{\mathtt{MSE}}$, decreasing $h$ would lead to a reduction in the approximation error and an increase in the variability of the point estimator, but the MSE reduction caused by the decrease in bias would be larger than the MSE increase caused by the variance increase, leading to a smaller MSE overall. In other words, when $h>h_{\mathtt{MSE}}$, it is possible to reduce the misspecification error without increasing the MSE. In contrast, when we set $h=h_{\mathtt{MSE}}$, both increasing and decreasing the bandwidth necessarily lead to a higher MSE. 

Given the quantities $\mathcal{V}$ and $\mathcal{B}$, increasing the sample size $n$ leads to a smaller optimal $h_{\mathtt{MSE}}$. This is also intuitive: as a larger sample becomes available, both bias and variance are reduced, because it is possible to reduce the error in the approximation by reducing the bandwidth without paying a penalty in added variability (as the larger number of available observations compensates for the bandwidth reduction).

In some applications, it may be useful to choose different bandwidths on each side of the cutoff. Since the RD treatment effect $\tau_{\mathtt{SRD}}=\mu_+ - \mu_-$ is simply the difference of two (one-sided) estimates, allowing for two distinct bandwidth choices can be accomplished by considering an MSE approximation for each estimate separately. In other words, two different bandwidths can be selected for $\hat{\mu}_+$ and $\hat{\mu}_-$, and then used to form the RD treatment effect estimator. Practically, this is equivalent to choosing an asymmetric neighborhood near the cutoff of the form $[\C-h_-,\C+h_+]$, where $h_-$ and $h_+$ denote the control (left) and treatment (right) bandwidths, respectively. In this case, the MSE-optimal choices are given by

\begin{eqnarray}
h_{\mathtt{MSE},-} &=& \left(\frac{\mathcal{V}_-}{2(p+1)\mathcal{B}_-^2}\right)^{1/(2p+3)} \; n_-^{-1/(2p+3)} \\
h_{\mathtt{MSE},+} &=& \left(\frac{\mathcal{V}_+}{2(p+1)\mathcal{B}_+^2}\right)^{1/(2p+3)} \; n_+^{-1/(2p+3)}.
\end{eqnarray}

These bandwidth choices will be most practically relevant when the bias and/or variance of the control and treatment groups differ substantially, for example, because of different curvature of the unknown regression functions, or different conditional variance of the outcome given the score near the cutoff.

In practice, the optimal bandwidth selectors described above (and variants thereof) are implemented by constructing preliminary plug-in estimates of the unknown quantities entering their formulas. For example, given a bandwidth choice and sample size, the misspecification biases $\mathscr{B}_+$ and $\mathscr{B}_-$ are estimated by forming preliminary ``curvature" estimates $\hat{\mu}_{-}^{(p+1)}$ and $\hat{\mu}_{+}^{(p+1)}$, which are constructed using a local polynomial of order $q\geq p+1$ with bias bandwidth $b$, not necessarily equal to $h$. The resulting estimators take the form
\[\hat{\mathscr{B}} = h^{2(p+1)}\; \hat{\mathcal{B}}, \qquad
  \hat{\mathcal{B}} = \hat{\mathcal{B}}_+ - \hat{\mathcal{B}}_-, \qquad
  \hat{\mathcal{B}}_+ = \hat{\mu}_{+}^{(p+1)}B_{+}, \qquad
  \hat{\mathcal{B}}_- = \hat{\mu}_{-}^{(p+1)}B_{-},
\]
where the quantities $B_{-}$ and $B_{+}$ are readily implementable given the information available (e.g., data, bandwidth choices, kernel choice, etc.). Similarly, a variance estimator is
\[\hat{\mathscr{V}} = \frac{1}{nh}\;\hat{\mathcal{V}}, \qquad
  \hat{\mathcal{V}} = \hat{\mathcal{V}}_+ + \hat{\mathcal{V}}_-,
\]
where the estimators $\hat{\mathcal{V}}_-$ and $\hat{\mathcal{V}}_+$ are usually constructed using plug-in pre-asymptotic formulas capturing the asymptotic variance of the estimates on the left and right of the cutoff, respectively. Natural choices are some version of heteroskedasticity-consistent standard error formulas or modifications thereof allowing for clustered data, all of which are implemented in the \texttt{rdrobust} software

Given these ingredients, data-driven MSE-optimal bandwidth selectors are easily constructed for the RD treatment effect (i.e., one common bandwidths on both sides of the cutoff) or for each of the two regression function estimators at the cutoff (i.e., two distinct bandwidths). For example, once a preliminary bandwidth choice is available to construct the above estimators, the MSE-optimal bandwidth choice is
\[\hat{h}_{\mathtt{MSE}} = \left(\frac{\hat{\mathcal{V}}}{2(p+1)\hat{\mathcal{B}}^2}\right)^{1/(2p+3)} \; n^{-1/(2p+3)},\]
and similarly for $\hat{h}_{\mathtt{MSE},+}$ and $\hat{h}_{\mathtt{MSE},-}$.

A potential drawback of the MSE bandwidth selection approach is that in some applications the estimated biases may be close to zero, leading to poor behavior of the resulting bandwidth selectors. To handle this computational issue, it is common to include a ``regularization'' term $\mathcal{R}$ to avoid small denominators in small samples. For example, in the case of a common bandwidth, the alternative formula is
\[h_{\mathtt{MSE}} = \left(\frac{\mathcal{V}}{2(p+1)\mathcal{B}^2 + \mathcal{R}}\right)^{1/(2p+3)} \; n^{-1/(2p+3)},\]
where the extra term $\mathcal{R}$ can be justified theoretically but requires additional preliminary estimators when implemented. Empirically, since $\mathcal{R}$ is in the denominator, including a regularization term will always lead to a smaller $h_{\mathtt{MSE}}$. This idea is also used in the case of $h_{\mathtt{MSE},-}$ and $h_{\mathtt{MSE},+}$, and other related bandwidth selection procedures. We discuss how to include and exclude a regularization term in practice in Section \ref{subsec:localpolyMeyersson}.

\subsubsection{Optimal Point Estimation}
Given the choice of polynomial order $p$ and kernel function $K(\cdot)$, the local polynomial RD point estimator $\hat\tau_{\mathtt{SRD}}$ is implemented for a choice of bandwidth $h$. Selecting either a common MSE-optimal bandwidth for $\hat \tau_{\mathtt{SRD}} = \hat{\mu}_{-} - \hat{\mu}_{+}$, or two distinct MSE-optimal bandwidths for its ingredients $\hat{\mu}_{-}$ and $\hat{\mu}_{+}$, leads to an RD point estimator that is both consistent and MSE-optimal, in the sense that it achieves the fastest rate of decay in an MSE sense. Furthermore, it can be shown that the triangular kernel is the MSE-optimal choice for point estimation. Because of these optimality properties, and the fact that the procedures are data driven and objective, modern RD empirical work routinely employs some form of automatic MSE-optimal bandwidth selection with triangular kernel, and reports the resulting MSE-optimal point estimator of the RD treatment effect.

\subsubsection{Point Estimation in Practice}
\label{subsec:localpolyMeyersson}

We now return to the Meyersson application to illustrate RD point estimation using local polynomials. First, we use standard least-squares commands to emphasize that local polynomial point estimation is simply a weighted least-squares fit. 

We start by choosing an ad hoc bandwidth $h = 20$, postponing the illustration of optimal bandwidth selection until the following section. Within this arbitrary bandwidth choice, we can construct the local linear RD point estimation with a uniform kernel using standard least-squares routines. As mentioned above, a uniform kernel simply means that all observations outside $[\C-h, \C+h]$ are excluded, and all observations inside this interval are weighted equally.

\labelsnippet{snippetLrdcontA}
\rsnip{Vol-1-R_meyersson_manualreg_tworegs_uniform_adhoc_p1.txt}{\Rlink{\thesnippetLrdcontA}}
\statasnip{Vol-1-STATA_meyersson_manualreg_tworegs_uniform_adhoc_p1}{\Slink{\thesnippetLrdcontA}}

The results indicate that within this ad hoc bandwidth of 20 percentage points, the percentage of women aged 15 to 20 who completed high school increases by about 2.927 percentage points with an Islamic victory: about 15.55\% of women in this age group had completed high school by 2000 in municipalities where the Islamic party barely won the 1994 mayoral elections, while the analogous percentage in municipalities where the Islamic party was barely defeated is about 12.62\%.
    
We now show that the same point estimator can be obtained by fitting a single linear regression that includes an interaction between the treatment indicator and the score---both approaches are algebraically equivalent.

\labelsnippet{snippetLrdcontB}
\rsnip{Vol-1-R_meyersson_manualreg_onereg_uniform_adhoc_p1.txt}{\Rlink{\thesnippetLrdcontB}}
\statasnip{Vol-1-STATA_meyersson_manualreg_onereg_uniform_adhoc_p1}{\Slink{\thesnippetLrdcontB}}

The coefficient on the treatment indicator is 2.92708, the same value we obtained by subtracting the intercepts in the two separate regressions.

To produce the same point estimation with a triangular kernel instead of a uniform kernel, we simply use a least-squares routine with weights. First, we create the weights according to the triangular kernel formula.

\labelsnippet{snippetLrdcontC}
\rsnip{Vol-1-R_meyersson_manualreg_weights_triangular_adhoc_p1.txt}{\Rlink{\thesnippetLrdcontC}}
\statasnip{Vol-1-STATA_meyersson_manualreg_weights_triangular_adhoc_p1}{\Slink{\thesnippetLrdcontC}}

Then, we use the weights in the least-squares fit.
\labelsnippet{snippetLrdcontD}
\rsnip{Vol-1-R_meyersson_manualreg_tworegs_triangular_adhoc_p1.txt}{\Rlink{\thesnippetLrdcontD}}
\statasnip{Vol-1-STATA_meyersson_manualreg_tworegs_triangular_adhoc_p1}{\Slink{\thesnippetLrdcontD}}

Note that, with $h$ and $p$ fixed, changing the kernel from uniform to triangular alters the point estimator only slightly, from about 2.9271 to 2.9373. This is typical; point estimates tend to be relatively stable with respect to the choice of kernel.

Although using standard least-squares estimation routines is useful to clarify the algebraic mechanics behind local polynomial point estimation, the confidence intervals and standard errors provided by these routines will be generally invalid for our purposes, a point we discuss extensively in the upcoming sections. Thus, from this point on, we employ the \texttt{rdrobust} software package, which is specifically tailored to RD designs and includes several functions to conduct local polynomial bandwidth selection, RD point estimation, and RD inference using a fully non-parametric and internally coherent methodology. 

To replicate the previous point estimators using the command \texttt{rdrobust}, we use the options \texttt{p} to set the order of the polynomial, \texttt{kernel} to set the kernel, and \texttt{h} to choose the bandwidth manually. By default, \texttt{rdrobust} sets the cutoff value to zero, but this can be changed with the option \texttt{c}. We first use \texttt{rdrobust} to implement a local linear RD point estimator with $h=20$ and uniform kernel.

\labelsnippet{snippetLrdcontE}
\rsnip{Vol-1-R_meyersson_rdrobust_uniform_adhoc_p1_rho1_regterm1.txt}{\Rlink{\thesnippetLrdcontE}}
\statasnip{Vol-1-STATA_meyersson_rdrobust_uniform_adhoc_p1_rho1_regterm1}{\Slink{\thesnippetLrdcontE}}

The output includes many details. The four uppermost rows indicate that the total number of observations is 2,629, the bandwidth is chosen manually, and the observations are weighed with a uniform kernel. The final line indicates that the variance-covariance estimator (\texttt{VCE}) is constructed using nearest-neighbor (NN) estimators instead of sums of squared residuals (this default behavior can be changed with the option \texttt{vce}); we discuss details on variance estimation further below in the context of RD inference.

The middle rows resemble the output of \texttt{rdplot} in that they are divided in two columns that give information separately for the observations above (\texttt{Right}) and below (\texttt{Left}) the cutoff. The first row shows that the $2,629$ observations are split into $2,314$ (control) observations below the cutoff, and $315$ (treated) observations above the cutoff. The second row shows the effective number of observations that are used for estimation of the RD effect, that is, the number of observations whose scores are within distance $h$ from the cutoff, $X_i \in [\C-h,\C+h]$. The output indicates that there are $608$ observations with $X_i \in [\C-h, \C)$, and $280$ observations with $X_i \in [\C, \C+h]$. The third line shows the order of the local polynomial used to estimate the main RD effect, $\tau_{\mathtt{SRD}}$, which in this case is equal to $p=1$. The bandwidth used to estimate $\tau_{\mathtt{SRD}}$ is shown on the fifth line, \texttt{BW est. (h)}, where we see that the same bandwidth $h=20$ was used to the left and right of the cutoff. We defer discussion of \texttt{Order Bias (q)}, \texttt{BW bias (b)}, and \texttt{rho (h/b)} until we discuss inference methods. 

The bottom rows show the estimation results. The RD point estimator, reported in the first row of the \texttt{Coef.} column, is $\hat{\tau}_\mathtt{SRD}=2.927$, indicating that in municipalities where the Islamic party barely won, the educational attainment of women is roughly 3 percentage points higher than in municipalities where the party barely lost. As expected, this number is identical to the number we obtained with the least-squares function \texttt{lm} in \texttt{R} or the command \texttt{reg} in \texttt{Stata}.

The \texttt{rdrobust} routine also allows us to easily estimate the RD effect using triangular instead of uniform kernel weights.

\labelsnippet{snippetLrdcontF}
\rsnip{Vol-1-R_meyersson_rdrobust_triangular_adhoc_p1_rho1_regterm1.txt}{\Rlink{\thesnippetLrdcontF}}
\statasnip{Vol-1-STATA_meyersson_rdrobust_triangular_adhoc_p1_rho1_regterm1}{\Slink{\thesnippetLrdcontF}}

Once again, this produces the same coefficient of $2.937$ that we found when we used the weighted least-squares command with triangular weights. We postpone the discussion of standard errors, confidence intervals, and the distinction between the \texttt{Conventional} versus \texttt{Robust} results until we discuss inference methods.

Finally, if we wanted to reduce the approximation error in the estimation of the RD effect, we could increase the order of the polynomial and use a local quadratic fit instead of a local linear one. This can be implemented in \texttt{rdrobust} setting \texttt{p=2}.

\labelsnippet{snippetLrdcontG}
\rsnip{Vol-1-R_meyersson_rdrobust_triangular_adhoc_p2_rho1_regterm1.txt}{\Rlink{\thesnippetLrdcontG}}
\statasnip{Vol-1-STATA_meyersson_rdrobust_triangular_adhoc_p2_rho1_regterm1}{\Slink{\thesnippetLrdcontG}}

Note that the estimated effect changes from $2.937$ with $p=1$, to $2.649$ with $p=2$. It is not unusual to observe a change in the point estimate as one changes the polynomial order used in the estimation. Unless the higher-order terms in the approximation are exactly zero, incorporating those terms in the estimation will reduce the approximation error and thus lead to changes in the estimated effect. The relevant practical question is whether such changes in the point estimator change the conclusions of the study. For that, we need to consider inference as well as estimation procedures, a topic we discuss in the upcoming sections.

In general, choosing an ad hoc bandwidth (as done in the previous commands) is not advisable. It is unclear what the value $h=20$ means in terms of bias and variance properties, or whether this is the best approach for estimation and inference. The command \texttt{rdbwselect}, which is part of the \texttt{rdrobust} package, implements optimal, data-driven bandwidth selection methods. We illustrate the use of \texttt{rdbwselect} by selecting an MSE-optimal bandwidth for the local linear estimator of $\tau_{\mathtt{SRD}}$.

\labelsnippet{snippetLrdcontH}
\rsnip{Vol-1-R_meyersson_rdbwselect_triangular_mserd_p1_regterm1_all.txt}{\Rlink{\thesnippetLrdcontH}}
\statasnip{Vol-1-STATA_meyersson_rdbwselect_triangular_mserd_p1_regterm1_all}{\Slink{\thesnippetLrdcontH}}

The MSE-optimal bandwidth choice depends on the choice of polynomial order and kernel function, which is why both have to be specified in the call to \texttt{rdbwselect}. The first output line indicates the type of bandwidth selector; in this case, it is MSE-optimal (\texttt{mserd}).  The type of kernel used is also reported, as is the total number of observations. The middle rows report the number of observations on each side of the cutoff, and the order of polynomial chosen for estimation of the RD effect, the \texttt{Order est. (p)} row.

In the bottom rows, we see the estimated optimal bandwidth choices. The bandwidth \texttt{h} refers to the bandwidth used to estimate the RD effect $\tau_{\mathtt{SRD}}$; we sometimes refer to it as the \textit{main bandwidth}. The bandwidth \texttt{b} is an additional bandwidth used to estimate a bias term that is needed for robust inference; we omit discussion of \texttt{b} until the following sections. 

The estimated MSE-optimal bandwidth for the local-linear RD point estimator with triangular kernel weights is 17.239. The option \texttt{bwselect = "mserd"} imposes the same bandwidth \texttt{h} on each side of the cutoff, that is, uses the neighborhood $[\C-h, \C+h]$. This is why the columns \texttt{Left of c} and \texttt{Right of c} have the same value 17.239. If instead we wish to allow the bandwidth to be different on each side of the cutoff, we can choose two MSE-optimal bandwidths by using the \texttt{bwselect = "msetwo"} option. This leads to a bandwidth of 19.967 on the control side, and a bandwidth of 17.359 on the treated side, as shown below.

\labelsnippet{snippetLrdcontI}
\rsnip{Vol-1-R_meyersson_rdbwselect_triangular_msetwo_p1_regterm1_all.txt}{\Rlink{\thesnippetLrdcontI}}
\statasnip{Vol-1-STATA_meyersson_rdbwselect_triangular_msetwo_p1_regterm1_all}{\Slink{\thesnippetLrdcontI}}

Once we select the MSE-optimal bandwidth(s), we can pass them to the function \texttt{rdrobust} using the option \texttt{h}. But it is much easier to use the option \texttt{bwselect} in \texttt{rdrobust}. When we use this option, \texttt{rdrobust} calls \texttt{rdbwselect} internally, selects the bandwidth as requested, and then uses the optimally chosen bandwidth to estimate the RD effect. 

We now use the \texttt{rdrobust} command to perform bandwidth selection and point estimation in one step, using $p=1$ and triangular kernel weights.

\labelsnippet{snippetLrdcontJ}
\rsnip{Vol-1-R_meyersson_rdrobust_triangular_mserd_p1_rhofree_regterm1.txt}{\Rlink{\thesnippetLrdcontJ}}
\statasnip{Vol-1-STATA_meyersson_rdrobust_triangular_mserd_p1_rhofree_regterm1}{\Slink{\thesnippetLrdcontJ}}

As we can see, when the same MSE-optimal bandwidth is used on both sides of the cutoff, the effect of a bare Islamic victory on the educational attainment of women is $3.020$, slightly larger  than the $2.937$ effect that we found above when we used the ad hoc bandwidth of 20.

We can also explore the \texttt{rdrobust} output to obtain the estimates of the average outcome at the cutoff separately for treated and control observations.

\labelsnippet{snippetLrdcontK}
\rsnip{Vol-1-R_meyersson_rdrobust_triangular_mserd_p1_rhofree_regterm1_namescoefsout_all.txt}{\Rlink{\thesnippetLrdcontK}}
\statasnip{Vol-1-STATA_meyersson_rdrobust_triangular_mserd_p1_rhofree_regterm1_namescoefsout_all}{\Slink{\thesnippetLrdcontK}}

We see that the RD effect of 3.020 percentage points in the female high school attainment percentage is the difference between a percentage of 15.6649438\% in municipalities where the Islamic party barely wins and a percentage of 12.6454218\% in municipalities where the Islamic party barely loses, that is, $15.6649438 - 12.6454218 \approx 3.020$. By accessing the control mean at the cutoff in this way, we learn that the RD effect represents an increase of $(3.020/12.6454218) \times 100=23.88\%$ relative to the control mean. 

\begin{figure}[H]
\centering
\includegraphics[scale=0.60]{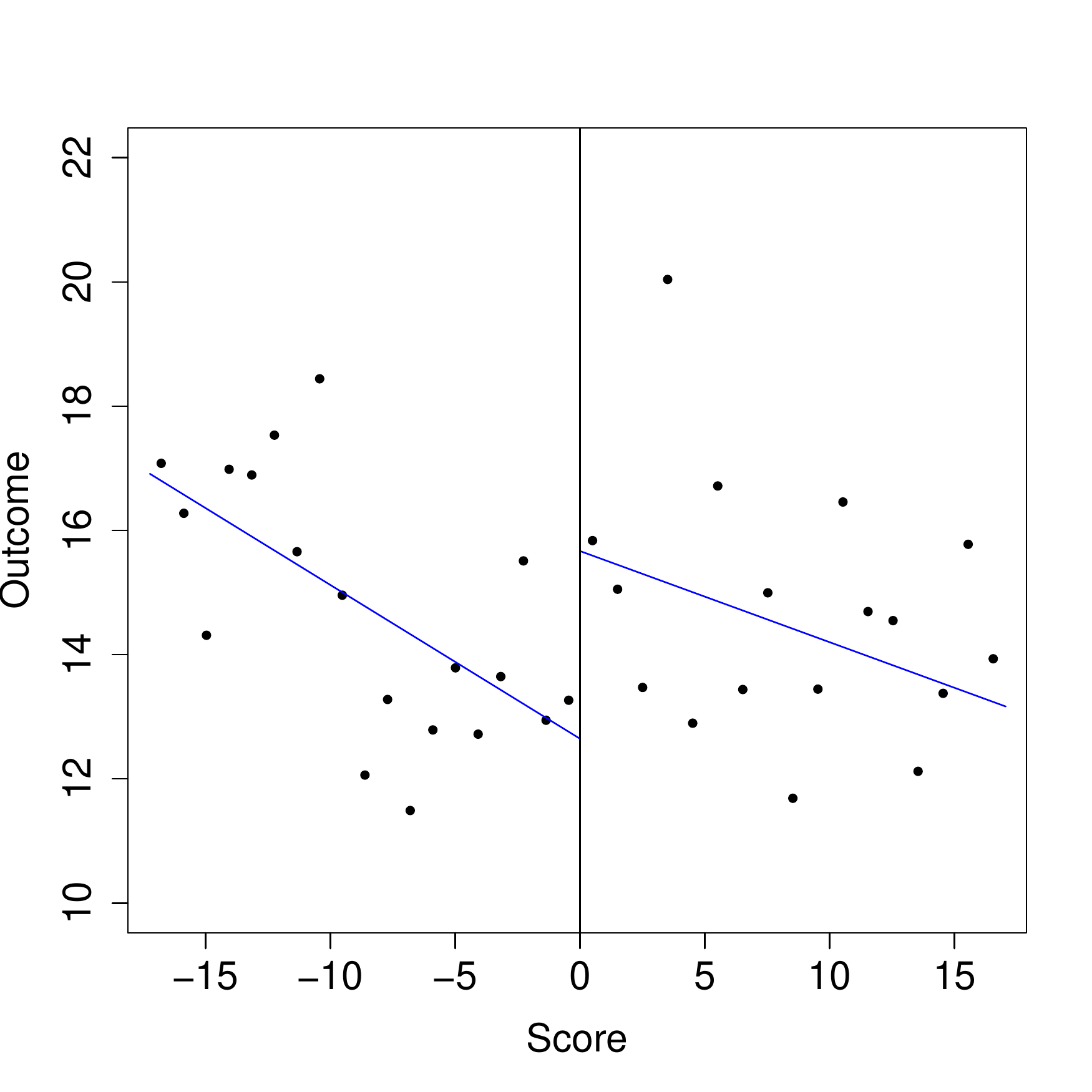}
\caption{Local Polynomial RD Effect Illustrated with \texttt{rdplot} (Meyersson Data)}\label{fig:Meyersson_rdplot_maineffect}
\end{figure}

This effect, together with the means at either side of the cutoff, can be easily illustrated with \texttt{rdplot}, using the options \texttt{h}, \texttt{p}, and \texttt{kernel}, to set exactly the same specification used in \texttt{rdrobust} and produce an exact illustration of the RD effect. We illustrate the commands below, and show the resulting plot in Figure \ref{fig:Meyersson_rdplot_maineffect}. 

\labelsnippet{snippetLrdcontL}
\rsnip{Vol-1-R_meyersson_rdplot_maineffect.txt}{\Rlink{\thesnippetLrdcontL}}
\statasnip{Vol-1-STATA_meyersson_rdplot_maineffect}{\Slink{\thesnippetLrdcontL}}

\labelfiguras{figO}

Finally, we note that by default, all MSE-optimal bandwidth selectors in \texttt{rdrobust} include the regularization term that we discussed in subsection \ref{subsec:bw}. We can exclude the regularization term with the option \texttt{scaleregul=0} in the \texttt{rdrobust} (or \texttt{rdbwselect}) call.

\labelsnippet{snippetLrdcontM}
\rsnip{Vol-1-R_meyersson_rdrobust_triangular_mserd_p1_rhofree_regterm0.txt}{\Rlink{\thesnippetLrdcontM}}
\statasnip{Vol-1-STATA_meyersson_rdrobust_triangular_mserd_p1_rhofree_regterm0}{\Slink{\thesnippetLrdcontM}}

In this application, excluding the regularization term has a very large impact on the estimated $h_\mathtt{MSE}$. With regularization,  $\hat{h}_\mathtt{MSE}$ is 17.239, while excluding regularization increases it to 34.983, an increase of roughly 100\%. Nevertheless, the point estimate remains relatively stable, moving from 3.020 with regularization to 2.843 without regularization. 

\subsection{Local Polynomial Inference}

In addition to providing a local polynomial point estimator of the RD treatment effect, we are interested in testing hypotheses and constructing confidence intervals. Although, at first glance, it seems that we could employ ordinary least-squares (OLS) inference methods, these methods would treat the local polynomial regression model as correctly specified (i.e., parametric), and de facto disregard its fundamental approximation (i.e., non-parametric) nature. Thus, it would be intellectually and methodologically incoherent to simultaneously select a bandwidth according to a bias-variance trade-off and then proceed as if the bias were zero, that is, as if the local polynomial fit were exact and no misspecification error existed.

These considerations imply that valid inference should take into account the effect of misspecification. In particular, the MSE-optimal bandwidths discussed previously ($h_\mathtt{MSE}$, $h_{\mathtt{MSE},-}$, and $h_{\mathtt{MSE},+}$) result in an RD point estimator that is both consistent and optimal in an MSE sense, but inferences based on these bandwidth choices pose a problem. The challenge is that these bandwidths are not ``small'' enough to remove the leading bias term in the standard distributional approximations used to conduct statistical inference. The root of the problem is that these bandwidth choices are developed for point estimation purposes, and as such they pay no attention to their effects in terms of distributional properties of typical t-tests or related statistics. Thus, constructing confidence intervals using standard OLS large-sample results using the data with $X_i\in[\C -h_\mathtt{MSE},\C+h_\mathtt{MSE}]$ will generally result in invalid inferences. 
 
There are two general approaches that can be used to address this key problem. One approach is to use the bandwidth $h_\mathtt{MSE}$ for both estimation and inference, but modify the usual t-statistic to account for the effects of misspecification due to the large bandwidth, as well as for the additional sampling error introduced by such modification. The other is to use $h_\mathtt{MSE}$ only for point estimation, and then choose a different bandwidth for inference purposes.  We elaborate on these issues next.

\subsubsection{Using the MSE-Optimal Bandwidth for Inference}

We first discuss how to make valid inferences when the bandwidth choice is $h_\mathtt{MSE}$ (or some data-driven implementation thereof). The local polynomial RD point estimator $\hat \tau_{\mathtt{SRD}}$ has an approximate large-sample distribution
\[ \frac{\hat\tau_{\mathtt{SRD}} -  \tau_{\mathtt{SRD}} - \mathscr{B}}{\sqrt{\mathscr{V}}} \overset{a}{\thicksim} \mathcal{N}(0, 1)\]
where $\mathscr{B}$ and $\mathscr{V}$ are, respectively, the asymptotic bias and variance of the RD local polynomial estimator of order $p$, discussed previously in the context of MSE expansions and bandwidth selection. This distributional result is similar to those encountered, for example, in standard linear regression problems---with the important distinction that now the bias term $\mathscr{B}$ features explicitly; this term highlights the trade-off between bandwidth choice and misspecification bias locally to the cutoff. The variance term $\mathscr{V}$ can be calculated as in (weighted) least-squares problems, for instance accounting for heteroskedasticity and/or clustered data. We do not provide the exact formulas for variance estimation, to save space and notation, but they can be found in the references given at the end of this section and are all implemented in \texttt{rdrobust}.

Given the distributional approximation for the RD local polynomial estimator, an asymptotic $95\%$ confidence interval for $\tau_{\mathtt{SRD}}$ is approximately given by
\[\mathtt{CI} = \left[\left(\hat\tau_{\mathtt{SRD}}-\mathscr{B}\right) \pm 1.96 \cdot \sqrt{\mathscr{V}} \right]. \]
This confidence interval depends on the unknown bias or misspecification error $\mathscr{B}$, and any practical procedure that ignores it will lead to incorrect inferences unless this term is negligible (i.e., unless the local linear regression model is close-to-correctly specified). The bias term arises because the local polynomial approach is a non-parametric approximation: instead of \textit{assuming} that the underlying regression functions are $p$th order polynomials (as would occur in OLS estimation), this approach uses the polynomial to \textit{approximate} the unknown regression functions.

We now discuss different strategies that are often employed to make inferences for $\tau_{\mathtt{SRD}}$ based on asymptotic distributional approximations in the presence of non-parametric misspecification biases, and explain why some of them are invalid. Our recommendation is to use a robust bias correction approach, which is theoretically valid, enjoys some optimality properties, and performs well in practice.

\subsubsection*{Conventional Inference and Undersmoothing}

A strategy sometimes found in RD empirical work is to ignore the misspecification error even when an MSE-optimal bandwidth is used. This is not only invalid but also methodologically incoherent: an MSE-optimal bandwidth cannot be selected in the absence of misspecification error (zero bias), and statistical inference based on standard OLS methods (ignoring the bias) cannot be valid when an MSE-optimal bandwidth is employed.

This na\"ive approach to statistical inference treats the local polynomial approach as parametric within the neighborhood around the cutoff and de facto ignores the bias term, a procedure that leads to invalid inferences in all cases except when the approximation error is so small that it can be ignored. When the bias term is zero, the approximate distribution of the RD estimator is $\hat\tau_{\mathtt{SRD}} -  \tau_{\mathtt{SRD}}/\sqrt{\mathscr{V}} \overset{a}{\thicksim} \mathcal{N}(0, 1)$ and the confidence interval is 
\[ \mathtt{CI}_{\mathtt{us}} = \left[~\hat{\tau}_\mathtt{SRD}\pm 1.96 \cdot \sqrt{\mathscr{V}} ~\right] \text{.} \]

Since this is the same confidence interval that follows from parametric least-squares estimation, we refer to it as \textit{conventional}. Using the conventional confidence interval $\mathtt{CI}_{\mathtt{us}}$ implicitly assumes that the chosen polynomial gives an exact approximation to the true functions $\E[Y_i(1) | X_i]$ and $\E[Y_i(0) | X_i]$. Since these functions are unknown, this assumption is not verifiable and will rarely be credible. If researchers use $\mathtt{CI}_{\mathtt{us}}$ when in fact the approximation error is non-negligible, all inferences will be incorrect, leading to under-coverage of the true treatment effect or, equivalently, over-rejection of the null hypothesis of zero treatment effect. For this reason, we strongly discourage researchers from using conventional inference when using local polynomial methods, unless the misspecification bias can credibly be assumed small (ruling out, in particular, the use of MSE-optimal bandwidth choices).

A theoretically sound but ad hoc alternative procedure is to use these conventional confidence intervals with a smaller or ``undersmoothed'' bandwidth relative to the MSE-optimal one used for construction of the point estimator $\hat\tau_{\mathtt{SRD}}$. Practically, this procedure involves first selecting the MSE-optimal bandwidth, then selecting a bandwidth smaller than the MSE-optimal choice, and finally constructing the conventional confidence intervals $\mathtt{CI}_{\mathtt{us}}$ with this smaller bandwidth---note that the latter step requires estimating both a new point estimator and a new standard error with the smaller bandwidth. The theoretical justification is that, for bandwidths smaller than the MSE-optimal choice, the bias term will become negligible in the large-sample distributional approximation. (This is why we use the subscript ``$\mathtt{us}$'' to refer to the conventional confidence interval.)

The main drawback of this undersmoothing procedure is that there are no clear and transparent criteria for shrinking the bandwidth below the MSE-optimal value: some researchers might estimate the MSE-optimal choice and divide by two, others may choose to divide by three, etc. Although these procedures can be justified in a strictly theoretical sense, they are all ad hoc and can result in lack of transparency and specification searching. Moreover, this general strategy leads to a loss of statistical power because a smaller bandwidth results in fewer observations used for estimation and inference. Finally, from a substantive perspective, some researchers prefer to avoid using different observations for estimation and inference, which is required by the undersmoothing approach.

\subsubsection*{Standard Bias Correction}

As an alternative to undersmoothing (i.e., to choosing a bandwidth smaller than the MSE-optimal bandwidth), inference could be based on the MSE-optimal bandwidth so long as the induced misspecification error is manually estimated and removed from the distributional approximation. This approach, known as \textit{bias correction}, first estimates the bias term $\mathscr{B}$ with the estimator $\hat{\mathscr{B}}$ (which in fact is already estimated for implementation of MSE-optimal bandwidth selection), and then constructs confidence intervals that are centered at the bias-corrected point estimate:
\[\mathtt{CI}_{\mathtt{bc}}
  = \left[ ~ \big(\hat{\tau}_\mathtt{SRD} - \hat{\mathscr{B}}\big) \pm 1.96 \cdot \sqrt{\mathscr{V}} ~ \right]. \]
As explained above, the bias term depends on the ``curvature'' of the unknown regression functions captured via their derivative of order $p+1$ at the cutoff. These unknown derivatives can be estimated with a local polynomial of order $q=p+1$ or higher, which requires another choice of bandwidth, denoted $b$. Therefore, the RD point estimate $\hat{\tau}_\mathtt{SRD}$ employs the bandwidth $h$, while the bias estimate  $\hat{\mathscr{B}}$ employs the additional bandwidth $b$. The ratio $\rho=h/b$ is important, as it relates to the variability of the bias correction estimate relative to the RD point estimator. Standard bias correction methods require $\rho=h/b\to0$, that is, a small $\rho$. In particular, note this rules out $\rho=h/b=1$, that is, standard bias correction does not allow $h=b$.

The bias-corrected confidence intervals $\mathtt{CI}_{\mathtt{bc}}$ allow for a wider range of bandwidths $h$ and, in particular, result in valid inferences when the MSE-optimal bandwidth is used. However, they typically  have poor performance in applications because the variability introduced in the bias estimation step is not incorporated in the variance term used. Despite employing the additional estimated term $\hat{\mathscr{B}}$, $\mathtt{CI}_{\mathtt{bc}}$ employs the same variance as $\mathtt{CI}_{\mathtt{us}}$, essentially ignoring the variability that is introduced when $\mathscr{B}$ is estimated. This results in a poor distributional approximation and hence considerable coverage distortions in practice.

\subsubsection*{Robust Bias Correction}

A superior strategy that is both theoretically sound and leads to improved coverage in finite samples is to use \textit{robust} bias correction for constructing confidence intervals. This approach leads to demonstrably superior inference procedures, with smaller coverage error and shorter average length than those associated with either $\mathtt{CI}_{\mathtt{us}}$ or $\mathtt{CI}_{\mathtt{bc}}$. Furthermore, the robust bias correction approach delivers valid inferences even when the MSE-optimal bandwidth for point estimation is used---no undersmoothing is necessary---and remains valid even when $\rho=h/b=1$ ($h=b$), which implies that exactly the same data can be used for both point estimation and inference.

Robust bias-corrected confidence intervals are based on the bias correction procedure described above, by which the estimated bias term $\hat{\mathscr{B}}$ is removed from the RD point estimator. However, in contrast to $\mathtt{CI}_{\mathtt{bc}}$, the derivation allows the estimated bias term to converge in distribution to a random variable and thus contribute to the distributional approximation of the RD point estimator. This results in a new asymptotic variance $\mathscr{V}_{\mathtt{bc}}$ that, unlike the variance $\mathscr{V}$ used in $\mathtt{CI}_{\mathtt{us}}$ and $\mathtt{CI}_{\mathtt{bc}}$, incorporates the contribution of the bias correction step to the variability of the bias-corrected point estimator. Because the new variance $\mathscr{V}_{\mathtt{bc}}$ incorporates the extra variability introduced in the bias estimation step, it is larger than the conventional OLS variance $\mathscr{V}$ when the same bandwidth is used. 

This approach leads to the robust bias-corrected confidence interval:
\[\mathtt{CI}_{\mathtt{rbc}}
  = \left[ ~ \big(\hat{\tau}_\mathtt{SRD}-\hat{\mathscr{B}}\big) \pm 1.96 \cdot \sqrt{\mathscr{V}_{\mathtt{bc}}} ~ \right], \]
which is constructed by subtracting the bias estimate from the local polynomial estimator and using the new variance formula for Studentization. Note that, like $\mathtt{CI}_{\mathtt{bc}}$, $\mathtt{CI}_{\mathtt{rbc}}$ is centered around the bias-corrected point estimate, $\hat{\tau}_\mathtt{SRD}-\hat{\mathscr{B}}$, not around the uncorrected estimate $\hat{\tau}_\mathtt{SRD}$. This robust confidence interval results in valid inferences when the MSE-optimal bandwidth is used, because it has smaller coverage errors and is therefore less sensitive to tuning parameter choices. In practice, the confidence interval can be implemented by setting $\rho=h/b=1$ ($h=b$) and choosing $h=h_\mathtt{MSE}$, or by selecting both $h$ and $b$ to be MSE-optimal for the corresponding estimators, in which case $\rho$ is set to $h_\mathtt{MSE}/b_\mathtt{MSE}$ or their respective data-driven implementations.

We summarize the differences between the three types of confidence intervals discussed in Table \ref{tab:CIs}. The conventional OLS confidence interval $\mathtt{CI}_\mathtt{us}$ ignores the bias term and is thus centered at the local polynomial point estimator $\hat{\tau}_\mathtt{SRD}$, and uses the conventional standard error $\sqrt{\hat{\mathscr{V}}}$. The bias-corrected confidence interval $\mathtt{CI}_\mathtt{bc}$ removes the bias estimate from the conventional point estimator, and is therefore centered at $\hat{\tau}_\mathtt{SRD} - \hat{\mathscr{B}}$; this confidence interval, however, ignores the variability introduced in the bias correction step and thus continues to use the standard error $\sqrt{\hat{\mathscr{V}}}$, which is the same standard error used by $\mathtt{CI}_\mathtt{us}$. The robust bias-corrected confidence interval $\mathtt{CI}_\mathtt{rbc}$ is also centered at the bias-corrected point estimator $\hat{\tau}_\mathtt{SRD} - \hat{\mathscr{B}}$ but, in contrast to $\mathtt{CI}_\mathtt{bc}$, it employs a different standard error, $\sqrt{\hat{\mathscr{V}}_\mathtt{bc}}$, which is larger than the conventional standard error $\sqrt{\hat{\mathscr{V}}}$ when the same bandwidth $h$ is used. Thus, relative to the conventional confidence interval, the robust bias-corrected confidence interval is both recentered and rescaled. As discussed above, when $h=h_\mathtt{MSE}$, the conventional confidence interval $\mathtt{CI}_\mathtt{us}$ is invalid. 

\labeltablas{tableC}
\begin{table}[H]
\centering
\caption{Local Polynomial Confidence Intervals \label{tab:CIs}} 
\begin{tabular}{lcc}
& Centered at & Standard Error \\
\midrule
Conventional: $\mathtt{CI}_\mathtt{us}$           & $\hat{\tau}_\mathtt{SRD}$                      & $\sqrt{\hat{\mathscr{V}}}$   \\
Bias-Corrected: $\mathtt{CI}_\mathtt{bc}$         & $\hat{\tau}_\mathtt{SRD} - \hat{\mathscr{B}}$  & $\sqrt{\hat{\mathscr{V}}}$   \\
Robust bias-corrected: $\mathtt{CI}_\mathtt{rbc}$ & $\hat{\tau}_\mathtt{SRD} - \hat{\mathscr{B}}$  & $\sqrt{\hat{\mathscr{V}}_\mathtt{bc}}$   \\
\bottomrule
\end{tabular}
\end{table}

From a practical perspective, the most important feature of the robust bias-corrected confidence interval $\mathtt{CI}_\mathtt{rbc}$ is that it can be used with the MSE-optimal point estimator $\hat{\tau}_\mathtt{SRD}$ when this estimator is constructed using the MSE-optimal bandwidth choice $h_\mathtt{MSE}$. In other words, using the robust bias-corrected confidence interval allows researchers to use the same observations with score $X_i\in[\C-h_\mathtt{MSE},\C+h_\mathtt{MSE}]$ for both optimal point estimation and valid statistical inference. 

\subsubsection{Using Different Bandwidths for Point Estimation and Inference}

Conceptually, the invalidity of the conventional confidence interval $\mathtt{CI}_\mathtt{us}$ based on the MSE-optimal bandwidth $h_\mathtt{MSE}$ stems from using for inference a bandwidth that is optimally chosen for point estimation purposes. Using $h_\mathtt{MSE}$ for estimation of the RD effect $\tau_{\mathtt{SRD}}$ results in a point estimator $\hat\tau_{\mathtt{SRD}}$ that is not only consistent but also has minimal asymptotic MSE. Thus, from a point estimation perspective, $h_\mathtt{MSE}$ leads to highly desirable properties. In contrast, serious methodological challenges arise when researchers attempt to use $h_\mathtt{MSE}$ for building confidence intervals and making inferences in the standard parametric way, because the MSE-optimal bandwidth choice is not designed with the goal of ensuring good (or even valid) distributional approximations. As shown above, robust bias correction restores a valid standard normal distributional approximation when $h_\mathtt{MSE}$ is used by recentering and rescaling the usual t-statistic, allowing researchers to use the same bandwidth $h_\mathtt{MSE}$ for both point estimation and inference.

While employing the MSE-optimal bandwidth for both optimal point estimation and valid statistical inference is very useful in practice, it may be important to also consider statistical inference that is optimal. A natural optimality criterion associated with robustness properties of confidence intervals is the minimization of their coverage error, that is, the discrepancy between the empirical coverage of the confidence interval and its nominal level. For example, if a $95\%$ confidence interval contains the true parameter $80\%$ of the time, the coverage error is $15$ percentage points. Minimization of coverage error for confidence intervals is an idea analogous to minimization of MSE for point estimators. 

Thus, an alternative approach to RD inference is to decouple the goal of point estimation from the goal of inference, using a different bandwidth for each case. In particular, this strategy involves estimating the RD effect with $h_\mathtt{MSE}$, and constructing confidence intervals using a different bandwidth, where the latter is specifically chosen to minimize an approximation to the coverage error (CER) of the confidence interval $\mathtt{CI}_\mathtt{rbc}$, leading to the choice $h =h_\mathtt{CER}$. Just like $h_\mathtt{MSE}$ minimizes the asymptotic MSE of the point estimator $\hat{\tau}_\mathtt{SRD}$, the CER-optimal bandwidth $h_\mathtt{CER}$ minimizes the asymptotic coverage error rate of the robust bias-corrected confidence interval for $\tau_{\mathtt{SRD}}$. This bandwidth cannot be obtained in closed form, but it can be shown to have a faster rate of decay than $h_\mathtt{MSE}$, which implies that for all practically relevant sample sizes $h_\mathtt{CER} < h_\mathtt{MSE}$. By design, constructing $\mathtt{CI}_\mathtt{rbc}$ using the CER-optimal bandwidth choice $h_\mathtt{CER}$ leads to confidence intervals that are not only valid but also have the fastest rate of coverage error decay.

Note that using $h_\mathtt{CER}$ for point estimation will result in an RD point estimator that has too much variability relative to its bias and is therefore not MSE-optimal (but is nonetheless consistent). Thus, we recommend that practitioners continue to use $h_\mathtt{MSE}$ for point estimation of $\tau_{\mathtt{SRD}}$, and use either $h_\mathtt{MSE}$ or $h_\mathtt{CER}$ to build the robust bias-corrected confidence interval $\mathtt{CI}_\mathtt{rbc}$ for inference purposes, where $\mathtt{CI}_\mathtt{rbc}$ will be either valid (if $h_\mathtt{MSE}$ is used) or valid and CER-optimal (if $h_\mathtt{CER}$ is used).

\subsubsection{RD Local Polynomial Inference in Practice}

We can now discuss the full output of our previous call to \texttt{rdrobust} with $p=1$ and triangular kernel, which we reproduce below.

\labelsnippet{snippetLrdcontN}
\rsnip{Vol-1-R_meyersson_rdrobust_triangular_mserd_p1_rhofree_regterm1.txt}{\Rlink{\thesnippetLrdcontN}}
\statasnip{Vol-1-STATA_meyersson_rdrobust_triangular_mserd_p1_rhofree_regterm1}{\Slink{\thesnippetLrdcontN}}

As reported before, the local linear RD effect estimate is $3.020$, estimated within the MSE-optimal bandwidth of $17.239$. The last output provides all the necessary information to make inferences. The row labeled \texttt{Conventional} reports, in addition to the point estimator $\hat{\tau}_\mathtt{SRD}$, the conventional standard error $\sqrt{\hat{\mathscr{V}}}$, the standardized test statistic $(\hat{\tau}_\mathtt{SRD} - \tau_{\mathtt{SRD}})/\sqrt{\hat{\mathscr{V}}}$, the corresponding p-value, and the 95\% conventional confidence interval $\mathtt{CI}_{\mathtt{us}}$. This confidence interval ranges from 0.223 to 5.817 percentage points, suggesting a positive effect of an Islamic victory on the educational attainment of women. Note that $\mathtt{CI}_{\mathtt{us}}$ is centered around the conventional point estimator $\hat{\tau}_\mathtt{SRD}$:
\begin{align*}
3.020 + 1.427 \times 1.96 = 5.81692 \approx 5.817 \\
3.020 - 1.427 \times 1.96 = 0.22308 \approx 0.223 \text{.}
\end{align*}

The row labeled \texttt{Robust} reports the robust bias-corrected confidence interval $\mathtt{CI}_{\mathtt{rbc}}$. In contrast to $\mathtt{CI}_{\mathtt{us}}$, $\mathtt{CI}_{\mathtt{rbc}}$ is centered around the point estimator $\hat{\tau}_\mathtt{SRD} - \hat{\mathscr{B}}$ (which is by default not reported), and scaled by the robust standard error $\sqrt{\hat{\mathscr{V}}_\mathtt{bc}}$ (not reported either). $\mathtt{CI}_{\mathtt{rbc}}$ ranges from -0.309 to 6.276; in contrast to the conventional confidence interval, it does include zero. As expected, $\mathtt{CI}_{\mathtt{rbc}}$ is not centered at $\hat{\tau}_\mathtt{SRD}$. 

For a fixed common bandwidth, the length of $\mathtt{CI}_{\mathtt{rbc}}$ is always greater than the length of $\mathtt{CI}_{\mathtt{us}}$  because  $\sqrt{\hat{\mathscr{V}}_\mathtt{bc}} > \sqrt{\hat{\mathscr{V}}}$.  We can see this in our example:
\begin{align*}
 \text{Length of  $\mathtt{CI}_{\mathtt{us}}$}& =  5.817 - 0.223 = 5.594\\
\text{Length of  $\mathtt{CI}_{\mathtt{rbc}}$}  & = 6.276 - (-0.309) = 6.585 \text{.}
\end{align*}
However, this will not necessarily be true if different bandwidths are used to construct each confidence interval.

The omission of the bias-corrected point estimator that is at the center of  $\mathtt{CI}_{\mathtt{rbc}}$ from the \texttt{rdrobust} output is intentional: when the MSE-optimal bandwidth for $\hat{\tau}_\mathtt{SRD}$ is used, the bias-corrected estimator is suboptimal in terms of MSE relative to $\hat{\tau}_\mathtt{SRD}$. (Although the bias-corrected estimator is consistent and valid whenever $\hat{\tau}_\mathtt{SRD}$ is.) Practically, it is usually desirable to report an MSE-optimal point estimator and then form valid confidence intervals either with the same MSE-optimal bandwidth or with some other optimal choice specifically tailored for inference.

In order to see all the ingredients that go into building the robust confidence interval, we can use the \texttt{all} option in \texttt{rdrobust}. 

\labelsnippet{snippetLrdcontO}
\rsnip{Vol-1-R_meyersson_rdrobust_triangular_mserd_p1_rhofree_regterm1_all.txt}{\Rlink{\thesnippetLrdcontO}}
\statasnip{Vol-1-STATA_meyersson_rdrobust_triangular_mserd_p1_rhofree_regterm1_all}{\Slink{\thesnippetLrdcontO}}

The three rows at the bottom of the output are analogous to the the rows in Table \ref{tab:CIs}: the \texttt{Conventional} row reports  $\mathtt{CI}_{\mathtt{us}}$, the \texttt{Bias-Corrected} row reports  $\mathtt{CI}_{\mathtt{bc}}$, and the \texttt{Robust} row reports  $\mathtt{CI}_{\mathtt{rbc}}$. We can see that the standard error used by $\mathtt{CI}_{\mathtt{us}}$ and $\mathtt{CI}_{\mathtt{bc}}$ is the same ($\sqrt{\hat{\mathscr{V}}} = 1.427$), while $\mathtt{CI}_{\mathtt{rbc}}$ uses a different standard error ($\sqrt{\hat{\mathscr{V}}_\mathtt{bc}} = 1.680$). We also see that the conventional confidence interval is centered at the conventional, non-bias-corrected point estimator 3.020, while both $\mathtt{CI}_{\mathtt{bc}}$ and $\mathtt{CI}_{\mathtt{rbc}}$ are centered at the bias-corrected point estimator 2.983. Since we know that $\hat{\tau}_\mathtt{SRD}= 3.020$ and $\hat{\tau}_\mathtt{SRD} - \hat{\mathscr{B}}= 2.983$, we can deduce that the bias estimate is $\hat{\mathscr{B}} = 3.020- 2.983 = 0.037$.

Finally, we investigate the properties of robust bias-corrected inference when employing a CER-optimal bandwidth choice. This is implemented via \texttt{rdrobust} with the option \texttt{bwselect="cerrd"}.

\labelsnippet{snippetLrdcontP}
\rsnip{Vol-1-R_meyersson_rdrobust_triangular_cerrd_p1_rhofree_regterm1.txt}{\Rlink{\thesnippetLrdcontP}}
\statasnip{Vol-1-STATA_meyersson_rdrobust_triangular_cerrd_p1_rhofree_regterm1}{\Slink{\thesnippetLrdcontP}}

The common CER-optimal bandwidth for both control and treatment units is $h_\mathtt{CER}=11.629$, which is smaller than the MSE-optimal bandwidth calculated previously, $h_\mathtt{MSE}=17.239$. The results are qualitatively similar, but now with a larger p-value as the nominal $95\%$ robust bias-corrected confidence interval changes from $[-0.309,6.276]$ with MSE-optimal bandwidth to $[-1.158,5.979]$ with CER-optimal bandwidth. The RD point estimator changes from the MSE-optimal value $3.020$ to the undersmoothed value $2.43$, where the latter RD estimate can be interpreted as having less bias but more variability than the former.

Since both the change in bandwidth choice from MSE-optimal to CER-optimal and the change from one common bandwidth to two different bandwidths are practically important, we conclude this section with a report of all the bandwidth choices. This is obtained using the \texttt{all} option in the \texttt{rdbwselect} command.

\labelsnippet{snippetLrdcontQ}
\rsnip{Vol-1-R_meyersson_rdbwselect_triangular_all_p1_regterm1.txt}{\Rlink{\thesnippetLrdcontQ}}
\statasnip{Vol-1-STATA_meyersson_rdbwselect_triangular_all_p1_regterm1}{\Slink{\thesnippetLrdcontQ}}

There are five MSE-optimal bandwidths reported. The row labeled \texttt{mserd} reports the bandwidth that minimizes the MSE of the RD point estimator under the constraint that the bandwidth to the left of the cutoff is the same as the bandwidth to the right of it, while the row labeled \texttt{msetwo} reports the bandwidth that minimizes the same MSE but allowing the left and right bandwidths to be different. In contrast to the \texttt{mserd} and \texttt{msetwo} bandwidths, which optimize the MSE of the RD point estimator,  $\hat{\tau}_{\mathtt{SRD}} =\hat{\mu}_+ - \hat{\mu}_-$, the \texttt{msesum}  row  reports the common bandwidth that minimizes the MSE of the sum of the regression coefficients, not their difference, that is, the MSE of $\hat{\mu}_+ + \hat{\mu}_-$. The fourth and fifth rows report a combination of the prior MSE-optimal bandwidths: \texttt{msecomb1} is the minimum between \texttt{mserd} and \texttt{msesum}, while  \texttt{msecomb2} is the median of \texttt{msetwo}, \texttt{mserd}, and \texttt{msesum}. The CER bandwidths reported in the last five rows are analogous to the prior five, with the only difference that the bandwidths  reported are optimal with respect to the CER of the confidence interval for $\tau_{\mathtt{SRD}}$, not its MSE.

\subsection{Extensions}

Up to this point, our discussion has considered local polynomials that included only the running variable as a regressor, in a setting where all the observations were assumed to be independent. We now discuss how local polynomial methods can be generalized to accommodate both additional covariates in the model specification, and clustering of observations. 

\subsubsection{Adding Covariates to the Analysis}

The simplest way to implement RD local polynomial analysis is to fit the outcome on the score alone. Although this basic specification is sufficient to analyze most applications, some researchers may want to augment it by including other covariates in addition to the score. Local polynomial methods can easily accommodate additional covariates, but the latter must satisfy an important condition. Unless researchers are willing to invoke parametric assumptions or redefine the parameter of interest, the covariates used to augment the analysis must be balanced at the cutoff. In general, covariate adjustment cannot be used to restore identification of standard RD design treatment effects when treated and control observations differ systematically at the cutoff. When the empirical evidence shows that important pre-determined covariates differ systematically at the cutoff, the assumption of continuity of the potential outcomes is implausible, and thus the non-parametric continuity-based RD framework discussed in this Element is no longer appropriate without further (restrictive) assumptions about the data generating process.

We let $\mathbf{Z}_i(1)$ and $\mathbf{Z}_i(0)$ denote two vectors of potential covariates, where $\mathbf{Z}_i(1)$ represents the value taken by the covariates above the cutoff (i.e., under treatment), and $\mathbf{Z}_i(0)$ represents the value taken below the cutoff (i.e., under control).  We assume that these covariates are predetermined, that is, that their values are determined prior to, or independently from, the treatment assignment and therefore that the treatment effect on them is zero by construction. For adjustment, researchers use the observed covariates, $\mathbf{Z}_i$, defined as
\[\mathbf{Z}_i
=\begin{cases} \mathbf{Z}_i(0) & \text{if } X_i < \C \\ 
\mathbf{Z}_i(1) & \text{if } X_i \geq \C.
\end{cases} \]

Predetermined covariates can be included in different ways to augment the basic RD estimation and inference methods. The two most natural approaches are (i) conditioning or subsetting, which makes the most sense when only a few discrete covariates are used, and (ii) partialling out via local polynomial methods. The first approach amounts to employing all the methods we discussed so far, after subsetting the data along the different subclasses generated by the interacted values of the covariates being used. For example, researchers may want to conduct separate analyses for men and women in the sample to study whether the estimated effects and confidence intervals differ between the two subgroups. The implementation of this conditioning approach does not require any modifications to the methods discussed above; they can be applied directly.

The second approach is based on augmenting the local polynomial model to include several additional covariates, which can be discrete or continuous. In this case, the idea is to directly include as many predetermined covariates as possible without affecting the validity of the point estimator, while at the same time improving its efficiency.

Our recommended covariate adjustment strategy is to augment the local polynomial fit by adding the covariates in a linear and additive-separable way. This involves fitting a weighted least-squares regression of the outcome $Y_i$ on (i) a constant, (ii) the treatment indicator $\T_i$, (iii) a $p$-order polynomial on the running variable, $(X_i-\C), (X_i-\C)^2,\ldots, (X_i-\C)^p$, (iv) a $p$-order polynomial on the running variable interacted with the treatment, $\T_i(X_i-\C) , \T_i(X_i-\C)^2,\ldots, \T_i(X_i-\C)^p$, and (v) the covariates $\mathbf{Z}_i$, using the weights $K((X_i-\C)/h)$. This defines the covariate-adjusted RD estimator:
\begin{align}
\label{eq:RDcov}
\tilde{\tau}_\mathtt{SRD} :
\tilde{Y}_i  = \tilde{\alpha} + \tilde{\tau}_\mathtt{SRD}T_i &+ \tilde\mu_{-,1} (X_i-\C) + \cdots + \tilde \mu_{-,p} (X_i-\C)^p  \\
             & +  \tilde\mu_{+,1} T_i(X_i-\C)+ \cdots + \tilde \mu_{+,p} T_i(X_i - \C)^p + \mathbf{Z}_i'\tilde{\bgamma} \nonumber \text{.}
\end{align}
The estimator $\tilde{\tau}_\mathtt{SRD}$ captures the average outcome jump at the cutoff in a fully interacted local polynomial regression fit, after partialling out the effect of the covariates $\mathbf{Z}_i$.  This approach reduces to the standard RD estimation when no covariates are included. 

A very important question is whether the covariate-adjusted estimator $\tilde{\tau}_\mathtt{SRD}$ estimates the same parameter as the unadjusted estimator $\hat{\tau}_\mathtt{SRD}$. It can be shown that under mild regularity conditions, a sufficient condition for $\tilde{\tau}_\mathtt{SRD}$ to be consistent for the average treatment effect at the cutoff, $\tau_\mathtt{SRD}= \E[Y_i(1) - Y_i(0) | X_i=\C]$, is that the RD treatment effect on the covariates is zero, that is, that the averages of the covariates under treatment and control at the cutoff are equal to each other,  $\E[\mathbf{Z}_i(0)|X_i=\C] = \E[\mathbf{Z}_i(1)|X_i=\C]$. This condition is analogous to the ``covariate balance'' requirement in randomized experiments, and will hold naturally when the covariates are truly predetermined. 

Thus, when predetermined covariates are included in the estimation as in Equation (\ref{eq:RDcov}), the covariate-adjusted estimator estimates the standard RD treatment effect, $\tau_\mathtt{SRD}$. This result, however, depends crucially on the particular way in which the covariates are included in (\ref{eq:RDcov}): linearly, additive-separably, and without interacting the covariates with the treatment. If, instead of adding $ \mathbf{Z}_i'\tilde{\bgamma}$, we interacted the covariates with the treatment and included the terms $ (1-T_i)\mathbf{Z}_i'\check{\bgamma}_- + T_i\mathbf{Z}_i'\check{\bgamma}_+$ , a zero RD treatment effect on the covariates would no longer be sufficient for $\tilde{\tau}_\mathtt{SRD}$ to be a consistent estimator of $\tau_\mathtt{SRD}$. We therefore recommend including covariates without interacting them with the treatment indicator, as shown in (\ref{eq:RDcov}).

In sum, if the goal is to estimate the RD treatment effect $\tau_\mathtt{SRD}$, the covariate adjustment should only include predetermined covariates, as including posttreatment or imbalanced covariates will change the parameter being estimated. It follows that, in general, it is not possible to include imbalanced covariates in the estimation to ``fix'' an RD design in which predetermined covariates are discontinuous at the cutoff and the required continuity assumptions are called into question. For the inclusion of covariates to ``control for'' unexpected imbalances, researchers will either need to invoke parametric assumptions on the regression functions to enable extrapolation, or redefine the parameter of interest. Therefore, analogously to the case of randomized experiments, the generally valid justification for including covariates in RD analysis is the potential for efficiency gains, not the promise to fix implausible identification assumptions. In many RD applications, a covariate-adjusted local polynomial estimation strategy will lead to shorter confidence intervals for the RD treatment effect, increasing the precision of statistical inferences.\smallskip

\subsubsection*{Practical Implementation of Covariate-Adjusted RD Analysis}

Including covariates in a linear-in-parameters way as in Equation (\ref{eq:RDcov}) requires the same type of choices as in the standard, unadjusted case: a polynomial order $p$, a kernel function $K(\cdot)$, and a bandwidth $h$. Once again, the bandwidth is a crucial choice, and we recommend using optimal data-driven methods to select it.  Since the covariate-adjusted point estimator $\tilde{\tau}_\mathtt{SRD}$ is a function of the covariates, its MSE will also be a function of the covariates. Thus, the optimal bandwidth choices for  $\tilde{\tau}_\mathtt{SRD}$ will depend on the covariates and will be in general different from the previously discussed bandwidths $h_\mathtt{MSE}$ and $h_\mathtt{CER}$. As a consequence, the principled implementation of covariate-adjusted local polynomial methods requires employing an MSE-optimal bandwidth that accounts for the inclusion of covariates in the bandwidth selection step. Although we omit the technical details here, both the MSE-optimal and the CER-optimal bandwidth choices that account for covariate adjustment have been theoretically derived; and they are both implemented in the \texttt{rdrobust} software.

We illustrate the inclusion of covariates with the Meyersson application, using the predetermined covariates introduced in Section \ref{subsec:intro-Meyersson}: variables from the 1994 election (\texttt{vshr\_islam1994}, \texttt{partycount}, \texttt{lpop1994}), and the geographic indicators (\texttt{merkezi}, \texttt{merkezp}, \texttt{subbuyuk}, \texttt{buyuk}). In order to keep the same number of observations as in the analysis without covariates, we exclude the indicator for electing an Islamic party in 1989 (\texttt{i89}) because this variable has many missing values.

We start by using \texttt{rdbwselect} to choose an MSE-optimal bandwidth with covariates, using the default options: a polynomial of order one, a triangular kernel, and the same bandwidth on each side of the cutoff (\texttt{mserd} option). We include covariates using the option \texttt{covs}.

\labelsnippet{snippetLrdcontR}
\rsnip{Vol-1-R_meyersson_rdbwselect_triangular_mserd_p1_regterm1_covariates_noi89.txt}{\Rlink{\thesnippetLrdcontR}}
\statasnip{Vol-1-STATA_meyersson_rdbwselect_triangular_mserd_p1_regterm1_covariates_noi89}{\Slink{\thesnippetLrdcontR}}

The MSE-optimal bandwidth including covariates is 14.409, considerably different from the value of 17.239 that we found before in the absence of covariate adjustment. This illustrates the general principle that covariate adjustment will generally change the values of the optimal bandwidths, which in turn will change the point estimates. (Note, however, that the covariate-adjusted local polynomial RD estimate would be different from the unadjusted estimate even if the same bandwidth were employed, as in finite samples the inclusion of covariates will change the estimated coefficients in the local polynomial fit.)

To perform covariate-adjusted local polynomial estimation and inference, we use the \texttt{rdrobust} command using the \texttt{covs} option.

\labelsnippet{snippetLrdcontS}
\rsnip{Vol-1-R_meyersson_rdrobust_triangular_mserd_p1_regterm1_covariates_noi89.txt}{\Rlink{\thesnippetLrdcontS}}
\statasnip{Vol-1-STATA_meyersson_rdrobust_triangular_mserd_p1_regterm1_covariates_noi89}{\Slink{\thesnippetLrdcontS}}

The estimated RD effect is now 3.108, similar to the unadjusted estimate of 3.020 that we found before. This similarity is reassuring because, if the included covariates are truly predetermined, the unadjusted estimator and the covariate-adjusted estimator are estimating the same parameter and should result in similar estimates. In terms of inference, with the inclusion of covariates, the 95\% robust confidence interval is now $[0.194,6.132]$. The unadjusted robust confidence interval we estimated in the previous section is $[-0.309,6.276]$. Thus, including covariates reduces the length of the confidence interval from $6.276 - (-0.309) =6.585 $ to $ 6.132 - 0.194 = 5.938 $, a reduction of $(|5.938 - 6.585 |/6.585) \times 100 = 9.82\%$. The shorter confidence interval obtained with covariate adjustment (and the slight increase in the point estimate) results in the robust p-value decreasing from 0.076 to 0.037. 

This exercise illustrates the main benefit of covariate adjustment in local polynomial RD estimation: when successful, the inclusion of covariates in the analysis decreases the length of the confidence interval while simultaneously leaving the point estimate (roughly) unchanged.

\subsubsection{Clustering the Standard Errors}

Another issue commonly encountered by practitioners is the clustering of observations in groups, such as individuals inside households, municipalities inside counties, or households inside villages. When the units of analysis are clustered into groups and the researcher suspects that the errors are correlated within (but not across) groups, it may be appropriate to employ variance estimators that are robust to the clustered nature of the data.

Using ideas from least-squares estimation and inference methods, it is possible to adjust the local polynomial variance estimators to account for clustering. Since the CER- and MSE-optimal bandwidth selectors depend on the variance estimators, employing cluster-robust variance estimators changes the optimal bandwidth relative to the case of no clustering. Consequently, in the local polynomial RD setting (and in contrast to the ordinary least-squares setting) employing cluster-robust variance estimators leads not only to different estimated standard errors relative to the unclustered case, but also to different point estimates. In general, cluster-robust variance estimators can be smaller or larger than variance estimators that do not account for clustering. This fact, combined with the associated change in the point estimator that results from the change in the optimal bandwidth when cluster-robust variance estimators are employed, means that cluster-robust standard errors can lead to recentered confidence intervals that can be either shorter or longer in length.

The cluster-robust variance estimation formulas are beyond the scope of this practical guide, but we do illustrate how to employ these estimators in practice using \texttt{rdrobust}. We provide further illustration of these methods in the discussion of RD designs with discrete running variables in the accompanying Element (\citeauthor*{Cattaneo-Idrobo-Titiunik_2019_Vol2}, forthcoming).

In the Meyersson application, we estimate the effect of Islamic victory on the educational attainment of women, clustering each individual observation (which corresponds to a municipality) by province. In \texttt{rdrobust}, we use the option \texttt{cluster} to pass the variable that contains the cluster information for every observation.

\labelsnippet{snippetLrdcontT}
\rsnip{Vol-1-R_meyersson_rdrobust_triangular_mserd_p1_regterm1_clusters.txt}{\Rlink{\thesnippetLrdcontT}}
\statasnip{Vol-1-STATA_meyersson_rdrobust_triangular_mserd_p1_regterm1_clusters}{\Slink{\thesnippetLrdcontT}}

Using a cluster-robust variance estimator leads to a point estimator of 2.969 percentage points, slightly smaller than the point estimator of 3.020 that we obtained without clustering in Section \ref{subsec:localpolyMeyersson}. This change in the point estimate occurs because the MSE-optimal bandwidth is now 19.035, larger than the 17.239 bandwidth estimated in the absence of clustering. In addition, the cluster-robust variance estimator is larger than the unclustered variance estimator---for example, comparing to the prior results in the absence of clustering, the conventional standard error changes from 1.427 to 1.604 when a cluster-robust estimator is used. The decrease in the point estimator, together with the increase in the variance, lead to a wider confidence interval; with a cluster-robust variance estimator, the robust confidence interval is [-0.583,6.460], wider and with center closer to zero than the [-0.309,6.276] robust confidence interval in the absence of clustering.

We can also combine a cluster-robust variance estimator with covariate adjustment in the local polynomial fit, using the \texttt{covs} and \texttt{cluster} options simultaneously. 

\labelsnippet{snippetLrdcontU}
\rsnip{Vol-1-R_meyersson_rdrobust_triangular_mserd_p1_regterm1_covariates_noi89_clusters.txt}{\Rlink{\thesnippetLrdcontU}}
\statasnip{Vol-1-STATA_meyersson_rdrobust_triangular_mserd_p1_regterm1_covariates_noi89_clusters}{\Slink{\thesnippetLrdcontU}}

Relative to the unadjusted, unclustered case, adding covariates and employing cluster-robust variance estimators leads to a different optimal bandwidth of 15.675, which changes the point estimate to 3.146. Once again, these changes translate into  a different confidence interval, equal to [0.243,6.154]. Relative to the case of cluster-robust variance estimator without covariate adjustment, adding covariates reduces the length of the confidence interval and shifts it to the right. As a result, the cluster-robust covariate-adjusted 95\% confidence interval does not include zero.

\subsection{Further Reading}

A textbook discussion of non-parametric local polynomial methods can be found in \citet*{Fan-Gijbels_1996_Book}, and their application to RD estimation and inference is discussed by \citet*{Hahn-Todd-vanderKlaauw_2001_ECMA}. \citet*{Calonico-Cattaneo-Titiunik_2015_JASA} and \citet*{Gelman-Imbens_2019_JBES} discuss the role of global polynomial estimation for RD analysis. MSE-optimal bandwidth selection for the local polynomial RD point estimator is developed in \citet*{Imbens-Kalyanaraman_2012_REStud}, \citet*{Calonico-Cattaneo-Titiunik_2014_ECMA}, \citet*{Bartalotti-Brummet_2017_AIE}, \citet*{Calonico-Cattaneo-Farrell-Titiunik_2019_RESTAT}, and \citet*{Arai-Ichimura_2018_QE}. Robust bias corrected confidence intervals were proposed by \citet*{Calonico-Cattaneo-Titiunik_2014_ECMA}, and their higher-order properties as well as CER-optimal bandwidth selection were developed by \citet*{Calonico-Cattaneo-Farrell_2018_JASA,Calonico-Cattaneo-Farrell_2019_wp-optcer,Calonico-Cattaneo-Farrell_2019_wp-cerbw}. See also \citet{Cattaneo-VazquezBare_2016_ObsStud} for an overview of RD bandwidth selection methods. Bootstrap methods based on robust bias correction are developed in \citet*{Bartalotti-Gray-He_2017_AIE}. RD analysis with the inclusion of predetermined covariates and cluster-robust inference is discussed in \citet*{Calonico-Cattaneo-Farrell-Titiunik_2019_RESTAT}, and other extensions of estimation and inference using robust bias correction are discussed in \citet*{Xu_2017_JoE}, \citet*{Dong_2019_JBES}, and \citet*{Dong-Lee-Gou_2019_wp}. \citet*{Tukiainen-et-al-2018_QE} offer an empirical example assessing the performance of robust bias correction inference methods. \citet*{Cattaneo-Titiunik-VazquezBare_2019_Stata} discuss power calculations using local polynomial menthods in RD designs. Further related results and references are given in the edited volume by \citet{Cattaneo-Escanciano_2017_AIE}.

\newpage
\section{Validation and Falsification of the RD Design}
\label{sec:falsification}

A main advantage of the RD design is that the mechanism by which treatment is assigned is known and based on observable features, giving researchers an objective basis to distinguish pre-treatment from post-treatment variables, and to identify qualitative information regarding the treatment assignment process that can be helpful to justify assumptions. However, the known rule that assigns treatment based on the score and cutoff is not by itself enough to guarantee that the assumptions needed to recover the RD effect are met.

For example, a scholarship may be assigned based on whether students receive an exam grade above a cutoff, but if the cutoff is known to the students' parents and there are mechanisms to appeal the grade, the RD design may be invalid if systematic differences among students are present due to the appeal process. Formally, the presence of an appeal process might invalidate the assumption that the average potential outcomes are continuous at the cutoff. If the parents who are successful in appealing the grade when their child is barely below the cutoff are systematically different from the parents who choose not to appeal in ways that affect the outcome of interest, then the RD design based on the final grade assigned to each student would be invalid (while the RD design based on the original grade would not). For instance, if the outcome of interest is performance on a future exam and parent involvement is positively correlated  with students' future academic achievement, the average potential outcomes of students at or just above the cutoff will be much higher than the average potential outcomes of students just below the cutoff, leading to a discontinuity at the cutoff and thus invalidating the RD design.

If the RD cutoff is known to the units that will be the beneficiaries of the treatment, researchers must worry about the possibility of units actively changing or manipulating the value of their score when they miss the treatment barely. Thus, the first type of information that should be provided is whether an institutionalized mechanism to appeal the score exists, and if so, how often (and by whom) it is used to successfully change the score. Qualitative data about the administrative process by which scores are assigned, cutoffs determined and publicized, and treatment decisions appealed, is extremely useful to validate the design. For example, social programs are commonly assigned based on some poverty index; if program officers moved units with index barely below the cutoff to the treatment group in a systematic way (e.g., all households with small children), then the RD design would be invalid whenever the systematic differences between treated and control units near the cutoff were correlated with outcome differences. This type of behavior can typically be identified collecting qualitative information (such as interviews, internal rules and memos, etc.) from the program administration officers.

In many cases, however, qualitative information will be limited, and researchers will be unable to completely rule out the possibility of units manipulating their score. More importantly, the fact that there are no institutionalized or known mechanisms to appeal and change the score does not imply the absence of informal mechanisms by which this may happen. Thus, an essential step in evaluating the plausibility of the RD assumptions is to provide empirical evidence supporting the validity of the design. Naturally, the continuity assumptions that guarantee the validity of the RD design are about unobservable features and as such are inherently untestable. Nonetheless, the RD design offers an array of empirical methods that, under reasonable assumptions, can provide useful evidence about the plausibility of its assumptions. These so-called validation methods are based on various empirical implications of the unobservable RD assumptions that can be expected to hold in most cases, and can provide indirect evidence about its validity. 

We now discuss five empirical validation tests based on (i) the null treatment effect on predetermined covariates or placebo outcomes, (ii) the continuity of the score density around the cutoff, (iii) the treatment effect at artificial cutoff values, (iv) the exclusion of observations near the cutoff, and (v) the sensitivity to bandwidth choices.

\subsection{Predetermined Covariates and Placebo Outcomes}

One of the most important RD falsification tests involves examining whether, near the cutoff, treated units are similar to control units in terms of observable characteristics. The idea is simply that, if units lack the ability to precisely manipulate the score value they receive, there should be no systematic differences between units with similar values of the score. Thus, except for their treatment status, units just above and just below the cutoff should be similar in all variables that could not have been affected by the treatment. These variables can be divided into two groups: variables that are determined before the treatment is assigned---which we call \textit{predetermined covariates}; and variables that are determined after the treatment is assigned but, according to substantive knowledge about the treatment's causal mechanism, could not possibly have been affected by the treatment---which we call \textit{placebo outcomes}.

Note that predetermined covariates can be unambiguously defined, but placebo outcomes are always specific to each application. For example, any characteristic that is determined before the moment when treatment is assigned is generally a predetermined covariate. In contrast, whether a variable is a placebo outcome depends on the particular treatment under consideration. For example, if the treatment is access to clean water and the outcome of interest is child mortality, a treatment effect is expected on mortality due to water-bone illnesses but not on mortality due to other causes such as car accidents \citep*[see][]{GalianiGertlerSchargrodsky2005-JPE}. Thus, mortality from road accidents would be a reasonable placebo outcome in this example. However, child mortality from road accidents would not be an adequate placebo outcome to validate an RD design that studies the effects of a safety program aimed at increasing the use of car seats.

Regardless of whether the analysis is based on predetermined covariates or placebo outcomes, the fundamental principle behind this type of falsification analysis is always the same: all predetermined covariates and placebo outcomes should be analyzed in the same way as the outcome of interest. In the continuity-based approach, this principle means that for each predetermined covariate or placebo outcome, researchers should first choose an optimal bandwidth, and then use local polynomial techniques within that bandwidth to estimate the ``treatment effect'' and employ valid inference procedures such as the robust bias-corrected methods discussed previously. The fundamental idea behind this test is that, since the predetermined covariate (or placebo outcome) could not have been affected by the treatment, the null hypothesis of no treatment effect should not be rejected if the RD design is valid. The reasoning is that if covariates or placebo outcomes that are known to correlate strongly with the outcome of interest are discontinuous at the cutoff, the continuity of the potential outcome functions is unlikely to hold, and thus the validity of the design is called into question.

\begin{figure}[H]
	\vspace{-0.50in}
	\hspace{-0.70in}
	\centering
	\begin{subfigure}[t]{0.45\textwidth}
		\includegraphics[scale=0.45]{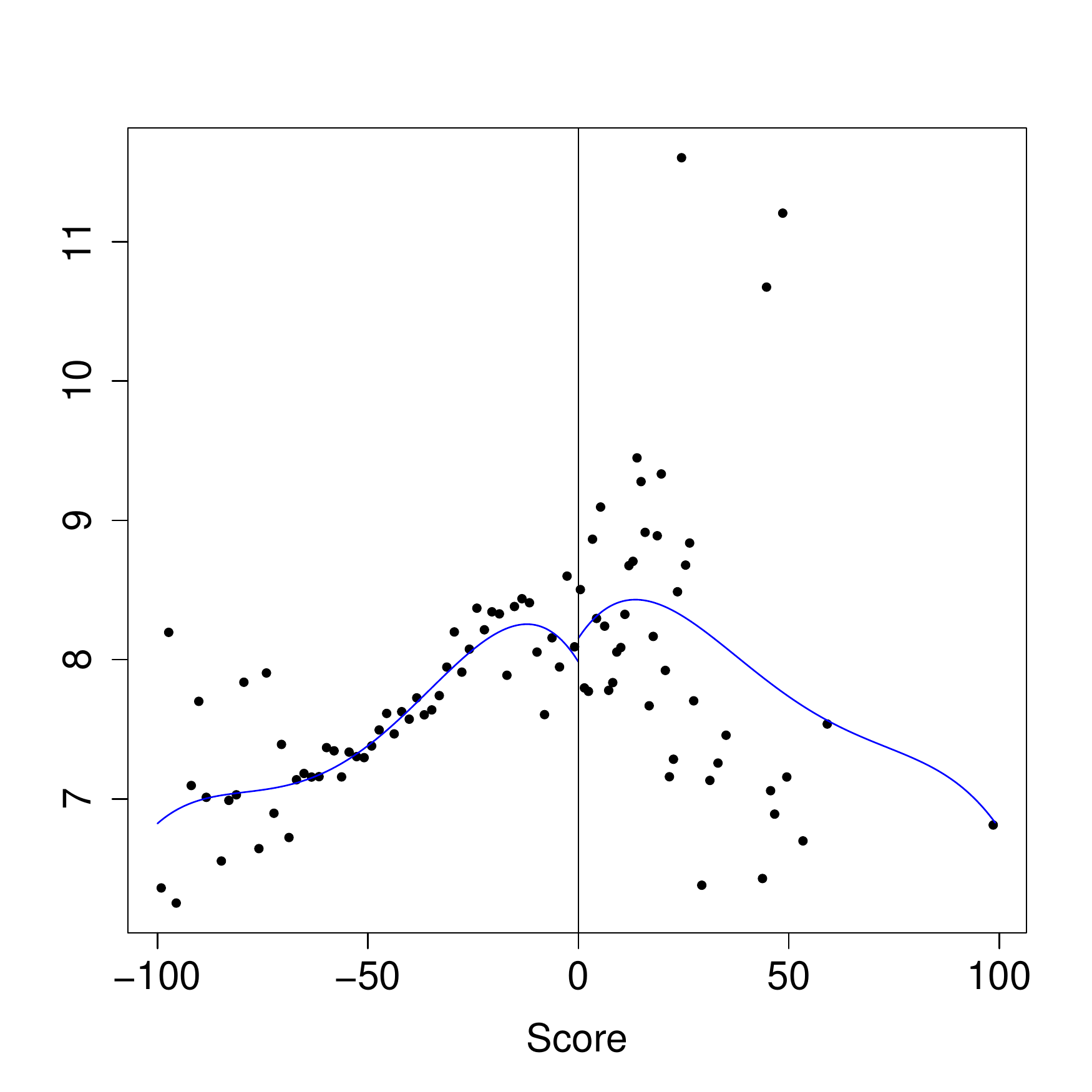}
		\vspace{-0.35in}
		\caption{Log Population in 1994}
	\end{subfigure}
	\hspace{0.20in}%
	\begin{subfigure}[t]{0.45\textwidth}
		\includegraphics[scale=0.45]{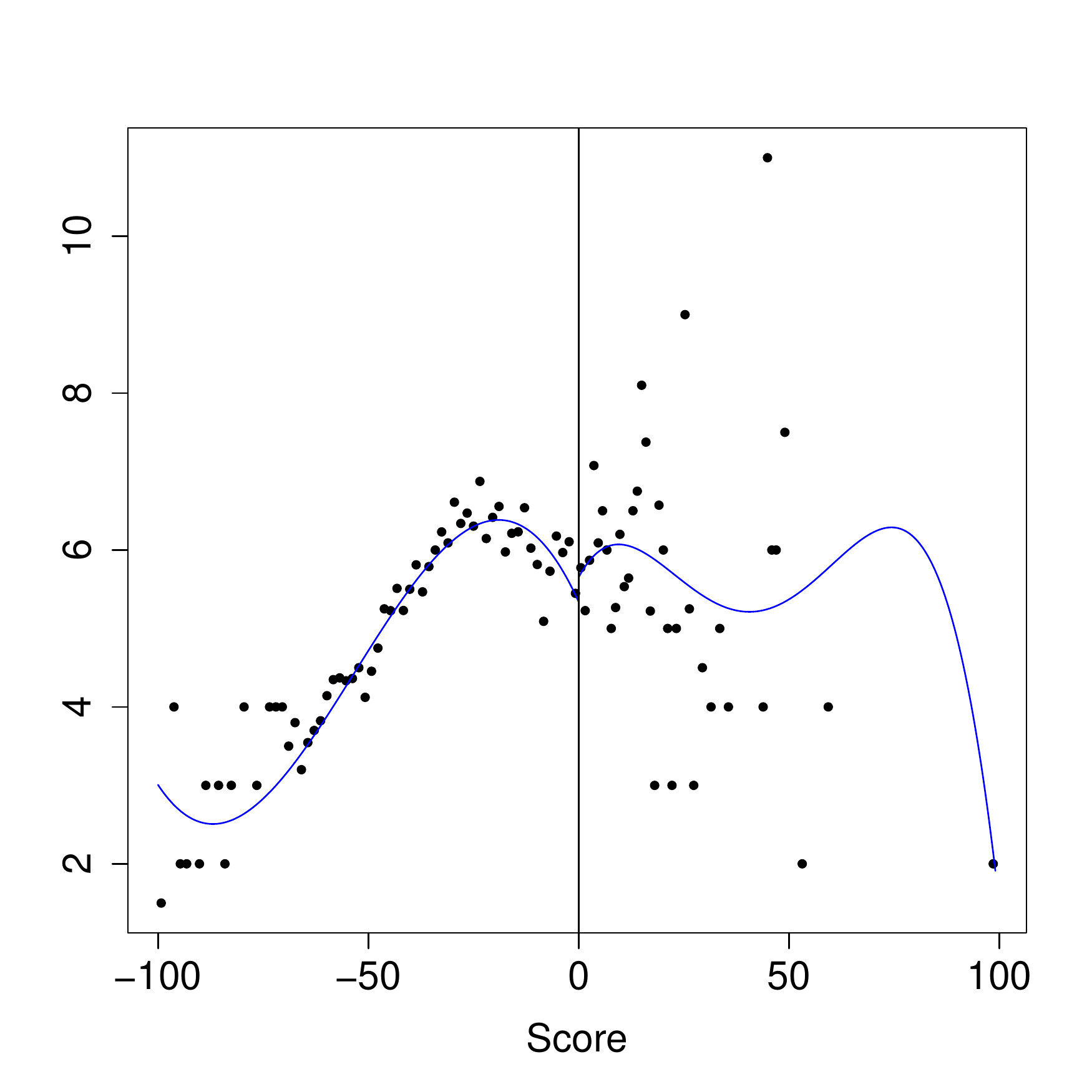}
		\vspace{-0.35in}
		\caption{Number of Parties Receiving Votes in 1994}
	\end{subfigure}\\
	\vspace{-0.3in}
	\hspace{-0.70in}        
	\begin{subfigure}[t]{0.45\textwidth}
		\includegraphics[scale=0.45]{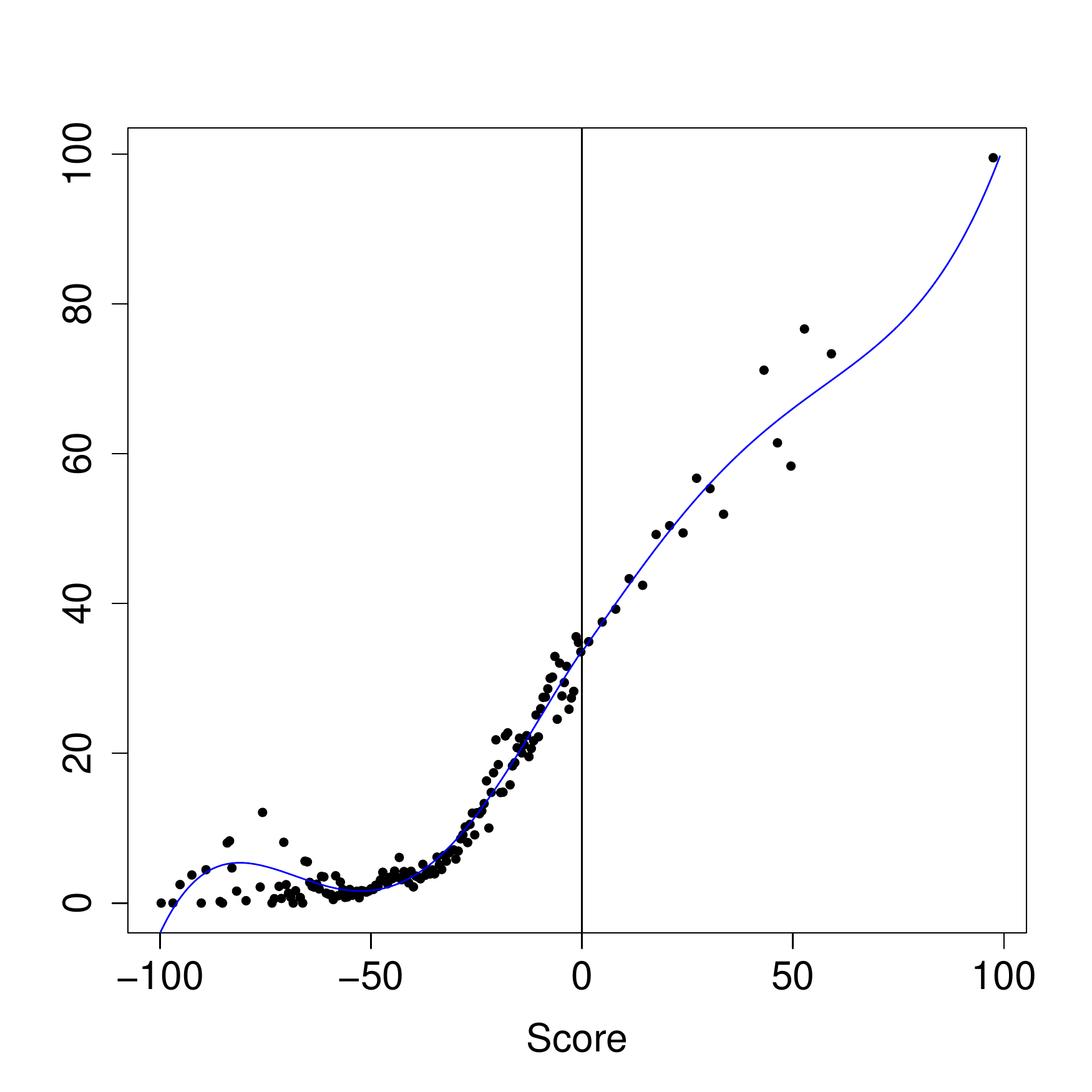}
		\vspace{-0.35in}
		\caption{Islamic Vote Percentage in 1994}
	\end{subfigure}
	\vspace{-0.3in}
	\hspace{0.20in}
	\centering
	\begin{subfigure}[t]{0.45\textwidth}
		\includegraphics[scale=0.45]{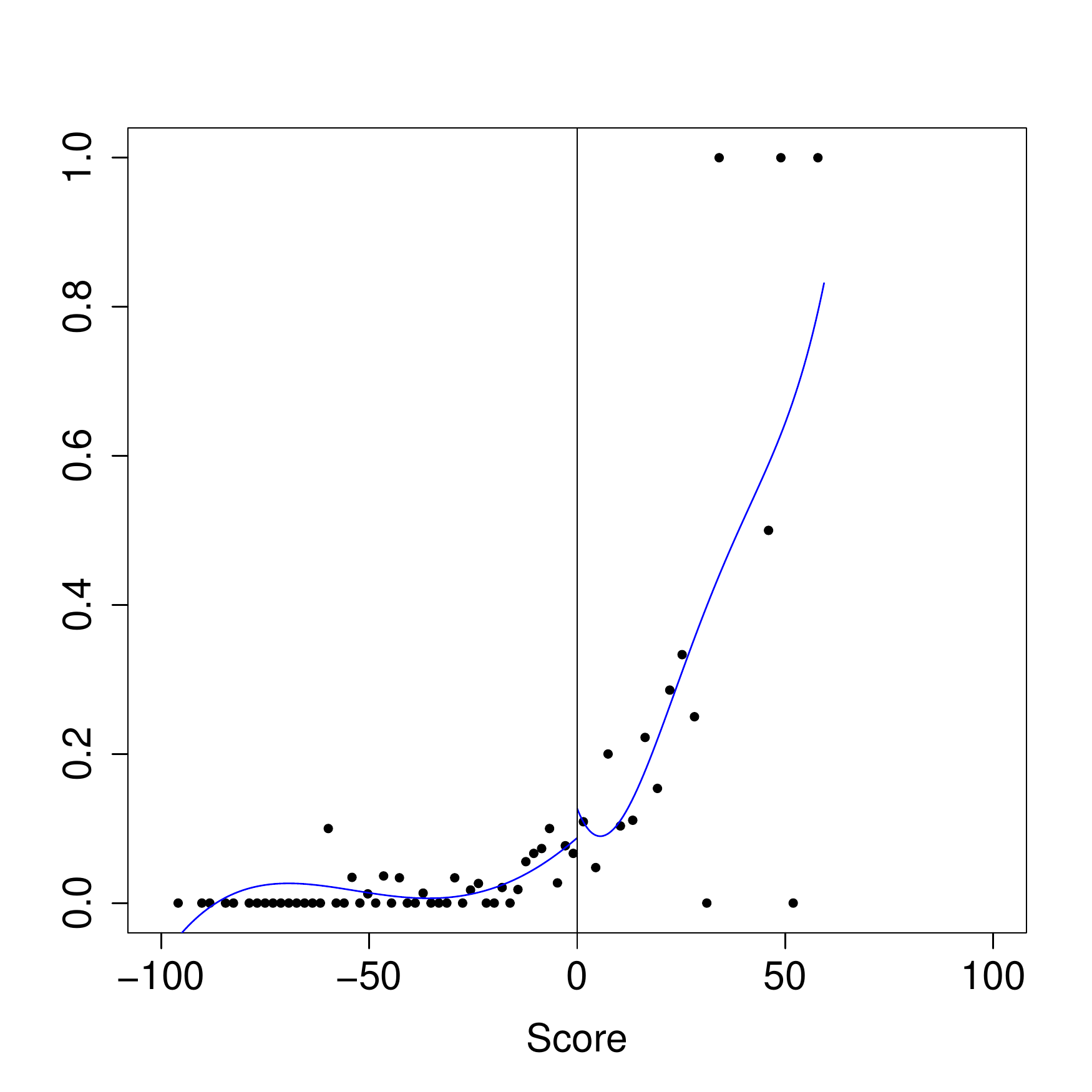}
		\vspace{-0.35in}
		\caption{Islamic Mayor in 1989}
	\end{subfigure}\\
	\hspace{-0.70in}        
	\begin{subfigure}[t]{0.45\textwidth}
		\includegraphics[scale=0.45]{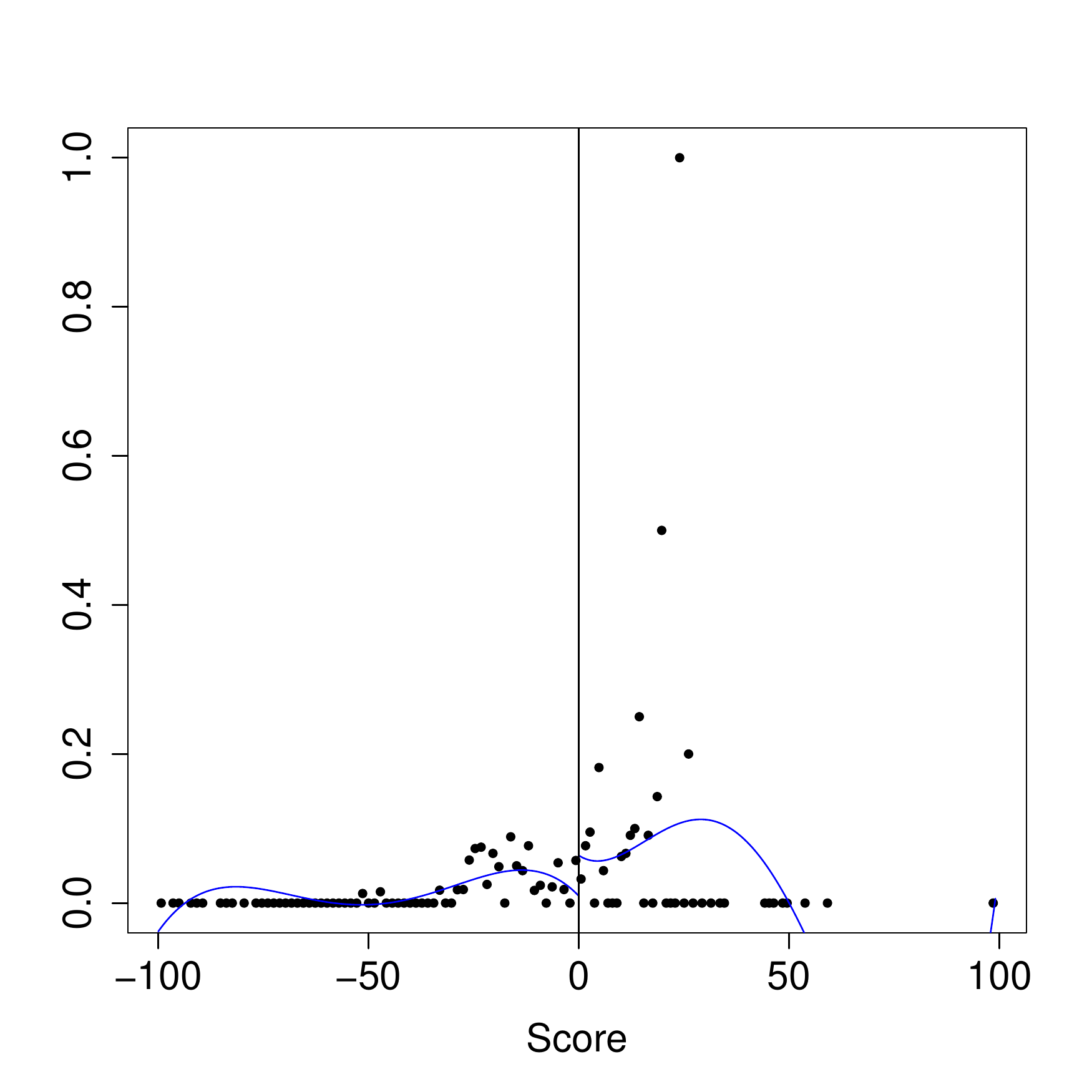}
		\vspace{-0.35in}
		\caption{Province Center Indicator}
	\end{subfigure}
	\vspace{-0.3in}
	\hspace{0.20in}
	\centering
	\begin{subfigure}[t]{0.45\textwidth}
		\includegraphics[scale=0.45]{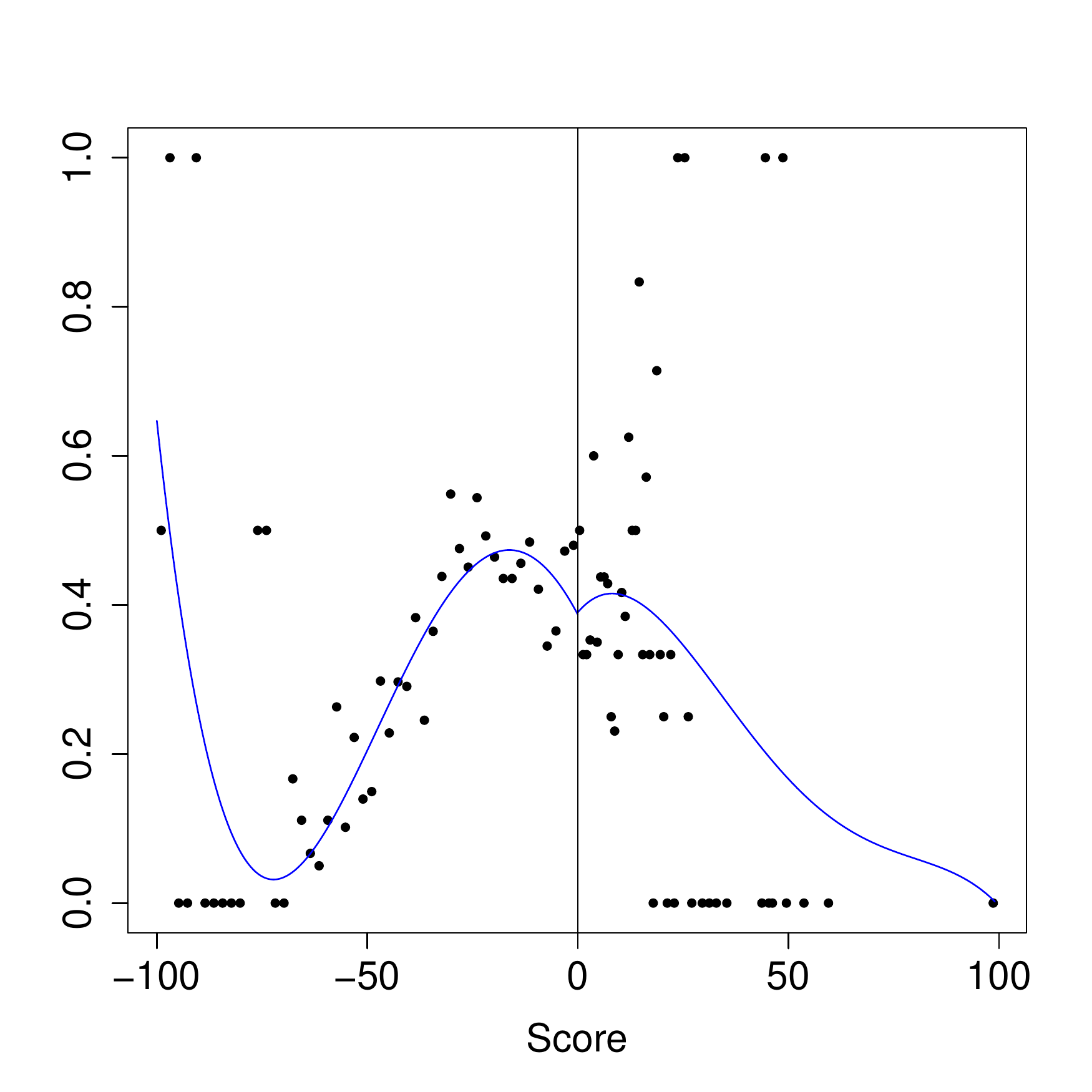}
		\vspace{-0.35in}
		\caption{District Center Indicator}
	\end{subfigure}
	\vspace{0.2in}
	\caption{RD Plots for Predetermined Covariates (Meyersson Application)}\label{fig:rdplot_covariates_AllSupport}
\end{figure}

When using the continuity-based approach to RD analysis, this falsification test employs the local polynomial techniques discussed in Section \ref{sec:localpoly} to test whether the predetermined covariates and placebo outcomes are continuous at the cutoff, in other words, to test whether the treatment has an effect on them. We illustrate with the Meyersson application, using the set of predetermined covariates used for covariate adjustment in Section \ref{subsec:localpolyMeyersson}. We start by presenting a graphical analysis, creating an RD plot for every covariate using \texttt{rdplot} with the default options (mimicking variance, evenly-spaced bins). The plots are presented in Figure \ref{fig:rdplot_covariates_AllSupport}. The specific commands are omitted to conserve space, but they are included in the online replication materials.

\labelfiguras{figP}

The graphical analysis does not reveal obvious discontinuities at the cutoff, but of course a statistical analysis is required before we can reach a formal conclusion. In order to implement the analysis, an optimal bandwidth must be chosen for each covariate. Crucially, these bandwidths will be generally different from the bandwidth used to analyze the original outcome of interest. As shown in the RD plots, each covariate exhibits a different estimated regression function, with different curvature and overall shape. As a result, the optimal bandwidth for local polynomial estimation and inference will be different for every variable, and must be re-estimated accordingly in each case. This implies that the statistical analysis must be conducted separately for each covariate, choosing a different optimal bandwidth for each covariate analyzed.

To implement this formal falsification test, we simply run \texttt{rdrobust} using each covariate of interest as the outcome variable. As an example, we analyze the covariate \texttt{lpop1994}, the logarithm of the municipality population in 1994. Since this covariate was measured in 1994, it could not have been affected by the treatment, that is, by the party that wins the 1994 election. We estimate a local linear RD effect with triangular kernel weights and common MSE-optimal bandwidth using the default options in \texttt{rdrobust}.

\labelsnippet{snippetLrdcontV}
\rsnip{Vol-1-R_meyersson_falsification_rdrobust_lpop1994.txt}{\Rlink{\thesnippetLrdcontV}}
\statasnip{Vol-1-STATA_meyersson_falsification_rdrobust_lpop1994}{\Slink{\thesnippetLrdcontV}}

The point estimate is very close to zero and the robust p-value is $0.999$, so we find no evidence that, at the cutoff, treated and control municipalities differ systematically in this covariate. In other words, we find no evidence that the population size of the municipalities is discontinuous at the cutoff. In order to provide a complete falsification test, the same estimation and inference procedure should be repeated for all important covariates, that is, for all available covariates that would be expected to be correlated with the treatment in the presence of manipulation. In a convincing RD design, these tests would show that there are no discontinuities in any variable. Table \ref{tab:rdrobust_covariates} contains the local polynomial estimation and inference results for several predetermined covariates in the Meyersson dataset. All results were obtained employing \texttt{rdrobust} with the default specifications, as shown for \texttt{lpop1994} above.

\labeltablas{tableD}
\begin{table}[H]
	\centering
	\caption{Formal Continuity-Based Analysis for Covariates}
	\resizebox{\textwidth}{!}{\begin{tabular}{lccccc}
			\toprule
			\multirow{2}{*}{Variable} & MSE-Optimal & RD        & \multicolumn{2}{c}{\underline{Robust Inference}} & Eff. Number\\
			& Bandwidth   & Estimator & p-value & Conf. Int.                             & Observations\\
			\midrule
			Percentage of men aged 15-20 with high school education & 12.055 & 1.561 & 0.358 & [-1.757, 4.862] & 590 \\
Islamic Mayor in 1989 & 11.782 & 0.053 & 0.333 & [-0.077, 0.228] & 418 \\
Islamic percentage of votes in 1994 & 13.940 & 0.603 & 0.711 & [-2.794, 4.095] & 668 \\
Number of parties receiving votes 1994 & 12.166 & -0.168 & 0.668 & [-1.357, 0.869] & 596 \\
Log population in 1994 & 13.319 & 0.012 & 0.999 & [-0.644, 0.645] & 633 \\
District center & 13.033 & -0.067 & 0.462 & [-0.285, 0.130] & 624 \\
Province center & 11.556 & 0.029 & 0.609 & [-0.064, 0.109] & 574 \\
Sub-metro center & 10.360 & -0.016 & 0.572 & [-0.114, 0.063] & 513 \\
Metro center & 13.621 & 0.008 & 0.723 & [-0.047, 0.068] & 642 \\

			\bottomrule
	\end{tabular}}
	\label{tab:rdrobust_covariates}
\end{table}

All point estimates are small and all $95\%$ robust confidence intervals contain zero, with p-values ranging from $0.333$ to $0.999$. In other words, there is no empirical evidence that these predetermined covariates are discontinuous at the cutoff. Note that the number of observations used in the analysis varies for each covariate; this occurs because the MSE-optimal bandwidth is different for every covariate analyzed. Note also that we employ the default \texttt{rdrobust} options for simplicity, but for falsification purposes it may be more appropriate to use the CER-optimal bandwidth because, in this case, we are only interested in testing the null hypothesis of no effect, that is, we are mostly interested in inference and the point estimates are of no particular interest. These two alternative bandwidth choices give a natural trade-off between size and power of the falsification tests: the MSE-optimal bandwidth leads to more powerful hypothesis tests with possibly larger size distortions than tests implemented using the CER-optimal bandwidth. In this application, switching to \texttt{bwselect="cerrd"} does not change any of the empirical conclusions (results available in the replication files).

We complement these results with a graphical illustration of the RD effects for every covariate, to provide further evidence that in fact these covariates do not jump discretely at the cutoff. For this, we employ \texttt{rdplot} with the same options we used for inference in \texttt{rdrobust}: we plot each covariate inside their respective MSE-optimal bandwidth, using a polynomial of order one, and a triangular kernel function to weigh the observations. Below we illustrate the specific command for the \texttt{lpop1994} covariate.

\labelsnippet{snippetLrdcontW}
\rsnip{Vol-1-R_meyersson_falsification_rdplot_rdrobust_lpop1994.txt}{\Rlink{\thesnippetLrdcontW}}
\statasnip{Vol-1-STATA_meyersson_falsification_rdplot_rdrobust_lpop1994}{\Slink{\thesnippetLrdcontW}}

\begin{figure}[H]
	\vspace{-0.50in}
	\hspace{-0.70in}
	\centering
	\begin{subfigure}[t]{0.485\textwidth}
		\includegraphics[scale=0.465]{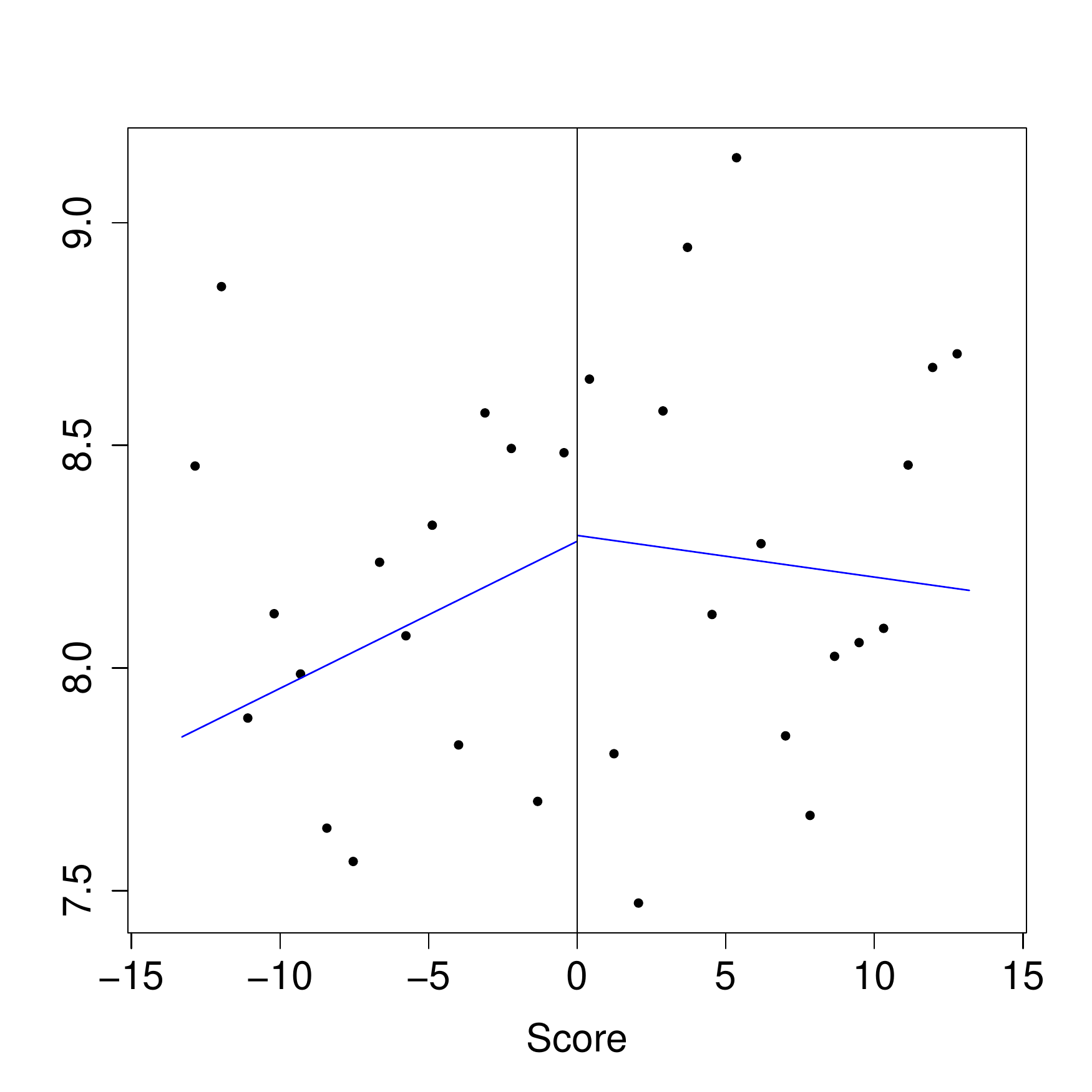}
		\vspace{-0.35in}
		\caption{Log Population in 1994}
	\end{subfigure}
	\hspace{0.20in}%
	\begin{subfigure}[t]{0.485\textwidth}
		\includegraphics[scale=0.465]{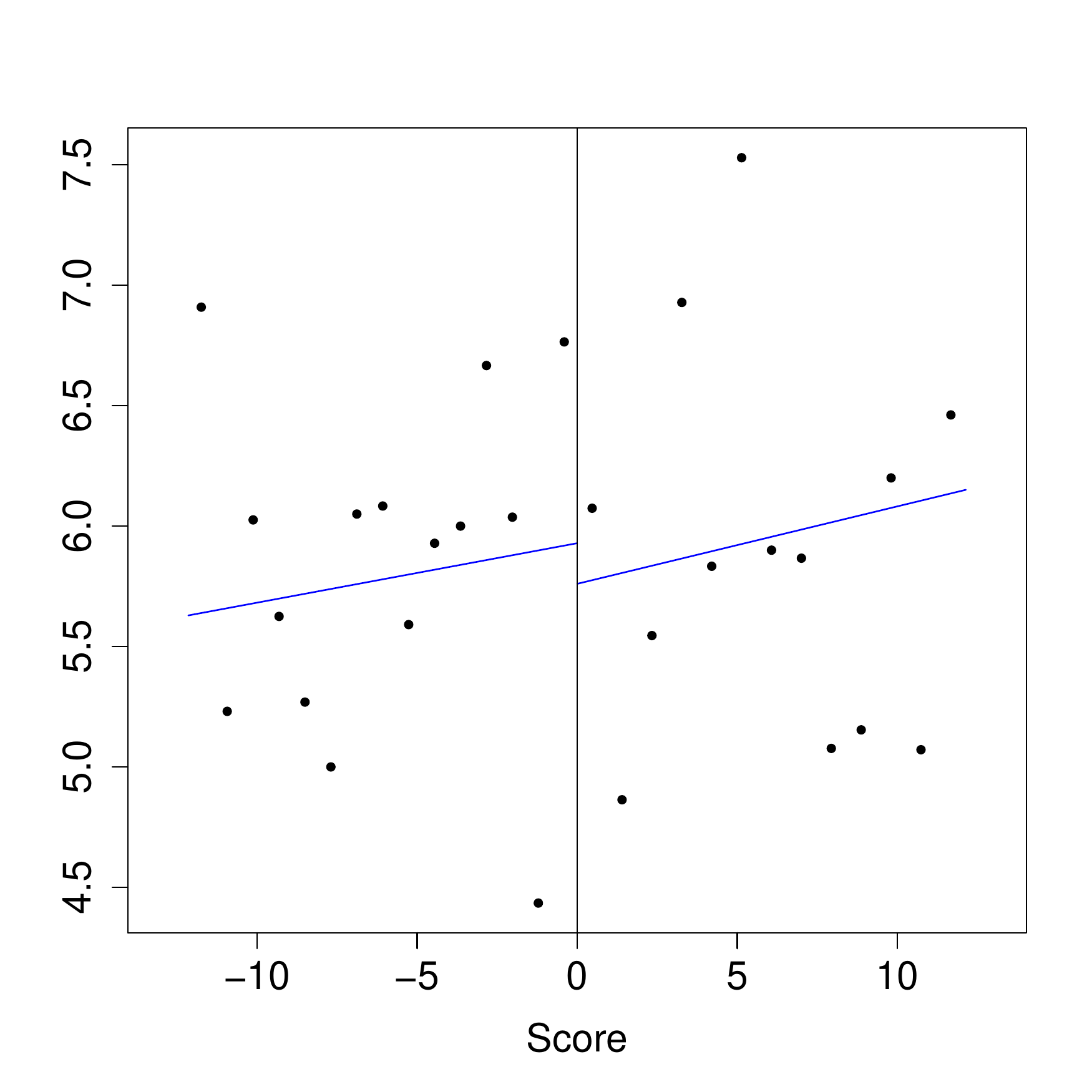}
		\vspace{-0.35in}
		\caption{Number of Parties Receiving Votes in 1994}
	\end{subfigure}\\
	\hspace{-0.70in}        
	\begin{subfigure}[t]{0.485\textwidth}
		\includegraphics[scale=0.465]{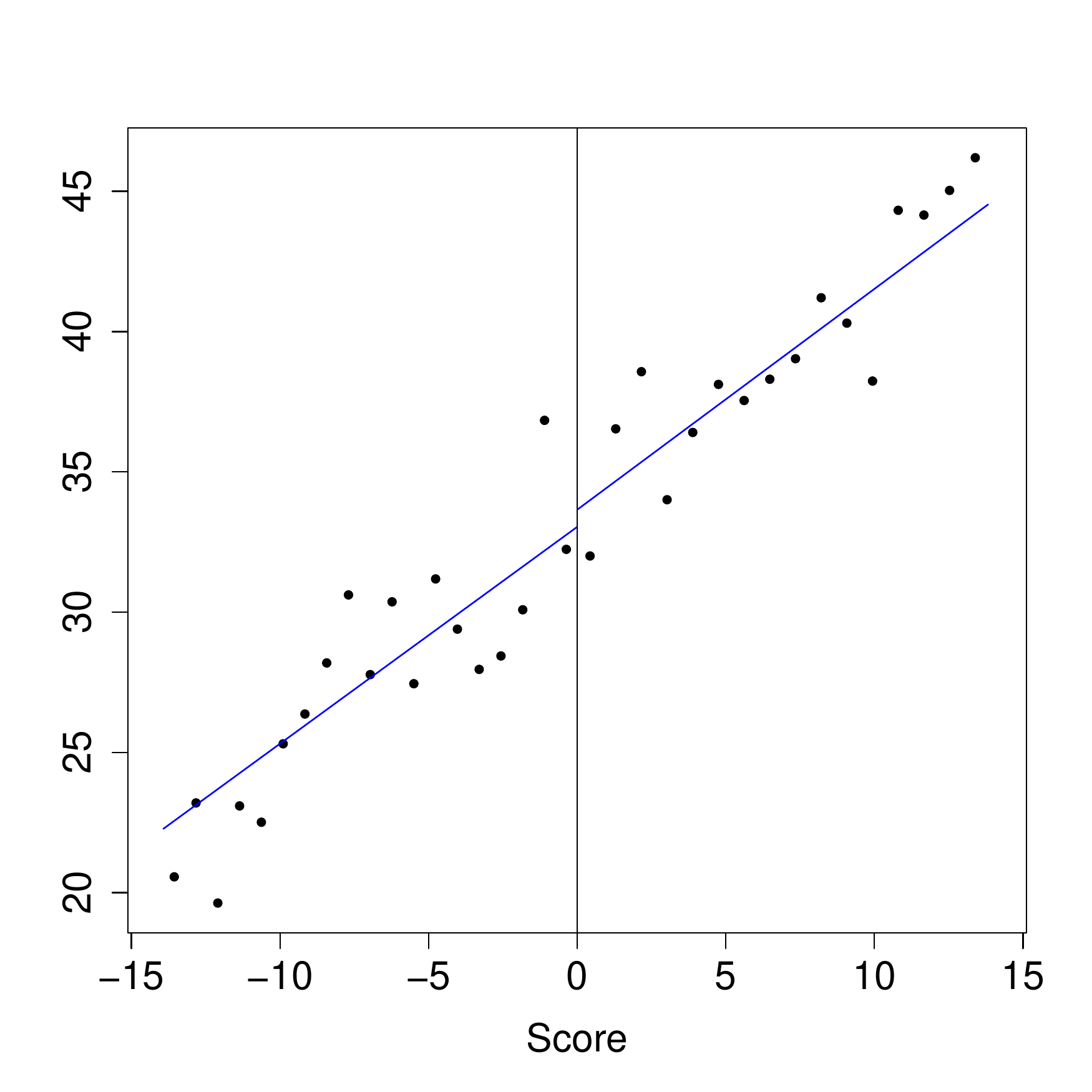}
		\vspace{-0.35in}
		\caption{Islamic Vote Percentage in 1994}
	\end{subfigure}
	\vspace{-0.3in}
	\hspace{0.20in}
	\centering
	\begin{subfigure}[t]{0.485\textwidth}
		\includegraphics[scale=0.465]{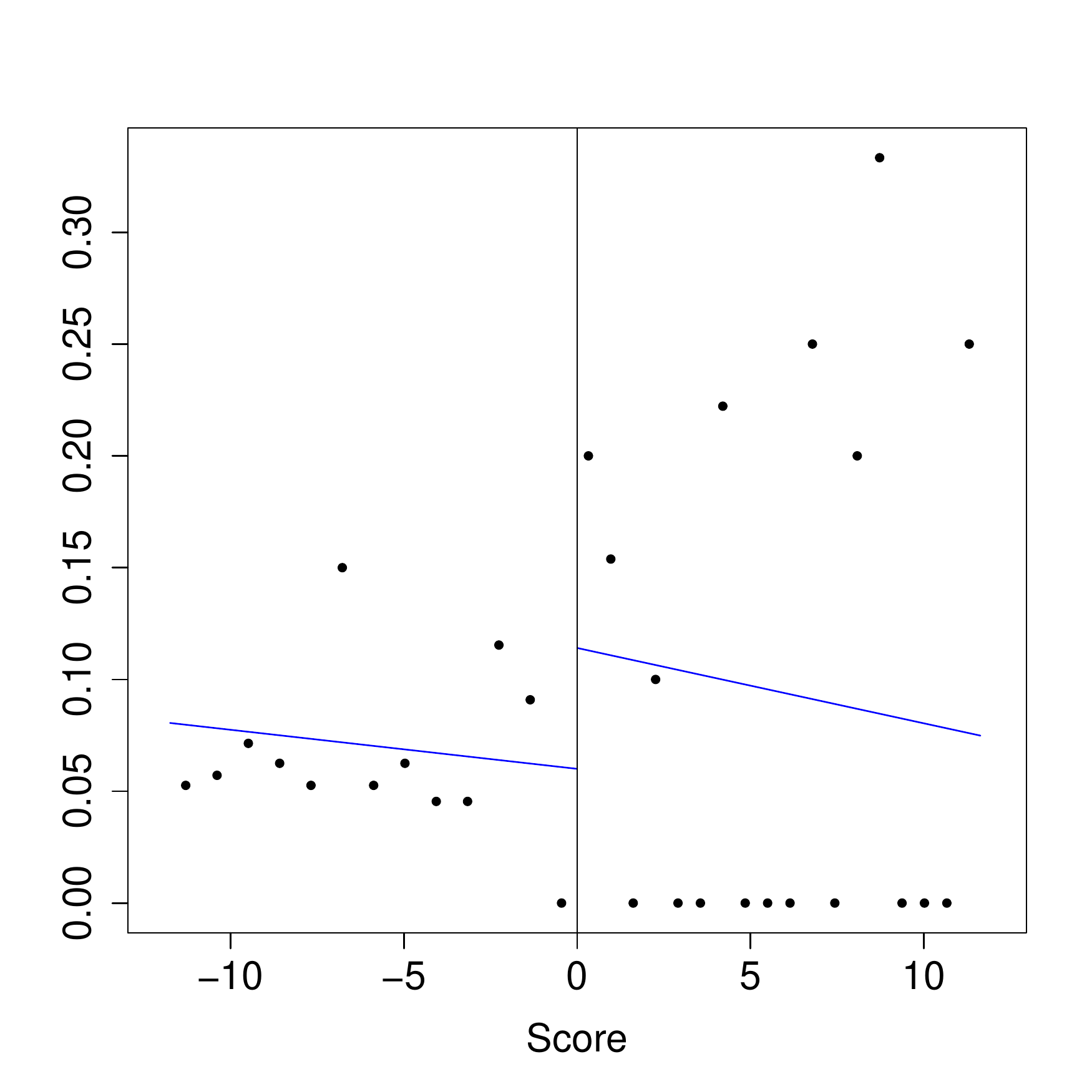}
		\vspace{-0.35in}
		\caption{Islamic Mayor in 1989}
	\end{subfigure}       
	\vspace{0.2in}
	\caption{Graphical Illustration of Local Linear RD Effects for Predetermined Covariates (Meyersson data)}\label{fig:rdrobust_covariates-RDeffects}
\end{figure}

We run the same commands for each covariate. A sample of the resulting plots is presented in Figure \ref{fig:rdrobust_covariates-RDeffects}. Consistent with the formal statistical results, the graphical analysis within the optimal bandwidth shows that the right and left intercepts in the local linear fits are very close to each other in most cases (the variable \texttt{Islamic Mayor in 1989} shows a more noticeable jump, but the formal analysis above indicates that this jump is not distinguishable from zero). 

\labelfiguras{figQ}

The plots and estimated effects for the covariates stand in contrast to the analogous results we reported for the outcome of interest in the previous sections, where, despite some variability, we saw a more noticeable jump at the cutoff. In general, a stark contrast between null effects for the covariates and a large nonzero effect for the outcome can be interpreted as evidence in favor of the validity of the RD design. However, the converse is not true, as it is possible to see a valid RD design where the treatment has no effect on the outcome, and thus where both covariate and outcome results are null.

\subsection{Density of Running Variable}

The second type of falsification test examines whether, in a local neighborhood near the cutoff, the number of observations below the cutoff is surprisingly different from the number of observations above it. The underlying assumption is that, if units do not have the ability to precisely manipulate the value of the score that they receive, the number of treated observations just above the cutoff should be approximately similar to the number of control observations below it. In other words, even if units actively attempt to affect their score, in the absence of precise manipulation, random change would place roughly the same amount of units on either side of the cutoff, leading to a continuous probability density function when the score is continuously distributed. RD applications where there is an abrupt change in the number of observations at the cutoff will tend to be less credible. 

\begin{figure}[ht]
\hspace{-0.2in}%
\begin{subfigure}[t]{0.42\textwidth}
		\includegraphics[scale=0.55]{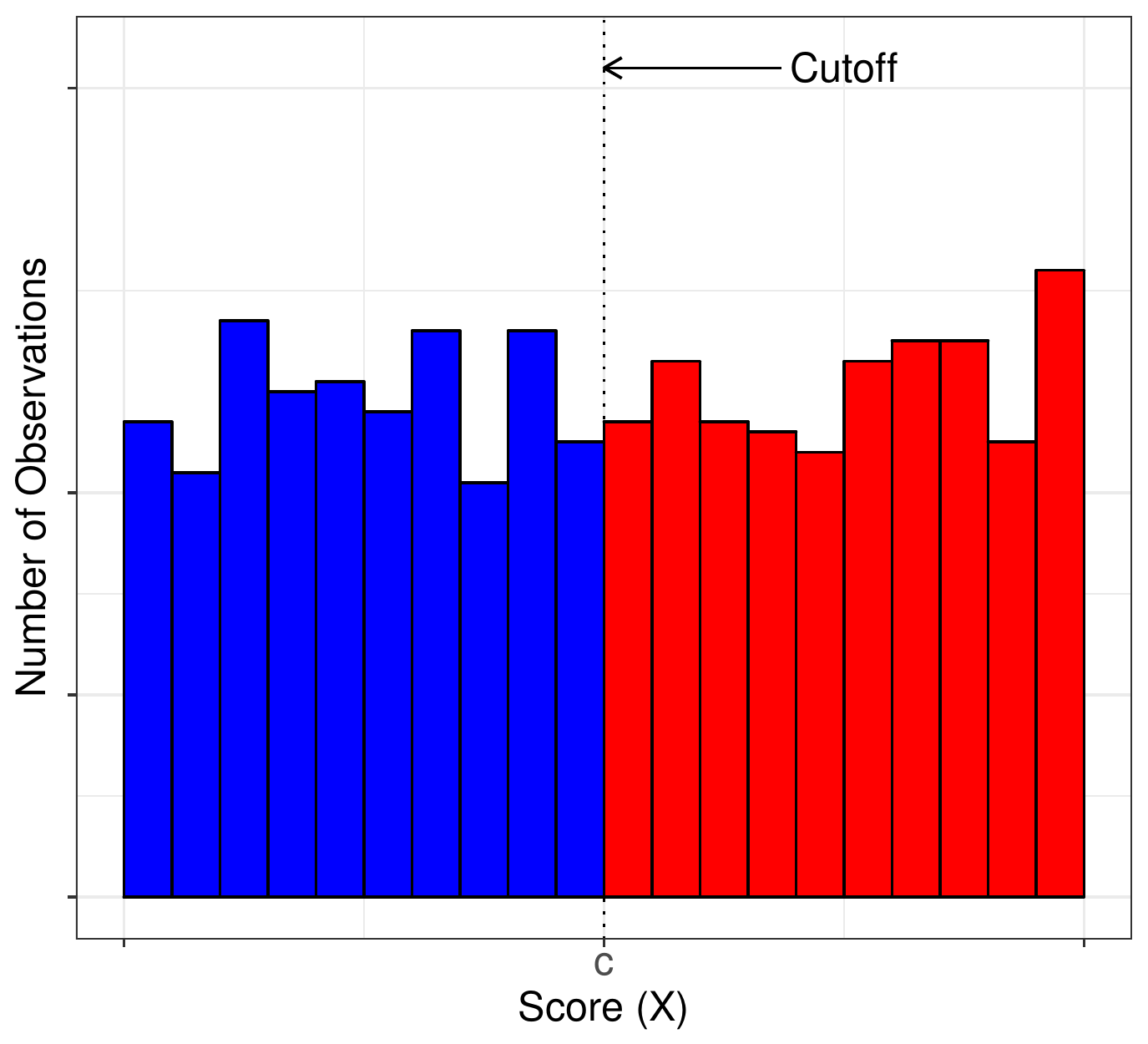}
		\caption{No Sorting}
        \label{nosort}		
	\end{subfigure}
		\hspace{0.5in}%
	\begin{subfigure}[t]{0.45\textwidth}
		\includegraphics[scale=0.55]{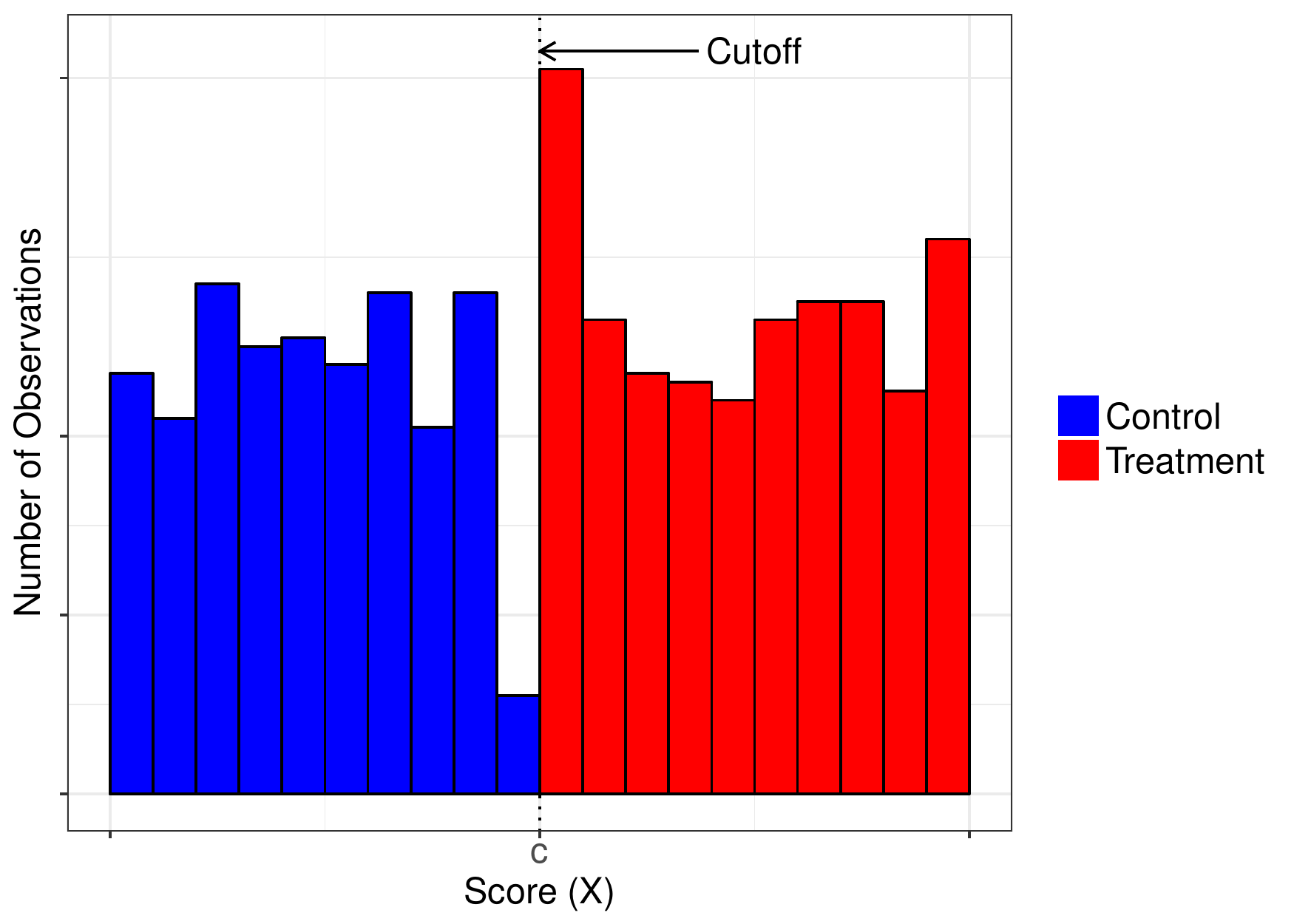}
		\caption{Sorting}
		\label{sort}
	\end{subfigure}
\caption{Histogram of Score}\label{fig:RD-density}
\end{figure}

Figure \ref{fig:RD-density} shows a histogram of the running variable in two hypothetical RD examples. In the scenario illustrated in Figure \ref{fig:RD-density}(a), the number of observations above and below the cutoff is very similar. In contrast, Figure \ref{fig:RD-density}(b) illustrates a case in which the density of the score right below the cutoff is considerably lower than just above it---a finding that suggests that units were able to systematically increase the value of their original score to be assigned to the treatment instead of the control group. 
	
\labelfiguras{figR}

In addition to a graphical illustration of the density of the running variable, researchers should explore the assumption more formally using a statistical test, often called a density test. One possible strategy is to choose a small neighborhood around the cutoff, and perform a simple Bernoulli test within that neighborhood with a probability of ``success'' equal to 1/2. This strategy tests whether the number of treated observations in the chosen neighborhood is compatible with what would have been observed if units had been assigned to the treatment group (i.e., to being above the cutoff) with a 50\% probability. The test is finite sample exact, under the assumptions imposed.

For example, if we keep only the observations with $X_i \in [-2,2]$ in the Meyersson application, we find that in this neighborhood there are 47 control observations and 53 treated observations. Using this information and setting a probability of success equal to 1/2, we can perform a binomial test using standard functions in \texttt{R} or \texttt{Stata}.

\labelsnippet{snippetLrdcontWII}
\rsnip{Vol-1-R_meyersson_falsification_binomial_byhand_adhoc.txt}{\Rlink{\thesnippetLrdcontWII}}
\statasnip{Vol-1-STATA_meyersson_falsification_binomial_byhand_adhoc}{\Slink{\thesnippetLrdcontWII}}

The p-value is 0.6173, so this simple test finds no evidence of ``sorting'' around the cutoff in this neighborhood: the numbers of treated and control observations are consistent with what would be expected if municipalities were assigned to an Islamic win or loss by the flip of an unbiased coin. 

In a continuity-based approach, however, there are often not clear guidelines about how to choose the neighborhood where the binomial test should be conducted. Nevertheless, it is natural to conduct this test for different (nested) neighborhoods around the cutoff. Furthermore, the use of this randomization-based test is also natural in the context of a local randomization approach to RD analysis, which we discuss extensively in the accompanying Element (\citeauthor*{Cattaneo-Idrobo-Titiunik_2019_Vol2}, forthcoming).

A complementary approach is to conduct a test of the null hypothesis that the density of the running variable is continuous at the cutoff, which fits naturally into the continuity-based framework adopted in this Element. The implementation of this test requires the estimation of the density of observations near the cutoff, separately for observations above and below the cutoff. We employ here an implementation based on a local polynomial density estimator that does not require pre-binning of the data and leads to size and power improvements relative to other approaches. The null hypothesis is that there is no ``manipulation'' of the density at the cutoff, formally stated as continuity of the density functions for control and treatment units at the cutoff. Therefore, failing to reject implies that there is no statistical evidence of manipulation at the cutoff, and offers evidence supporting the validity of the RD design.

We implement this density test using the Meyersson data using the \texttt{rddensity} command, which is part of the \texttt{rddensity} library/package. Its only required argument is the running variable.

\labelsnippet{snippetLrdcontX}
\rsnip{Vol-1-R_meyersson_falsification_rddensity.txt}{\Rlink{\thesnippetLrdcontX}}
\statasnip{Vol-1-STATA_meyersson_falsification_rddensity}{\Slink{\thesnippetLrdcontX}}

\labelfiguras{figS}
\begin{figure}[ht]
\caption{Histogram and Estimated Density of the Score}\label{fig:rddensity}
	\centering
	\begin{subfigure}[t]{0.48\textwidth}
		\includegraphics[scale=0.5]{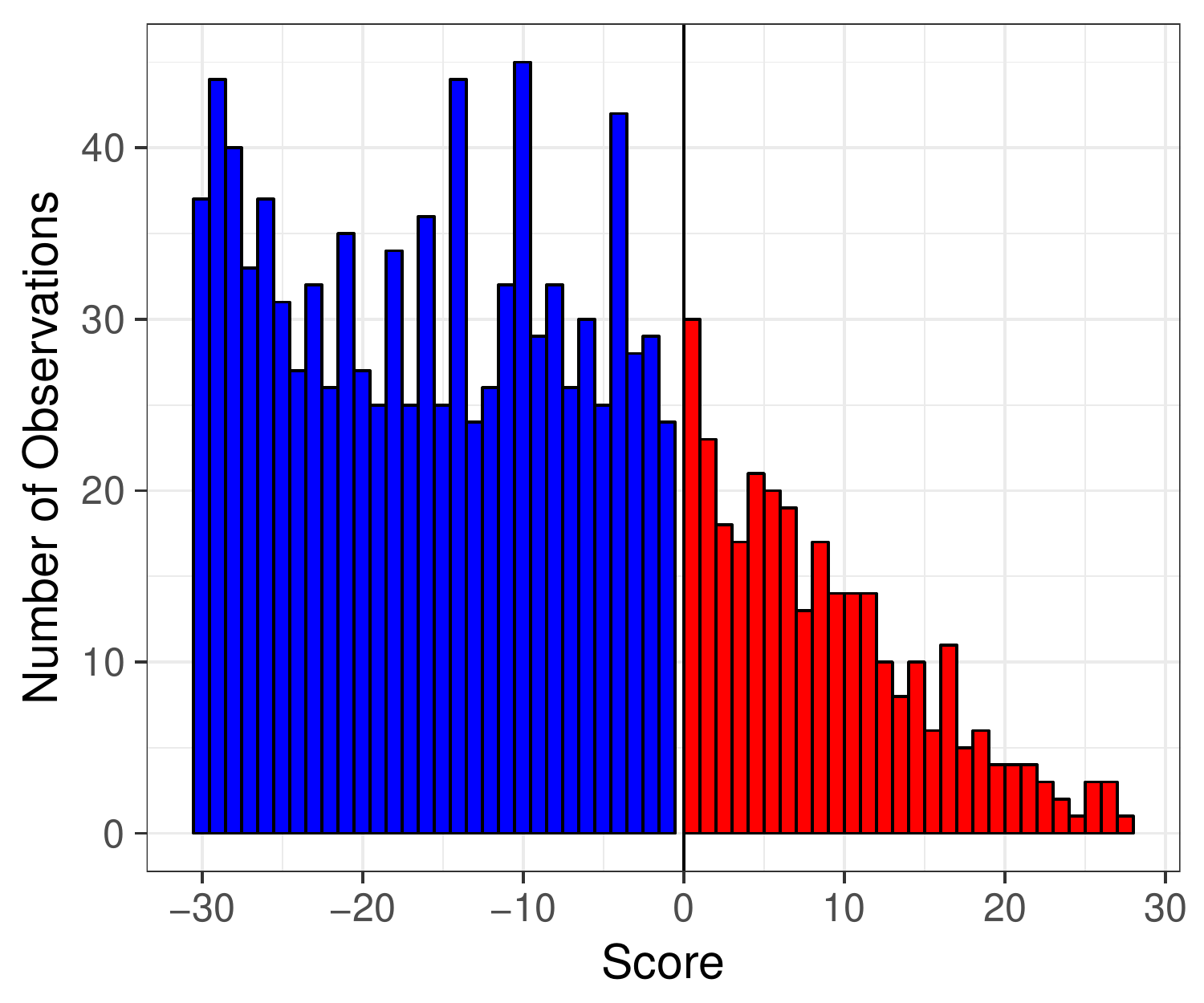}
		\caption{Histogram}\label{fig:histogram}		
	\end{subfigure}
\hspace{0.1in}%
	\begin{subfigure}[t]{0.48\textwidth}
		\includegraphics[scale=0.525]{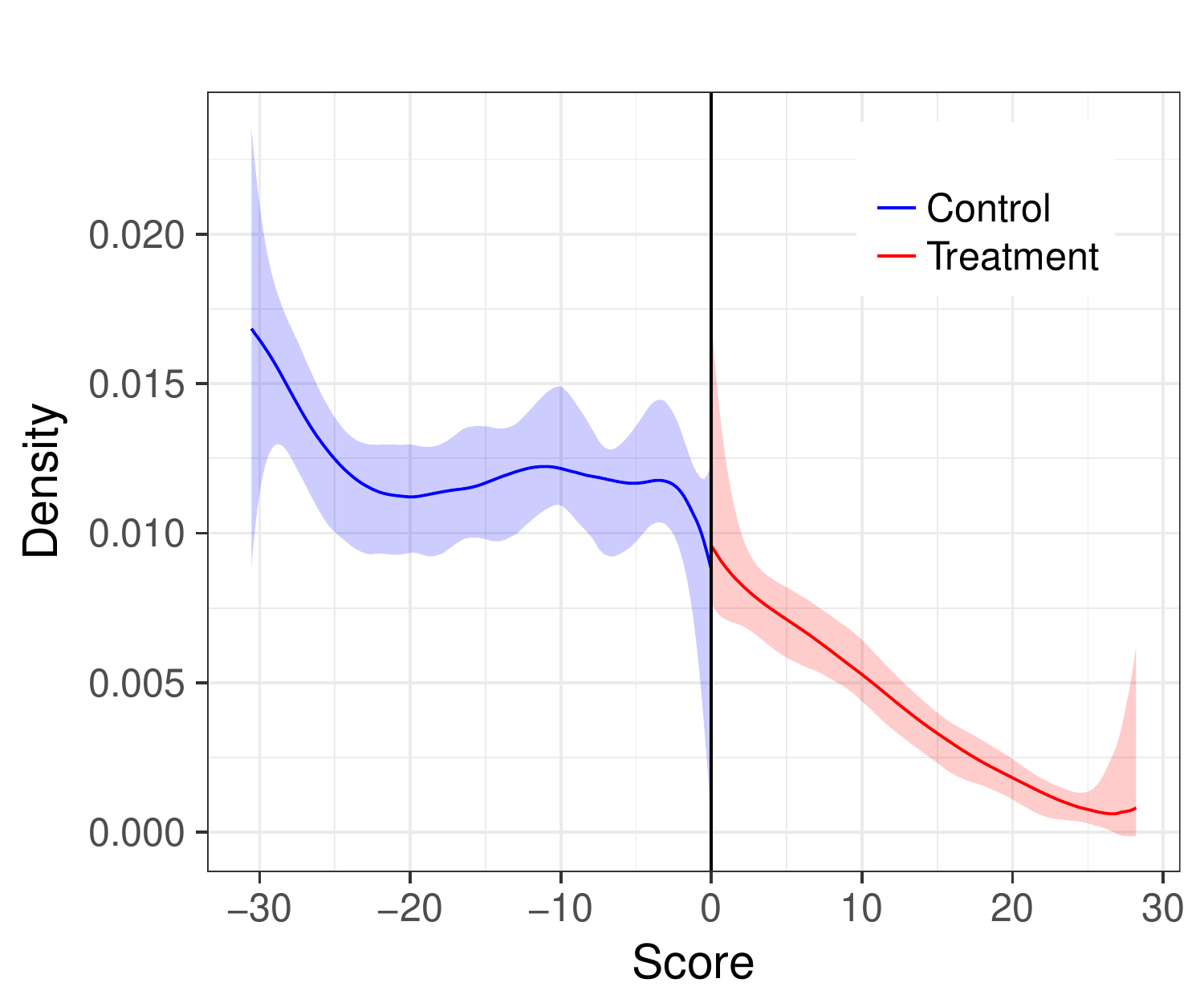}
		\caption{Estimated Density}\label{fig:lpdensity}
	\end{subfigure}
\end{figure}

The value of the statistic is $-1.394$ and the associated p-value is $0.1633$. This means that under the continuity-based approach, we fail to reject the null hypothesis of no difference in the density of treated and control observations at the cutoff. Figure \ref{fig:rddensity} provides a graphical representation of the continuity in density test approach, exhibiting both a histogram of the data and the actual density estimate with shaded 95\% confidence intervals. As we can see in \ref{fig:rddensity}(b), the density estimates for treated and control groups at the cutoff (the two intercepts in the figure) are very near each other, and the confidence intervals (shaded areas) overlap. This plot is consistent with the results from the formal test. 

\subsection{Placebo Cutoffs}
Another useful falsification analysis examines treatment effects at artificial or placebo cutoff values. To understand the motivation behind this falsification test, recall that the key RD identifying assumption is the continuity (or lack of abrupt changes) of the regression functions for treatment and control units at the cutoff in the absence of the treatment. While such a condition is fundamentally untestable at the cutoff, researchers can investigate empirically whether the estimable regression functions for control and treatment units are continuous at points other than the cutoff. Evidence of continuity away from the cutoff is, of course, neither necessary nor sufficient for continuity at the cutoff, but the presence of discontinuities away from the cutoff can be interpreted as potentially casting doubt on the RD design, at the very least in cases where such discontinuities can not be explained by substantive knowledge of the specific application. Another related use of this approach is to check whether the smoothness and other conditions needed for RD inference are supported by the data, at least in regions other than at the cutoff point.

This test replaces the true cutoff value by another value at which the treatment status does not really change, and performs estimation and inference using this artificial cutoff point. The expectation is that no significant treatment effect will occur at placebo cutoff values. A graphical implementation of this falsification approach follows directly from the RD plots discussed extensively in Section \ref{sec:graph}, by simply assessing whether there are jumps in the observed regression functions at points other than the true cutoff. A more formal implementation of this idea conducts statistical estimation and inference for RD treatment effects at artificial cutoff points, using control and treatment units separately. In the continuity-based framework adopted in this Element, we implement this approach using local-polynomial methods within an optimally-chosen bandwidth around the artificial cutoff to estimate treatment effects on the outcome, as we explained in Section \ref{sec:localpoly}.

In order to illustrate the procedure with the Meyersson data, we employ \texttt{rdrobust} after restricting to the appropriate group and specifying the artificial cutoff. To avoid ``contamination'' due to real treatment effects, for artificial cutoffs above the real cutoff we use only treated observations, and for artificial cutoffs below the real cutoff we use only control observations. Restricting the observations in this way guarantees that the analysis of each placebo cutoff uses only observations with the same treatment status. Thus, by construction, the treatment effect at each artificial cutoff should be zero.

We conduct estimation and inference at the artificial cutoff $\C = 1$ in the Meyersson application, using the option \texttt{c = 1} in \texttt{rdrobust} and including only treated observations. Our analysis thus compares the educational outcomes of municipalities where Islamic mayors won by a margin of $1\%$ or more, to municipalities where Islamic mayors won by less than $1\%$. Since there is an Islamic mayor on both sides of the cutoff, we expect to see no discontinuity in the outcome at $1\%$.

\labelsnippet{snippetLrdcontY}
\rsnip{Vol-1-R_meyersson_falsification_rdrobust_alternative-cutoff_c1.txt}{\Rlink{\thesnippetLrdcontY}}
\statasnip{Vol-1-STATA_meyersson_falsification_rdrobust_alternative-cutoff_c1}{\Slink{\thesnippetLrdcontY}}

\begin{figure}[ht]
\centering
\includegraphics[scale=0.85]{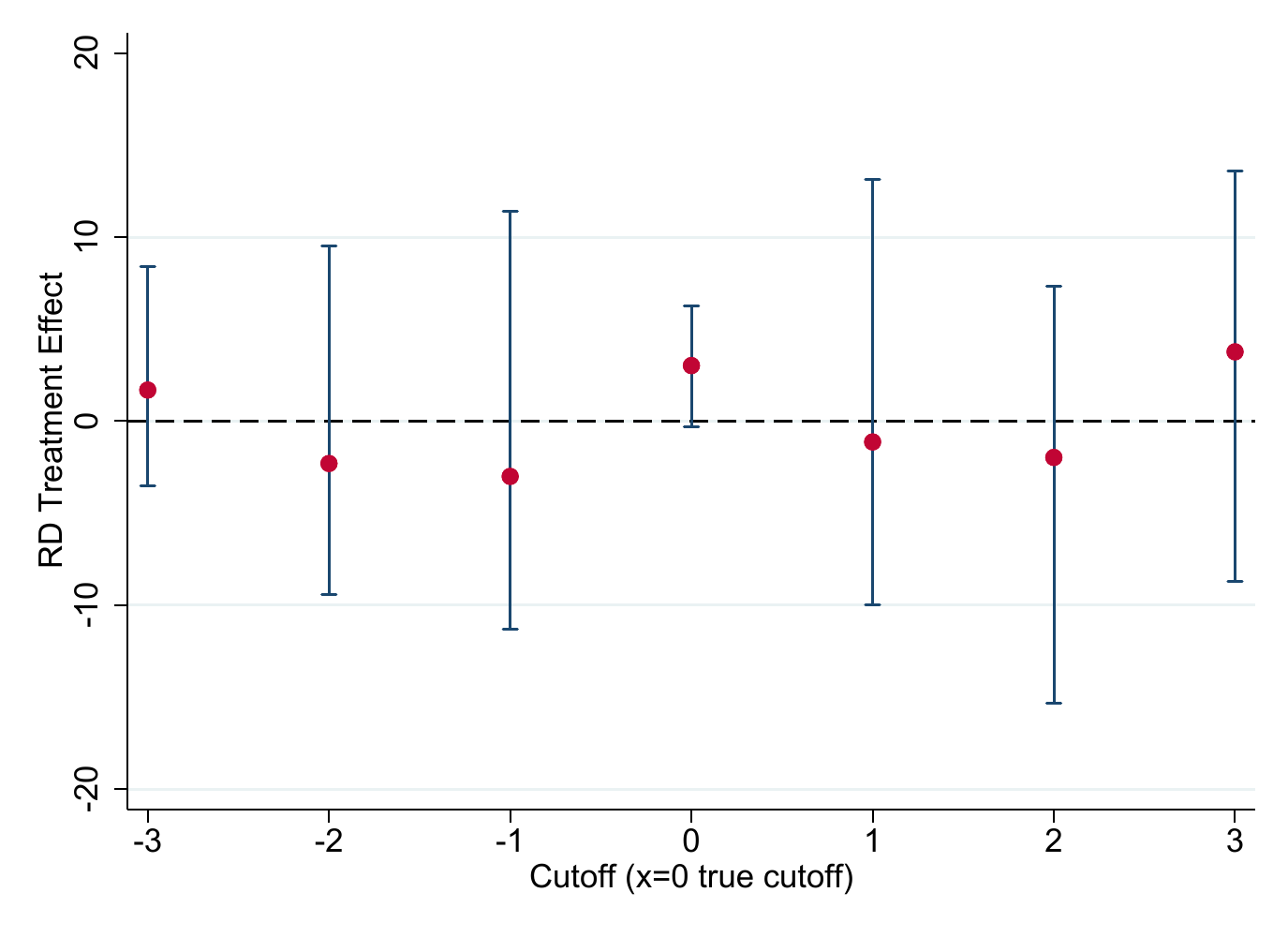}
\caption{RD Estimation for True and Artificial Cutoffs}
\label{fig:placebocutoffs}
\end{figure}

The robust p-value is 0.787, consistent with the conclusion that the outcome of interest does not jump at the artificial $1\%$ cutoff, and in contrast to the results at the true cutoff reported in Section \ref{sec:localpoly}. Table \ref{tab:rdrobust_alt-cutoffs} presents the results of similar analyses for other placebo cutoffs ranging from $-5\%$ to $5\%$ in increments of $1\%$. Figure \ref{fig:placebocutoffs} graphically illustrates the main results from this falsification test.

\labeltablas{tableE}
\begin{table}[H]
\centering
\caption{Continuity-Based Analysis for Alternative Cutoffs}
\resizebox{\textwidth}{!}{\begin{tabular}{ccccccc}
\toprule
Alternative & MSE-Optimal & RD        & \multicolumn{2}{c}{\underline{Robust Inference}} & \multicolumn{2}{c}{N. of Obs.} \\
Cutoff      & Bandwidth   & Estimator & p-value & Conf. Int.                             & Left & Right\\
\midrule
$-3$ & $3.934$ & $1.688$ & $0.421$ & [$-3.509$, $8.397$] & $135$ & $74$\\ 
$-2$ & $4.642$ & $-2.300$ & $0.991$ & [$-9.414$, $9.518$] & $152$ & $47$\\ 
$-1$ & $4.510$ & $-3.003$ & $0.992$ & [$-11.295$, $11.409$] & $139$ & $24$\\ 
$0$ & $17.239$ & $3.020$ & $0.076$ & [$-0.309$, $6.276$] & $529$ & $266$\\ 
$1$ & $2.362$ & $-1.131$ & $0.787$ & [$-9.967$, $13.147$] & $30$ & $49$\\ 
$2$ & $2.697$ & $-1.973$ & $0.488$ & [$-15.333$, $7.313$] & $53$ & $50$\\ 
$3$ & $2.850$ & $3.766$ & $0.668$ & [$-8.700$, $13.569$] & $68$ & $56$\\ 

\bottomrule
\end{tabular}}
\label{tab:rdrobust_alt-cutoffs}
\end{table}

\labelfiguras{figT}

The true cutoff of 0 is included in order to have a benchmark to compare---the particular results regarding the true cutoff were discussed at length in Section \ref{sec:localpoly}. All other cutoffs are artificial or placebo, in the sense that treatment did not actually change at those points. We find that in all but one of the artificial cutoff points, the RD point estimator is smaller in absolute value than the true RD estimate (3.020), and that all p-values are above 0.4. Therefore, we conclude that the outcome of interest does not jump discontinuously at the artificial cutoffs considered. 

\subsection{Sensitivity to Observations near the Cutoff}

Another falsification approach seeks to investigate how sensitive the results are to the response of units who are located very close to the cutoff. If systematic manipulation of score values has occurred, it is natural to assume that the units closest to the cutoff are those most likely to have engaged in manipulation. The idea behind this approach is to exclude such units and then  repeat the estimation and inference analysis using the remaining sample. This idea is sometimes referred to as a ``donut hole'' approach. Even when manipulation of the score  is not suspected, this strategy is also useful to assess the sensitivity of the results to the unavoidable extrapolation involved in local polynomial estimation, as the few observations closest to the cutoff are likely to be the most influential when fitting the local polynomials.

For implementation in the continuity-based approach, we use \texttt{rdrobust} after subsetting the data. For example, in the Meyersson application, we consider first the case where units with score $|X_i| < 0.3$ are excluded from the analysis, which requires us to engage in more extrapolation than before. The exclusion of observations implies that a new optimal bandwidth will be selected. 

\labelsnippet{snippetLrdcontZ}
\rsnip{Vol-1-R_meyersson_falsification_rdrobust_donuthole.txt}{\Rlink{\thesnippetLrdcontZ}}
\statasnip{Vol-1-STATA_meyersson_falsification_rdrobust_donuthole}{\Slink{\thesnippetLrdcontZ}}

The results show that the conclusions from the analysis are robust to excluding observations with $|X_i| < 0.3$. In the new analysis, we have 2307 total observations to the left of the cutoff, and 309 total observations to the right of it. As expected, these numbers are smaller than those employed in the original analysis (2314 and 315). Note that, although the total number of observations will always decrease when observations closest to the cutoff are excluded, the effective number of observations used in the analysis may increase or decrease, depending on how the bandwidth changes. In this case, the bandwidth changes from 17.239 in the original analysis to 16.043 in the analysis that excludes units with $|X_i| < 0.30$; this results in a loss of 65 effective observations, from 795 (529 + 266) to 730 (482 + 248), which is much larger than the decrease in total observations, which is only 13. The exclusion of these observations changes the point estimate from 3.020 to 3.414, and the robust confidence interval from $[-0.309,6.276]$ to $[-0.067,6.965]$. The conclusion of the analysis remains largely unchanged, however, since both the original and the new estimated effect are significant at 10\% level.

\begin{figure}[H]
\centering
\caption{RD Estimation for the Donut-Hole Approach}
\includegraphics[scale=0.85]{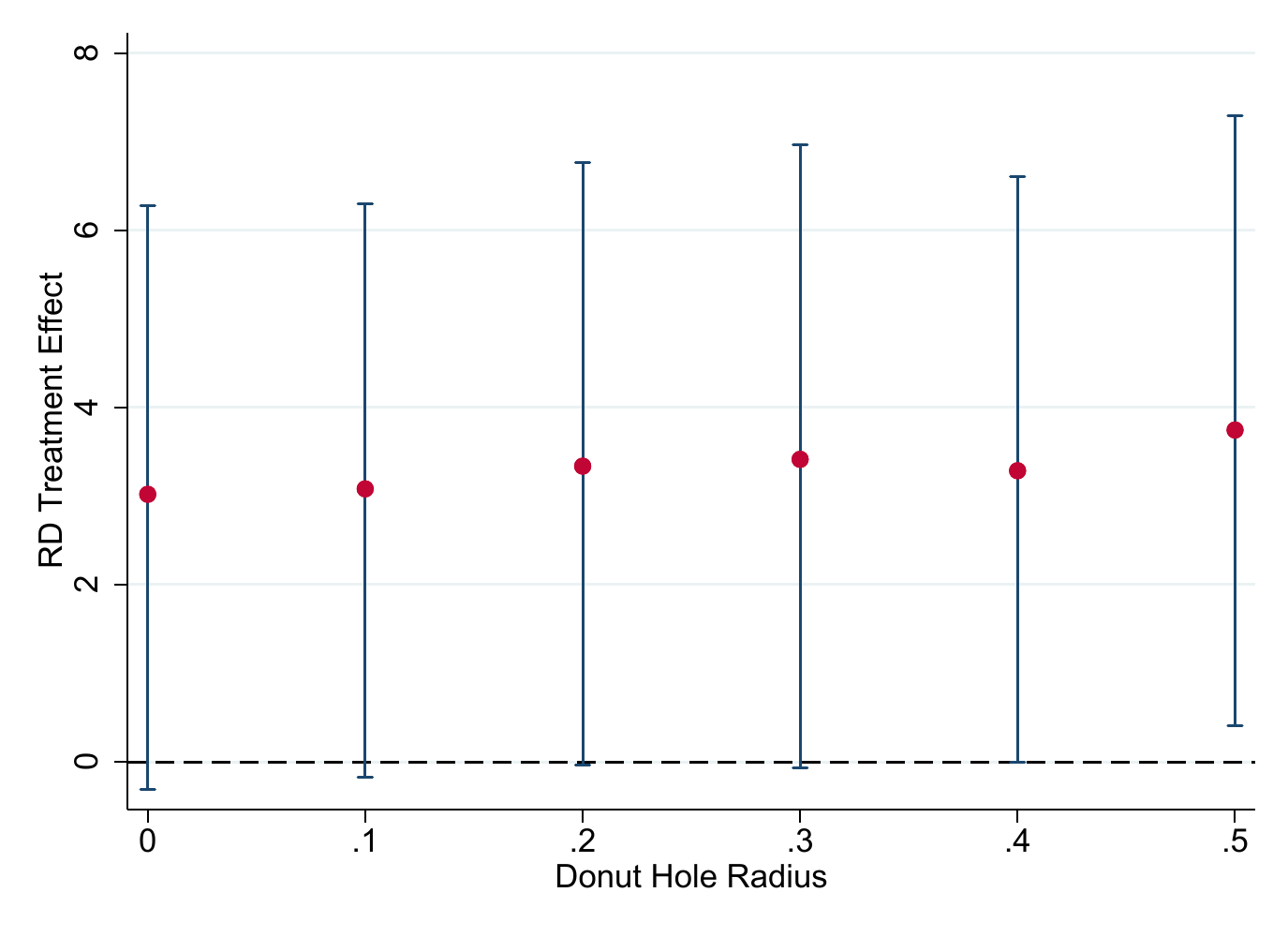}
\label{fig:donuthole}
\end{figure}

In practice, it is natural to repeat this exercise a few times to assess the actual sensitivity for different amounts of excluded units. Table \ref{tab:donuthole} illustrates this approach, and Figure \ref{fig:donuthole} depicts the results graphically. In all the cases considered, the conclusions remain unchanged. 

\labeltablas{tableF}
\begin{table}[H]
\centering
\caption{Continuity-Based Analysis for the Donut-Hole Approach}
\resizebox{\textwidth}{!}{\begin{tabular}{cccccccc}
\toprule
Donut-Hole  & MSE-Optimal & RD        & \multicolumn{2}{c}{\underline{Robust Inference}}  & Number of    & \multicolumn{2}{c}{Excluded Obs.}\\
Radius      & Bandwidth   & Estimator & p-value & Conf. Int.                              & Observations & Left & Right \\
\midrule
0.00 & 17.239 & 3.020 & 0.076 & [-0.309, 6.276] & 795 & 0 & 0\\ 
0.10 & 17.954 & 3.081 & 0.064 & [-0.175, 6.298] & 815 & 1 & 1\\ 
0.20 & 16.621 & 3.337 & 0.052 & [-0.033, 6.759] & 765 & 5 & 4\\ 
0.30 & 16.043 & 3.414 & 0.055 & [-0.067, 6.965] & 730 & 7 & 6\\ 
0.40 & 17.164 & 3.286 & 0.050 & [-0.001, 6.601] & 774 & 9 & 9\\ 
0.50 & 15.422 & 3.745 & 0.028 & [0.408, 7.292] & 697 & 13 & 14\\ 

\bottomrule
\end{tabular}}
\label{tab:donuthole}
\end{table}

\labelfiguras{figU}

\subsection{Sensitivity to Bandwidth Choice}
The last falsification method we discuss analyzes the sensitivity of the results to the bandwidth choice. In contrast to the donut hole approach, which investigates sensitivity as units from the center of the neighborhood around the cutoff are removed, the method we discuss now investigates sensitivity as units are added or removed at the end points of the neighborhood. The implementation of this method is also straightforward, as it requires employing local polynomial methods with different bandwidth choices. However, the interpretation of the results must be done with care. As we discussed throughout this Element, choosing the bandwidth is one of the most consequential decisions in RD analysis, because the bandwidth may affect the results and conclusions. 

In the continuity-based approach, this falsification test is implemented by changing the bandwidth used for local polynomial estimation. It is well understood how the bandwidth will affect the results: as the bandwidth increases, the bias of the local polynomial estimator increases and its variance decreases. Thus, it is natural to expect that, as we increase the bandwidth, the confidence intervals will decrease in length but will also be displaced (because of the bias).

The considerations above suggest that when the goal is to interpret point estimators, investigating the sensitivity to bandwidth choices is only useful over small ranges around the MSE-optimal bandwidth; otherwise, the results will be mechanically determined by the statistical properties of the estimation and inference methods. In other words, bandwidths much larger than the MSE-optimal bandwidth will lead to estimated RD effects that have too much bias, and bandwidths much smaller than the MSE-optimal choice will lead to RD effects with too much variance. In both cases, point estimation will be unreliable, and so will be the conclusions from the falsification test. Similarly, if the emphasis is on optimal inference, the sensitivity of the results should only be explored for bandwidth values near the CER-optimal choice.

\begin{figure}[ht]
\centering
\caption{Sensitivity to Bandwidth in the Continuity-Based Approach}
\includegraphics[scale=0.85]{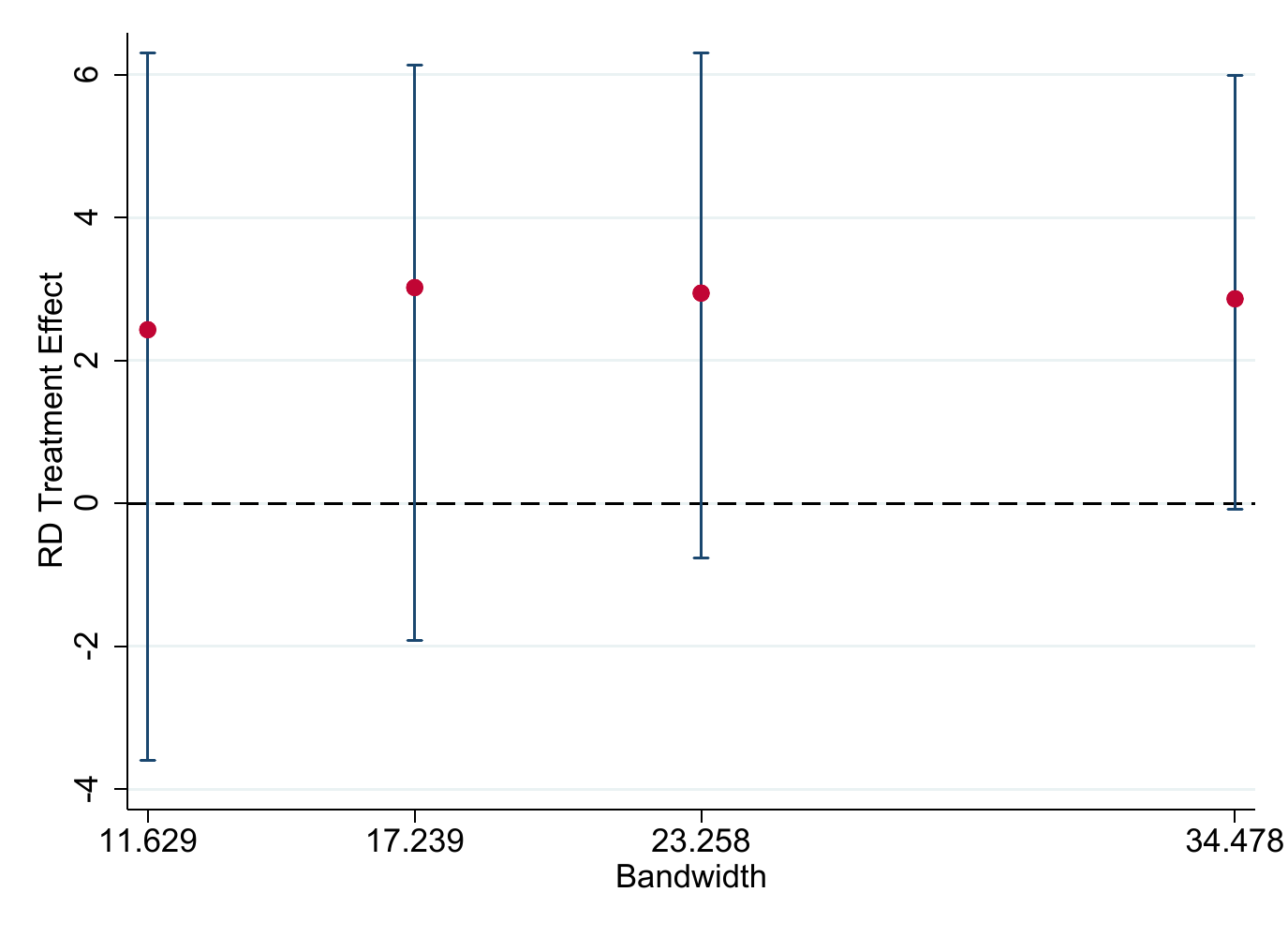}
\label{fig:sensitivity-BW}
\end{figure}

We illustrate this sensitivity approach with the Meyersson data for four bandwidth choices close to the MSE-optimal and CER-optimal choices: (i) the CER-optimal choice $h_\mathtt{CER}=11.629$, (ii) the MSE-optimal choice $h_\mathtt{MSE}=17.239$, (iii) $2\cdot h_\mathtt{CER}=23.258$, and (iv) $2\cdot h_\mathtt{MSE}=34.478$. Figure \ref{fig:sensitivity-BW} shows the local polynomial RD point estimators and robust 95\% confidence intervals for each bandwidth. The code is omitted, but is included in the replication files. 

\labelfiguras{figV}

The results based on the CER-optimal choice $h_\mathtt{CER}=11.629$ are consistent with the results based on the MSE-optimal choice $h_\mathtt{MSE}=17.239$ in that they both lead to a similar point estimate, but the CER-optimal choice results in a longer confidence interval according to which the effect cannot be distinguished from zero at conventional levels. The two largest bandwidths, $2\cdot h_\mathtt{CER}=23.258$ and $2\cdot h_\mathtt{MSE}=34.478$, lead to results that are broadly consistent with the empirical findings obtained with the MSE-optimal choice. 

\subsection{Further Reading}

The density test to detect RD manipulation was first proposed by \citet{McCrary_2008_JoE}. \citet*{Cattaneo-Jansson-Ma_2019_wp} develop the local polynomial density estimator implemented in \texttt{rddensity}; see also \citet*{Cattaneo-Jansson-Ma_2018_Stata} for details on this statistical package and further numerical evidence. \citet{Frandsen_2017_AIE} develops a related manipulation test for cases where the score is discrete. The importance of falsification tests and the use of placebo outcomes is generally discussed in the analysis of experiments literature \citep*[e.g.,][]{Imbens-Rubin_2015_Book,Rosenbaum_2002_Book,Rosenbaum_2010_Book}. \citet{Lee_2008_JoE} applies and extends these ideas to the context of RD designs, and \citet{Canay-Kamat_2018_ReStud} develop a permutation inference approach in the same context. \citet*{Ganong-Jager_2018_JASA} develop a permutation inference approach based on the idea of placebo RD cutoffs for the Kink RD designs, Regression Kink designs, and related settings. Finally, falsification testing based on donut hole specifications is discussed in \citet*{Bajari-Hong-Park-Town_2011_wp} and \citet*{Barreca-Lindo-Waddell_2016_EI}, among others.

\newpage
\section{Final Remarks}

We have discussed foundational aspects of identification, estimation, inference, and falsification in the Sharp RD design, when the parameter of interest is the average treatment effect at the cutoff. Because our goal in this Element was to discuss the conceptual foundations of RD methodology, we focused on the simplest possible case where (i) there is a single running variable, (ii) there is a single cutoff, (iii) compliance with treatment assignment is perfect, (iv) the running variable is continuous and hence has no mass points, (v) the object of interest is the average treatment effect at the cutoff, and (vi) results are based on continuity and smoothness assumptions. This canonical setup is the most standard and commonly encountered in empirical work dealing with RD designs.

In the accompanying Element (\citeauthor*{Cattaneo-Idrobo-Titiunik_2019_Vol2}, forthcoming), we discuss several departures from the canonical Sharp RD design setting. The first topic we consider is an alternative interpretation of the RD design based on the idea of local random assignment. In contrast to the continuity-based approach adopted in this Element, the local randomization approach assumes that there is a window around the cutoff where the treatment can be assumed to have been as-if randomly assigned, and the analysis proceeds by adopting the usual tools from the analysis of experiments. This approach is also well suited to analyze RD designs where the running variable is discrete with relatively few mass points, a situation that occurs often in practice and we also discuss in detail. Additional topics covered in the accompanying Element include the Fuzzy RD design, where compliance with treatment is imperfect, RD settings with multiple running variables, which have as an important special case the geographic RD design where treatment assignment depends on the spatial distance to the border between geographic regions, and RD setups where treatment assignment depends on multiple cutoffs instead of only one.

We hope that the discussion in this Element, together with the additional methods presented in the accompanying Element, will provide a useful and practical template to guide applied researchers in analyzing and interpreting RD designs in a principled, rigorous, and transparent way.

\newpage
\bibliographystyle{econometrica}
\bibliography{Cattaneo-Idrobo-Titiunik_2019_CUP}

\newpage
\appendix

\end{document}